\documentclass[acmsmall]{acmart}
\usepackage{makecell}
\usepackage{CJKutf8}
\usepackage[utf8]{inputenc}
\usepackage[T1]{fontenc}
\usepackage{textcomp}
\usepackage{multirow}
\usepackage{multicol}
\usepackage{threeparttable} 

\AtBeginDocument{%
  \providecommand\BibTeX{{%
    \normalfont B\kern-0.5em{\scshape i\kern-0.25em b}\kern-0.8em\TeX}}}

\setcopyright{acmcopyright}
\acmJournal{PACMHCI}
\acmYear{2024} \acmVolume{8} \acmNumber{CSCW2} \acmArticle{01} \acmMonth{01} \acmPrice{15.00}\acmDOI{XX.XXXX/XXXXXXX}

\begin{document}

\title[Images Connect Us Together]{Images Connect Us Together: Navigating a COVID-19 Local Outbreak in China Through Social Media Images}

\author{Changyang He}
\affiliation{%
  \institution{Department of Computer Science and Engineering, Hong Kong University of Science and Technology}
  \city{Hong Kong SAR}
  \country{China}
}
\email{cheai@cse.ust.hk}

\author{Lu He}
\affiliation{%
  \institution{Zilber College of Public Health, University of Wisconsin-Milwaukee}
  \city{Milwaukee}
  \country{USA}
}
\email{he32@uwm.edu}

\author{Wenjie Yang}
\affiliation{%
  \institution{Department of Computer Science and Engineering, Hong Kong University of Science and Technology}
  \city{Hong Kong SAR}
  \country{China}
}
\email{wyangbc@connect.ust.hk}

\author{Bo Li}
\affiliation{%
  \institution{Department of Computer Science and Engineering, Hong Kong University of Science and Technology}
  \city{Hong Kong SAR}
  \country{China}
}
\email{bli@cse.ust.hk}

\renewcommand{\shortauthors}{Changyang He et al.}

\begin{abstract}

Social media images, curated or casual, have become a crucial component of communicating situational information and emotions during health crises. Despite its prevalence and significance in informational dissemination and emotional connection, there lacks a comprehensive understanding of visual crisis communication in the aftermath of a pandemic which is characterized by uncertain local situations and emotional fatigue. To fill this gap, this work collected 345,423 crisis-related posts and 65,376 original images during the Xi'an COVID-19 local outbreak in China, and adopted a mixed-methods approach to understanding themes, goals, and strategies of crisis imagery. Image clustering captured the diversity of visual themes during the outbreak, such as text images embedding authoritative guidelines and ``visual diaries'' recording and sharing the quarantine life. Through text classification of the post that visuals were situated in, we found that different visual themes highly correlated with the informational and emotional goals of the post text, such as adopting text images to convey the latest policies and sharing food images to express anxiety. We further unpacked nuanced strategies of crisis image use through inductive coding, such as signifying authority and triggering empathy. We discuss the opportunities and challenges of crisis imagery and provide design implications to facilitate effective visual crisis communication.

\end{abstract}



\maketitle

\section{INTRODUCTION}

Social media has become a crucial information channel during crises. Characterized by de-centralized communication and crowdsourced information creation, social media affords a prominent place where people gather to cultivate situational awareness~\cite{kou2017conspiracy, leavitt2017role, dailey2015s}, route help-seeking requests~\cite{starbird2011voluntweeters, yang2022save, he2022help} and provide mutual emotional support~\cite{he2021beyond,yi2022depicting,huang2015connected}. As the development of camera-embedded smartphones and high-speed internet eases the creation and sharing of images online, \textbf{social media imagery} is an increasingly substantial component in crisis communication. Millions of images are generated across different disasters~\cite{alam2018crisismmd} and transmit informative crisis-related messages~\cite{sleigh2021qualitative}.

Among different modalities in crisis communication, imagery has manifested a unique capability in delivering certain kinds of knowledge, sentiments, and experiences during crises~\cite{perovich2022self,bica2019communicating,bica2017visual,li2018localizing,zhang2021mapping,sosea2021using,seo2014visual}. For example, crisis images in COVID-19, whether formal visualizations~\cite{zhang2021mapping} or non-formal memes~\cite{perovich2022self}, could effectively engage viewers with crisis information such as disease prevalence, and facilitate public decision making. Through specific visual elements, crisis imagery can also have an emotional impact on the audience~\cite{perovich2022self,joffe2008power}, with the power of emotional persuasion~\cite{joffe2008power} or emotional connection establishment~\cite{bica2017visual,porter2020visual}. A visual representation is also helpful for reaching and engaging the audience with relatively low literacy levels~\cite{guttman2017ethical} and could facilitate their sensemaking under crises. Nonetheless, the majority of work in crisis communication has exclusively focused on textual content for analysis. The understanding of visual crisis communication through social media imagery, and the interplay between crisis text and images, are still limited.

The informational and emotional richness of crisis imagery has great potential in supporting crisis response during \textit{local outbreaks in the aftermath of the pandemic} which are local situation-centered and emotionally fatigued~\cite{williams2021variant,brownson2020reimagining,abouk2021immediate}. In the aftermath of the pandemic, local outbreaks are common~\cite{localOutbreak,localOutbreakChina}, unpredictable, and affect local policies and residents' daily life~\cite{brownson2020reimagining}, which poses new challenges for crisis response. People not only need to cultivate situational awareness of the rapidly shifting local policies~\cite{abouk2021immediate,wu2021characterizing}, but also have to retract to more constrained everyday behaviors and potentially suffer from crisis fatigue~\cite{williams2021variant}. Further, local outbreaks are characterized by the information barrier, with segregated perceptions of crises between the affected local populations and the general public~\cite{tsai2021help}, which demands persuasive crisis communication to get attention from the outside world. Given the unique value of social media crisis imagery in information persuasion and emotional contagion, understanding the use of crisis imagery may shed light on how people utilize visual narratives to cope with these specific challenges of crisis communication in local outbreaks. In particular, unpacking which visual themes are presented in local outbreaks, and how people strategically adopt them to share situational information and vent emotions beyond textual narratives, may provide insight into efficient and effective multimodal crisis communication in the aftermath of the pandemic.

To fill this significant research gap, we propose the following research questions:

\begin{itemize}

  \item \textbf{RQ1}: \textit{What} themes of images do users share on social media during a COVID-19 local outbreak? (\textbf{Crisis Visual Themes})

  \item \textbf{RQ2}: \textit{For what} crisis communication goals do users share crisis images during the local outbreak? (\textbf{Crisis Visual Goals})
  
  \item \textbf{RQ3}: \textit{How} does the use of crisis images strategically facilitate crisis communication during the local outbreak? (\textbf{Crisis Visual Strategies})

\end{itemize}

To answer the research questions, we adopted mixed-methods and multimodal analysis, incorporating image clustering, text classification, and inductive coding, to uncover themes, goals, and strategies of crisis imagery. We focused on the event of COVID-19 outbreak in Xi'an City in China, and collected 345,423 crisis-related posts and 65,376 original images from Weibo, a popular Chinese social media platform. We first used image clustering to distinguish diverse crisis visual themes such as text images and posters. By identifying information and emotions of crisis-related posts with images, we then revealed how the use of different visual themes significantly correlated with the informational and emotional goals in text, such as the proliferation of text images in disseminating the latest policies and the association between food image and anxiety in lockdown. Through the inductive coding of image-attached posts, we further identified four types of strategic use of crisis images to facilitate crisis communication, including images as signs of authority, visual-based information enhancement, evidence to improve credibility, and triggers for empathy. We discuss opportunities (e.g., establishing emotional connections) and challenges (e.g., image-enhanced misinformation) of image-based crisis communication, reflect on the complementary roles of crisis images and language, and propose design implications to promote effective and accurate crisis communication through social media crisis images.

In summary, the contribution of this work to crisis communication literature in HCI and CSCW includes: (1) we propose an effective analytical structure to identify diverse crisis image themes, and capture a comprehensive taxonomy of visual representations in a COVID-19 local outbreak; (2) we uncover inter-modality correlations between text and images, demonstrating that crisis images of different types serve different informational and emotional goals; (3) we unveil nuanced and unique strategic use of social media crisis images to supplement or augment textual narratives, enhancing crisis communication as a whole; and (4) we provide design implications to facilitate accurate, efficient and effective visual crisis communication. This work sheds light on opportunities and challenges of visual-based crisis communication, and highlights the significance of understanding multimodal crisis communication as an organic whole.

\section{RELATED WORK}

\subsection{Crisis Communication on Social Media and Challenges in Local Outbreaks}

Crisis communication has been a crucial research topic that aims to understand and facilitate the communication of preparing for, responding to, and recovering from crises~\cite{coombs2020conceptualizing}. Traditional crisis communication on mass media largely highlights a one-way and top-down approach from authoritative agencies to the general public~\cite{veil2011work, zerman1995crisis}. The rapid development of social media has enabled two-way communication between the public and official agencies which emphasizes public participation and engagement in crisis response~\cite{eriksson2018lessons, zhang2021mapping, he2022help}. How users create, seek, and transmit crisis-related information on social media has attracted increasing research attention in the HCI and CSCW communities (e.g.,~\cite{li2021hello,olteanu2015expect, qu2011microblogging, vieweg2010microblogging, starbird2011voluntweeters}).

The two-way crisis communication on social media brings opportunities for crowdsourced information creation and dissemination~\cite{zade2018situational, he2021beyond}. Based on that, nuanced crisis communication patterns have been discovered and investigated by HCI and CSCW researchers. One line of work looks at how people cultivate situational awareness of a crisis through collective sensemaking~\cite{kou2017conspiracy, leavitt2014upvoting, leavitt2017role, dailey2015s, starbird2013delivering}. Under health crises like Zika and COVID-19 that are characterized by uncertainty, such a collective sensemaking process is substantial for risk assessment and decision making~\cite{gui2017managing}. Another strand of work focuses on self-organized collaboration in response to help-seeking requests when crises happen~\cite{starbird2011voluntweeters, yang2022save, he2022help, li2019using,zhao2020online, starbird2012promoting}. Online community members especially digital volunteers develop a set of strategies such as using specific hashtags~\cite{nishikawa2018time, yang2017harvey} and structuring posts with a specialized syntax~\cite{starbird2011voluntweeters,yang2022save,he2022help} to facilitate the timely routing of help-seeking posts to the intended target(s). Owing to misinformation and conspiracy theories often becoming widely spread during crises~\cite{starbird2014rumors,ferrara2020misinformation}, researchers have also investigated the debunking efforts and procedures to cope~\cite{yang2021know,he2022debunk}.

Local outbreaks in the aftermath of a pandemic, as a special type of public health crisis, pose unique challenges to citizens' crisis response. During local outbreaks, affected populations have to figure out and adapt to unpredictable local policies that influence their daily life~\cite{brownson2020reimagining, abouk2021immediate}. Also, local residents have to retract to constrain everyday behaviors, and may potentially suffer from fatigue with decreased motivation to comply with regulations~\cite{williams2021variant}. Moreover, in such local outbreaks, there are segregated perceptions of crises between the affected local populations and the general public, which demands persuasive crisis communication to cross the information barrier~\cite{tsai2021help}. On this note, multimodal crisis communication combining text and visual use may hold special value in transmitting situational information and building emotional connections. Nonetheless, little work has looked at multimodal crisis communication during local outbreaks with a holistic view of text and visuals. In this work, we aim to explore how people adopt visual narratives on social media along with text to satisfy these specific communication needs, given images' unique capacity for attention attraction, informational persuasion, and emotional contagion.

\subsection{Visual Crisis Communication}

Images inherently own distinguished power in delivering certain kinds of thoughts, experiences, emotions, and knowledge clearly and effectively, the role of which in communication could supplement written or oral language in many ways~\cite{gillies2005painting,powell2015clearer,schill2012visual}. For instance, images have the potential of arousing emotional resonance crossing the language and cultural divide~\cite{sawyer2012impact}. Recent technology development such as high-resolution mobile phone camera and low-cost digital image archives enables amateurs to curate and share images on social media conveniently~\cite{cooke2005visual,porter2020visual}, which make social media imagery a rich research source~\cite{manikonda2017modeling}. A plethora of work has examined imagery on social media in various areas such as mental health~\cite{manikonda2017modeling}, societal happiness~\cite{abdullah2015collective}, abusive behaviors~\cite{pang2015monitoring} and politics~\cite{porter2020visual}.

Online \textit{crisis imagery} has also attracted growing research attention in HCI and CSCW~\cite{perovich2022self,bica2019communicating,bica2017visual,li2018localizing,zhang2021mapping,sosea2021using,bowe2020learning,mortensen2017really,seo2014visual}. Some researchers have exploited the value of crisis imagery, as a special kind of data, to improve crisis-related decision-making, such as disaster type identification~\cite{asif2021automatic} and damage localization~\cite{li2018localizing}. Existing literature has also investigated how images with different visual characteristics are utilized for crisis communication. Among them, \textit{crisis visualizations}, one category of images that aims to systematically represent crisis information and augment the quality of risk communication, have been widely explored~\cite{zhang2021mapping,bowe2020learning,welhausen2015visualizing,preim2020survey}. For example, Zhang et al. centered on COVID-19 crisis visualizations and looked at how they helped to convey information on disease prevalence, epidemiological simulations, and economic and social change~\cite{zhang2021mapping}. \textit{Crisis memes}, one kind of informal crisis imagery that involves participatory creation, also manifests unique roles in crisis communication such as being a visual analogy and visual pun~\cite{perovich2022self}. 

Compared to crisis visualizations and memes as categories with special visual features, general crisis images that people post on social media tend to be more miscellaneous in visual topics and patterns, and are used for more diverse communication needs~\cite{bica2017visual}. For instance, they might be either a random photo recording the life under a crisis, or a screenshot of an article with the latest crisis-related policies. Such nature of the heterogeneity of social media crisis images complicates the analysis that aims to figure out their roles in crisis communication. Focusing on a local outbreak in the aftermath of the COVID-19 pandemic, we develop an image-clustering approach to effectively group different crisis visual themes (RQ1) as a preliminary step. Based on it, this work further contributes to visual crisis communication by illustrating how social media crisis images satisfy different informational and emotional needs in crises (RQ2), and what unique strategies they provide to augment crisis communication beyond textual narratives (RQ3).

\subsection{Social Media Imagery Analysis Techniques}

The growth of image use in everyday communication on social media has fueled the development of social media imagery analysis~\cite{zhang2021image,bica2017visual,manikonda2017modeling}. Researchers have associated image data with other post- or user-relevant information for analysis based on specific tasks, such as text~\cite{manikonda2017modeling,pang2015monitoring,abdullah2015collective}, geospatial data~\cite{bica2017visual}, social status of the user~\cite{stefanone2015image}, and social engagement~\cite{bakhshi2015we,manikonda2016tweeting}. For example, Bica et al. revealed the correlation between the severity of a disaster and the image tweet activity based on the geotag information~\cite{bica2017visual}.

Recent work has also begun to look specifically at the visual characteristics of social media imagery. For instance, Manikonda and Choudhury examined \textit{visual features}, such as color palettes and visual saliency, that were manifested through mental health images on Instagram~\cite{manikonda2017modeling}. \textit{Visual themes} are another significant topic of concern. Nonetheless, identifying meaningful and interpretable visual themes is a challenging task when images contain high-dimensional features~\cite{goldberger2002unsupervised}. Some studies applied qualitative content analysis to establish theme categories (e.g., ~\cite{bica2019communicating,perovich2022self,andalibi2017sensitive,brantner2020memes}), but content analysis suffers from scalability and reproducibility concerns. The recent progress of computer vision and deep learning enables more accurate quantitative thematic analysis at scale. Two representative approaches are supervised image classification~\cite{williams2020images,guerin2017cnn} and unsupervised image clustering~\cite{zhang2021image}. Image clustering has exhibited its great performance in discovering meaningful patterns that are less subject to human annotation~\cite{hu2014we,manikonda2017modeling,zhang2021image,peng2021makes,porter2020visual}. Nonetheless, \textit{extracting low-dimensional representations} of images remains a bottleneck for image clustering tasks~\cite{zhang2021image}. To present images as low-dimensional vectors, some prior work relied on human-assigned feature codes~\cite{porter2020visual}, and another line of scholarship leveraged visual feature descriptors like Scale-Invariant Feature Transform (SIFT)~\cite{daly2016mining} and Speeded Up Robust Features (SURF)~\cite{manikonda2017modeling}. Transfer learning-based feature extraction, with the development of pretrained deep learning models, has recently shown tremendous potential in social media imagery analysis. For example, Zhang and Peng compared different visual feature extraction methods for political image clustering, and revealed that transfer learning significantly outperformed other methods~\cite{zhang2021image}.

Despite the advancement in imagery analysis techniques~\cite{imran2020using,zhang2021image}, the setting of social media imagery during a COVID-19 local outbreak presents additional challenges to the analysis. First, emotional connections through visuals characterize the event, expressing both ``positive energy'' to establish public confidence~\cite{lu2021positive}, or anxiety and annoyance under ``lockdown'' fatigue~\cite{ross2021household,cho2023bright}. It necessitates multimodal sentiment analysis to understand and monitor public reactions~\cite{imran2020using}. Second, public discourse on local outbreaks naturally intersects with social and political issues under crisis management of the local government~\cite{marzouki2021understanding,cho2023bright}. Therefore, situating imagery analysis within the specific sociopolitical context is crucial in this setting. Finally, it is significant for the imagery analysis to consider the heterogeneity of visual narratives during the crisis on social media~\cite{perovich2022self,sleigh2021qualitative}.

This work customizes image clustering with visual features extracted by transfer learning to comprehensively unpack crisis image themes during a COVID-19 local outbreak. We also apply qualitative thematic and sentiment analysis to capture a more in-depth understanding of the goals and strategies of social media crisis imagery during the event.

\section{METHOD}

\begin{figure}
	\centering
		\includegraphics[scale=.43]{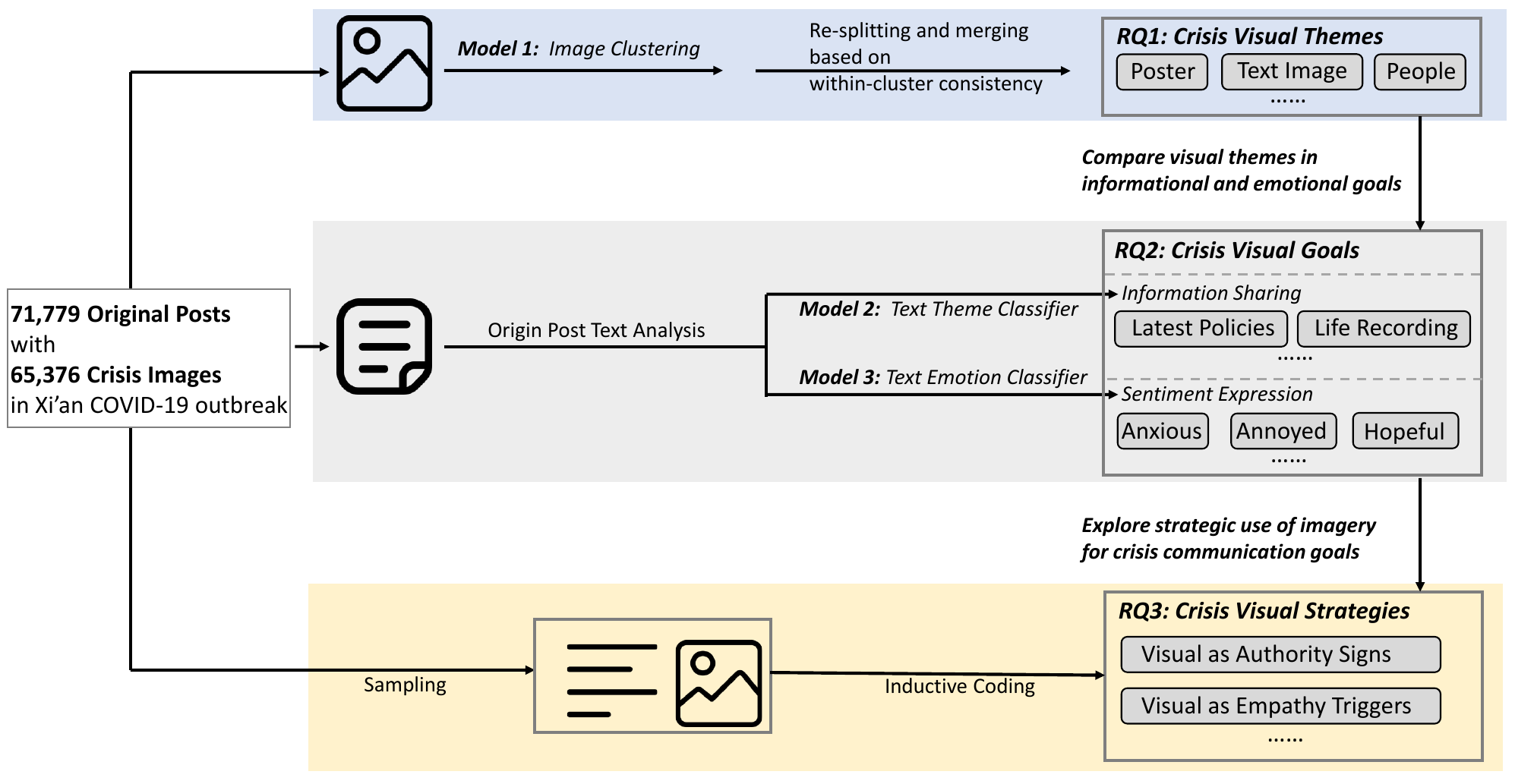}
	\caption{The analytic flow to understand visual crisis communication during a COVID-19 local outbreak.}
	\label{FIG: method}
\end{figure}

We adopted a mixed-methods approach to examine the characteristics of social media crisis imagery in a COVID-19 local outbreak. To systematically unpack crisis visual themes (RQ1, Section \ref{method: RQ1}), we used image clustering to discern different visual elements in crisis imagery. To understand their crisis communication goals (RQ2, Section \ref{method: RQ2}), we developed text classifiers to extract information and sentiments from original posts of crisis images, and made comparisons among different crisis visual themes. To further figure out the strategic use of images for these crisis communication goals (RQ3, Section \ref{method: RQ3}), we conducted inductive coding of crisis image samples in the post context, unearthing its unique values in enhancing crisis communication. The overall analytic flow of this study is described in Figure \ref{FIG: method}.

\subsection{Study Event: Xi'an COVID-19 Local Outbreak}\label{event}

The Xi'an COVID-19 local outbreak, caused mainly by SARS-CoV-2 Delta variant~\cite{XianSource}, is regarded as the most severe COVID-19 local outbreak in China after Wuhan outbreak as of January 2022~\cite{XianlocalOutbreakWiki}. The first COVID-19 case of this event was reported on December 9, 2021~\cite{XianlocalOutbreakWiki}, at which time COVID-19 had infected more than 260 million people worldwide and caused more than 5 million deaths~\cite{worldCases}. Following a dynamic zero-COVID strategy, the local government imposed a lockdown on December 23, 2021, which lasted for about one month till January 24, 2022~\cite{XianlocalOutbreakWiki}. The lockdown implemented a stay-at-home order, in which citizens needed permission to leave residential compounds or the city, affecting 13 million local residents~\cite{XianOutbreakSCMP}. A total of 2053 cases were reported in Xi'an during this event~\cite{XianlocalOutbreakWiki}. We focus on visual crisis communication in this event as a typical example of a COVID-19 local outbreak with lockdown management, which might contribute to the understanding of civic response in similar health crisis events with highly contagious viruses, varied and strict management measures, and uncertain local situations.

\subsection{Data Collection}

We chose Weibo as the research site, which is the largest Chinese microblogging website and a significant social media platform for crisis communication in China~\cite{yang2021know, chen2021exploring, qu2011microblogging}. To determine the data inclusion criteria, we first randomly browsed posts related to the Xi'an COVID-19 Local Outbreak. We finally chose the most frequent keyword ``\textit{Xi'an epidemic}'' (which sometimes appeared as a hashtag) for data collection. Applying the WeiboSuperSpider tool~\cite{SuperTopicCrawling}, we crawled posts whose text included the ``\textit{Xi'an epidemic}'' keyword and collected all the images within them. The data covered the period from December 9, 2021, when the first COVID-19 case was reported in this event, to January 24, 2022, the date when the Xi'an government lifted the lockdown. In total, we retrieved 345,423 relevant posts, and kept the 71,779 \textit{original} posts contributed by 39,866 distinct users as the target dataset. Among them, 29,075 posts (40.5\%) had one or more images, and the total image number was 66,183. We kept 65,376 images after excluding GIFs, which contained multiple frames and may produce unreliable results for static visual analysis.

\begin{figure}
	\centering
		\includegraphics[scale=.4]{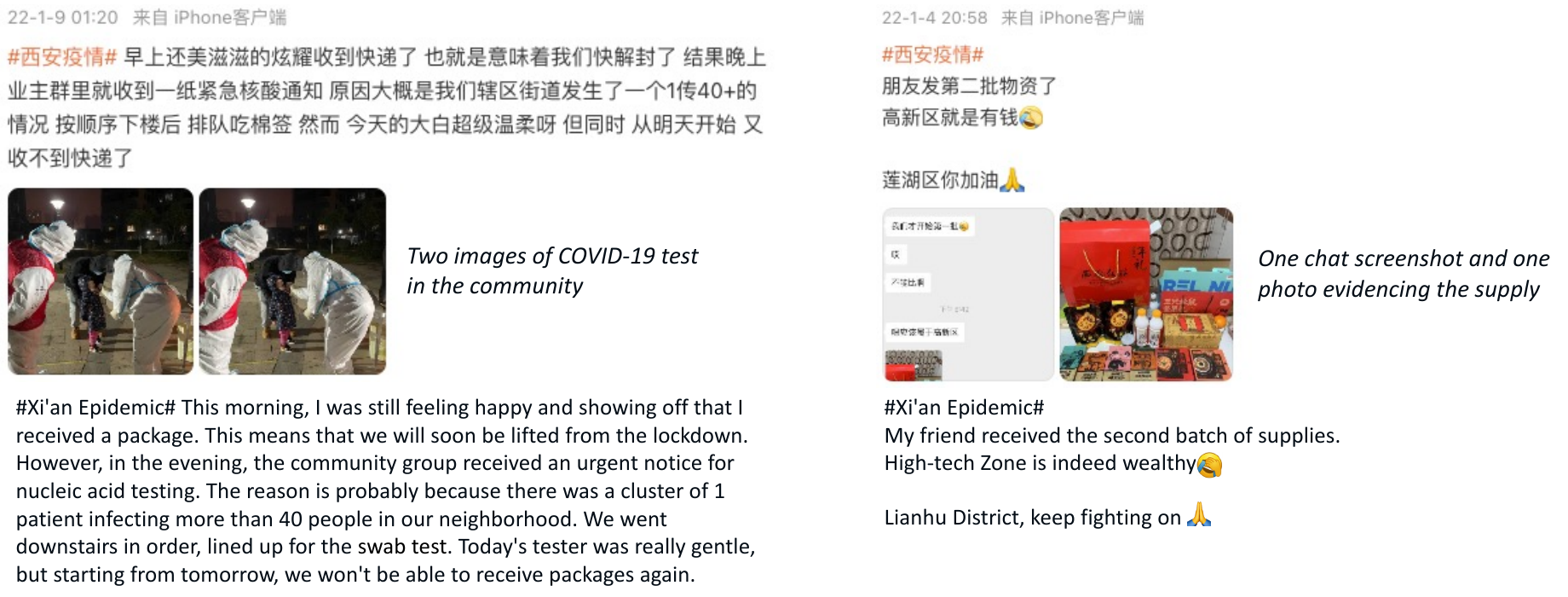}
	\caption{Two examples with visual use in the dataset of crisis-related posts during the Xi'an local outbreak.}
	\label{FIG: example_posts}
\end{figure}

Generally, posts in the collected dataset received an average of 48.5 likes (SD=2093.0), 7.2 comments (SD=130.7), and 2.9 shares (SD=59.5). On average, posts with images received slightly higher user engagement, with an average of 54.4 likes (SD=2007.8), 8.7 comments (SD=131.2), and 3.2 shares (SD=45.1), compared to posts without images which received an average of 44.5 likes (SD=2149.0), 6.1 comments (SD=130.4), and 2.6 shares (SD=67.6). The average post length was 164.6 (SD=246.5). In addition to the de-facto hashtag ``\#\textit{Xi'an epidemic}\#'' in the event (N=20,144, also as the keyword for collection), the other 5 most frequent hashtags included ``\#Xi'an\#'' (N=2028), ``\#Come on Xi'an\#'' (N=1605), ``\#Diary of Fighting Against the Epidemic in Xi'an\#'' (N=861), ``\#Epidemic prevention and control\#'' (N=560) and ``\#Xi'an Epidemic Updates\#'' (N=501). We provide two example posts with visuals in Figure ~\ref{FIG: example_posts} to familiarize readers with the dataset.

There were several considerations regarding the limitation of the dataset. First, we only used one most widely-adopted keyword ``\textit{Xi'an epidemic}'' for data collection. As also the de-facto hashtag/super-topic~\cite{chen2021exploring} of this local outbreak on Weibo, this keyword was representative to collect a sufficiently large dataset for quantitative analysis and capture different themes of user-generated posts regarding the event. We did not include other keywords such as ``\textit{Xi'an quarantine life}'' or ``\textit{Xi'an epidemic updates}'' that might introduce a bias towards specific aspects of the outbreak. However, this keyword could not ensure the comprehensiveness of data collection, as some pertinent posts might be omitted. Second, we collected data on March 14, 2022, i.e., 50 days after the event, so that the engagement indexes of posts (e.g., likes and shares) stabilized for a fair comparison. Nonetheless, it inevitably led to the loss of some posts, especially under factors such as self-deleting outdated crisis-related posts~\cite{he2022help}. Finally, different from Twitter, Weibo adopts a more flexible setting on post visibility, including ``public'', ``followers'', ``social circle (mutual followers)'', ``selected friends'' and ``private''~\cite{WeiboVisibility}. We only managed to collect the public posts, which limited the dataset size to some extent and was potentially skewed towards more public disclosure.

\subsection{RQ1: Uncovering Visual Content Themes with an Image Clustering Approach}\label{method: RQ1}

To have a preliminary understanding of what users disclosed and shared through images during the COVID-19 local outbreak, we first aimed to unpack the themes of social media crisis images. We chose unsupervised image clustering with transfer learning for feature extraction~\cite{zhang2021image} to categorize the data. Compared to the widely-adopted image content analysis of manually coding images and assigning theme labels~\cite{perovich2022self, zhang2021mapping, bica2019communicating}, image clustering can scale image analysis to larger datasets and is less subject to reproducibility concerns. Also, in contrast to supervised image classification, image clustering is not restricted by the size and quality of the human-labeled training dataset. The recent development of transfer learning and deep neural networks further brings great potential for image clustering, especially in categorizing miscellaneous social media images~\cite{zhang2021image}. Therefore, image clustering is well applicable for identifying visual themes of crisis imagery in the context of this work.

As our ultimate goal was to generate coherent clusters with pure and explainable content themes, we adopted \textit{within-cluster consistency}~\cite{zhang2021image} as a quantitative measure of whether our clusters were internally consistent. Specifically, \textit{within-cluster consistency} denoted the proportion of the most common theme in a human-annotated sample for a given cluster (similar to ``semantic validity'' in unsupervised topic modeling~\cite{grimmer2013text}). We averaged \textit{within-cluster consistency} of all clusters to evaluate the performance of each image clustering approach and selected models based on it.

\subsubsection{Transfer Learning for Low-Dimensional Feature Extraction}

We adopted transfer learning~\cite{torrey2010transfer} to extract low-dimensional features of high-dimensional pixel representations of images, which is the bottleneck of unsupervised image clustering~\cite{zhang2021image}. It outperforms visual-element-based extraction approaches such as bag-of-visual-words models~\cite{zhang2021image,sivic2003video,lowe2004distinctive}, and achieves good performances in a variety of imagery analysis tasks~\cite{zhang2021image,nguyen2017personalized}.

To find a suitable pretrained model, we performed a pilot experiment on 1000 image samples and tried various popular pretrained deep learning models, including VGG16~\cite{simonyan2014very}, VGG19~\cite{simonyan2014very}, ResNet~\cite{he2016deep}, MobileNet~\cite{howard2017mobilenets} and MobileNetV2~\cite{sandler2018mobilenetv2}, for feature extraction. All models were pretrained on ImageNet~\cite{deng2009imagenet}, a large image database covering 1000 general categories, and thus had the generalizability to fit into our dataset. Specifically, we first resized images to 224 $\times$ 224 pixel with RGB color mode, fed images into each candidate model with pre-trained weights on the ImageNet~\cite{deng2009imagenet}, and generated the last layer as the low-dimensional representations of images~\cite{zhang2021image}. Then, we applied a basic clustering method (K-means~\cite{lloyd1982least}) to the extracted features. We also experimented with a traditional feature extraction approach for comparison, i.e., the bag-of-visual-words model extracting scale-invariant visual features~\cite{sivic2003video}. When varying the value of K, we found 5 clusters generally yielded a high silhouette score (a metric measuring clustering quality~\cite{rousseeuw1987silhouettes}) as well as meaningful results across different feature extraction models. Therefore, we set $K=5$ for a fair comparison. After clustering, two coders manually coded visual themes of 100 samples (5 clusters $\times$ 20 samples in each cluster) for each candidate model, compared labels, and discussed to reach a consensus. They examined the performance with \textit{within-cluster consistency} (Appendix \ref{ModelPerformance}). The evaluation indicated that different pretrained models all outperformed the bag-of-visual-words model and did not exhibit substantially different performance among them. As such, we finally chose MobileNetV2~\cite{sandler2018mobilenetv2} with the highest feature extraction speed to facilitate further experiments when scaling up to the whole dataset. Assisted by MobileNetV2, we represented each image in the dataset (224 $\times$ 224 $\times$ 3) as a 1280-dimension vector, which was the last hidden layer (global average pooling layer~\cite{sandler2018mobilenetv2}).

\subsubsection{Image Clustering}\label{clustering}

After converting images to low-dimensional representations, we adopted clustering to uncover visual content themes that were accurate and interpretable.

First, we tried different types of clustering methods, including centroid-based (i.e., K-means~\cite{lloyd1982least}), density-based (i.e., DBSCAN~\cite{khan2014dbscan}) and distribution-based (i.e., Gaussian Mixture Model~\cite{reynolds2009gaussian}) approaches. Following a similar approach in determining the feature extraction model, we compared different clustering methods based on \textit{within-cluster consistency} and found that there was no substantial difference between them (Appendix \ref{ModelPerformance}). Therefore, we chose K-means to categorize crisis images for simplicity. To determine an optimal $K$ value, we defined the search space of $K$ in a range of 5 to 20, and computed silhouette scores~\cite{rousseeuw1987silhouettes} for each candidate; $K = 6$ was finally adopted with a maximal silhouette score. We noticed that most generated clusters were surprisingly pure and clear (e.g., text images) but two clusters were a mixture of sub-themes (i.e., diverse types of photos). Therefore, we performed a re-splitting and merging step as detailed below.

\begin{enumerate}
  \item \textbf{Sampling:} Randomly sampling 50 images for each cluster to measure the within-cluster consistency. It yields 300 image samples (50 images $\times$ 6 clusters) in total.
  \item \textbf{Coding:} Two coders independently coded images and assigned theme labels $l_j^i$ for image $i$ in cluster $j$. In particular, the coders first followed a descriptive coding process to analyze visual content~\cite{saldana2021coding}, focusing on visual types (e.g., chat screenshots and in-situ photos) and visual elements (e.g., medical staff and food). They then conducted pattern coding~\cite{saldana2021coding} to connect fine-grained codes and identify larger visual themes that potentially captured image clusters (e.g., assigning the high-level theme label ``text images'' for relevant codes such as ``chat screenshots'' and ``document photos''). They reached a consensus on theme labels through several rounds of meetings, comparisons, and discussions. 
  \item \textbf{Measuring Consistency:} For each theme $t$ that appeared in Cluster $C_j$, we computed its prevalence $P_t = \frac{\sum_{i\in C_j} l_j^i = t}{|C_j|}$, i.e., the percentage of images belonging to this theme in the cluster. We defined a cluster as ``\textit{consistent}'' when it had a dominant theme $t$ whose prevalence was larger than a dominance threshold $thld_d$, i.e., $P_t > thld_d$. A larger dominance threshold $thld_d$ denotes stricter consistency when the dominant theme is required to have a higher proportion in the cluster. In this work, we set $thld_d$ as 60\%.
  \item \textbf{Splitting:} For each inconsistent cluster, we re-split it according to how many significant themes it contained. Specifically, we selected significant themes $T_s$ whose prevalence $P_t$ was greater than a significance threshold $thld_s$ (We set $thld_s$ as 20\% in this work). Then, we set $K = |T_s|$ (number of significant themes) to separate the inconsistent cluster into $K$ sub-clusters using K-means. 
  \item \textbf{Merging:} After splitting the inconsistent clusters, we repeated sampling, coding, measuring consistency, and finding the dominant theme for each sub-cluster. No inconsistent sub-cluster was detected in this round. We merged all clusters and sub-clusters with the same dominant theme.
\end{enumerate}

In total, the image clustering yielded six clusters with distinct visual themes\footnote{Note that it is a coincidence that the number of clusters (six) equaled the initial $K$ value after the splitting and merging steps}, including \textit{posters}, \textit{text images}, \textit{indoor objects}, \textit{outdoor scenes}, \textit{people}, and \textit{food}, which will be detailed in Section \ref{findings: RQ1}. Two coders independently annotated themes of 600 label-assigned samples (100 samples $\times$ 6 clusters) as \textit{true} or \textit{false} (i.e., whether the actual theme of each image corresponded to the label assigned by image clustering) to evaluate the final performance. After annotation, the authors adopted Cohen's kappa to validate inter-rater reliability~\cite{mchugh2012interrater}. The high Cohen's kappa ($\kappa = 0.86$) suggested strong agreement between the two coders and strengthened the validity of the manual evaluation. The evaluation generated an average recall of 79.5\%, indicating the substantially good performance of image clustering in uncovering visual themes. 

Within all image clusters, we noticed that \textit{indoor objects} and \textit{outdoor scenes} were two relatively general categories, which communicated diverse indoor and outdoor activities valuable in the event (e.g., community nucleic acid test), and thus could benefit from a fine-grained classification to generate more meaningful results. Therefore, we applied deep learning models fine-tuned on specific scene-based datasets to predict scenes or items. To identify \textit{indoor objects}, we chose MIT Indoor Scenes dataset as one of the benchmark datasets for indoor scene recognition~\cite{quattoni2009recognizing}. To discern \textit{outdoor scenes}, we chose Places365 database as one of the authoritative datasets for scene classification~\cite{zhou2017places}. We adopted Vision Transformer (ViT) model~\cite{dosovitskiy2020image} fine-tuned on MIT Indoor Scenes dataset and VGG16~\cite{simonyan2014very} fine-tuned on Places365 database, both of which achieved good performance in the specific task~\cite{ViT-indoor, gkallia2017keras_places365}.

\subsection{RQ2: Investigating Visual Crisis Communication Goals with Text Analysis of Original Posts}\label{method: RQ2}

RQ1 focused on visual features and generated a comprehensive taxonomy of crisis image themes during a COVID-19 local outbreak. It captured the substantial diversity of crisis imagery, ranging from text-embedded images with the latest policies to photos recording daily quarantine life. This led to a further question: \textit{which goals do these diverse types of images serve for crisis communication}, and more importantly, \textit{how different visual themes are adopted for different crisis communication goals}? To investigate this question, we (1) performed quantitative text analysis of original posts to unpack crisis communication goals, and (2) compared different visual themes in the informational and emotional objectives. The quantitative text analysis and inter-image comparison particularly focused on \textit{information sharing} and \textit{sentiment expression}, two significant communication objectives in crisis settings~\cite{he2021beyond, zade2018situational, vieweg2010microblogging,qu2009online, qu2011microblogging}.

\subsubsection{Text Analysis of Original Posts}\label{OPAnalysis}

We developed information and emotion classifiers for original post analysis as a preliminary step to figure out the visual goals, i.e., how different types of crisis images were used to fulfill different crisis communication needs.

\textit{\textbf{Information Theme Classifier}}: To identify which types of information were disseminated in the local outbreak-related posts, we developed a codebook through inductive thematic analysis, and leveraged a text classifier to generalize the information themes to the whole corpus.

To establish the codebook of information themes, we used inductive thematic analysis to code a sample of posts ~\cite{fereday2006demonstrating}, capturing the codes naturally reflecting the topics of crisis-related posts. Specifically, two coders carefully read through 200 post samples independently, and generated codes of information themes (e.g., ``\textit{latest policies and measures}'') that described the data. Through several rounds of discussions and comparisons, they reached a consensus on the theme codes of crisis-related posts as shown in Table \ref{tab:taxonomy}, and returned to annotating every post in the 200 samples based on the four type labels. The Cohen's Kappa ($\kappa = 0.91$) suggested substantial inter-rater agreement between the two coders~\cite{mchugh2012interrater}. Finally, the two coders separately annotated another 400 posts each, generating 1,000 theme-assigned samples as the training dataset. An additional sample of 200 posts was further labeled as the test dataset. The annotation process also helped the two coders validate that the four information types were inclusive to describe user-generated posts during the event with no new code emerging. Generally, the information themes aligned with the public response on Chinese social media during the early COVID-19 outbreak such as the discussion on management measures~\cite{wang2020concerns,he2021beyond,li2020data}. However, as a COVID-19 resurgence, little discourse focused on the causative agent, prevention knowledge, and epidemiological characteristics~\cite{li2020data,xu2020characterizing}. Meanwhile, \textit{life recording during lockdown} characterized the event with prolonged abnormal living conditions.

\begin{table*}
  \small
  \caption{The codebook of information themes of Weibo posts related to the Xi'an COVID-19 local outbreak.}
  \label{tab:taxonomy}
  \begin{tabular}{p{2cm}p{3.3cm}p{5.8cm}p{1.5cm}}
    \toprule
    Type & Definition & Example & Percentage in the Sample \\
    \midrule
    Situational Information & Posts communicating situational COVID-19-related information, such as recent cases, social events, and scientific suggestions on new variants  & \textit{Up to now, a total of 1,451 cases have been diagnosed in Xi'an, including 2 critical illness cases and 11 serious illness cases. The critical illness rate was 0.14\%, and the serious illness rate was 0.76\%.} & 34.9\% \\
    \hline
    Attitude Disclosure & Expressing attitudes towards the COVID-19 local outbreak and relevant issues & \textit{I'm not in the mood to eat meals and go to class. I want to go home... I beg COVID-19 to stop troubling Xi'an. Don't let Xi'an people be unable to go back hometown.} & 29.8\% \\
    \hline
    Life Recording under Lockdown & Recording personal life, status, and challenges under lockdown & \textit{This is the first time I feel that the epidemic is so close to me... The community downstairs of my company has raised a cordon.} & 23.4\% \\
    \hline
    Latest Policies and Measures & Announcement or adjustment of the COVID-19 management policies and measures & \textit{[IMPORTANT! Xi'an starts a new round of nucleic acid screening on December 27th] According to the news from the Xi'an Epidemic Prevention and Control Headquarters: From 12:00 on December 27th, Xi'an will start a new round of COVID-19 nucleic acid screening. Reminder: Keep social distance and protect yourself.} & 11.9\% \\
    
    \bottomrule
  \end{tabular}
\end{table*}

We applied Bidirectional Encoder Representations from Transformers (BERT)~\cite{devlin2018bert} to classify the local-outbreak-related posts for its good performance and generalizability in text classification tasks. Specifically, we adopted BERT-wwm, a Chinese BERT model pretrained on Chinese Wikipedia~\cite{cui2021pre}, and fine-tuned it with our training dataset (N=1000) to adjust it to our specific tasks. The micro f1-score achieved 82.1\% in the test dataset (N=200), indicating its substantially good performance. Finally, we leveraged the information theme classifier to assign theme labels to the whole dataset of original posts (N=71,779).

\textit{\textbf{Emotional Type Classifier}}: To identify emotional types of text posts, we first tried some widely-adopted Chinese emotion prediction models (e.g., Jingdong Sentiment API~\footnote{https://neuhub.jd.com/ai/api/nlp/sentiment} and SnowNLP~\footnote{https://github.com/isnowfy/snownlp}), yet we noticed that these models were not well-applicable to our context through manual evaluation. Therefore, we followed an emotion identification method similar to our information theme classification approach, including (1) codebook establishment through open coding~\cite{corbin2014basics} on 100 samples by two authors, capturing emotions in this specific scenario (five emotion types shown below); (2) validation of inter-rater reliability on assigning the five emotion types with Cohen's Kappa (N=100, $\kappa = 0.91$); (3) annotation on training dataset (N=1000) and test dataset (N=200); and (4) BERT-based emotional type classification (micro f1-score = 78.6\% on 5-class emotion classification). The open coding indicated that a single label could well represent each post, and five emotion classes were identified to effectively describe emotions in our samples:

\begin{itemize}
    \item \textbf{Positive - Hopeful (19.9\%\footnote{Percentage of each emotion denotes its proportion in the 1000 training dataset.}):} Mutual encouragement and wishes to overcome the difficulty;
    \item \textbf{Positive - Appreciative (4.3\%):} Gratitude to medical staff, community workers, and other social connections in the local outbreak;
    \item \textbf{Neutral (52.2\%):} Neutral posts such as news, latest policies, and unemotional life recording;
    \item \textbf{Negative - Annoyed (11.9\%):} Feeling of annoyance when normal life was disturbed by the local outbreak and relevant measures;
    \item \textbf{Negative - Anxious (11.7\%):} The nervousness and worry due to the uncertainty during the local outbreak (e.g., uncertain lockdown time and food supply).
\end{itemize}

\subsubsection{Comparison of Crisis Visual Goals}

After identifying the information theme and emotion type of each post, we extracted the crisis image category of the post established in RQ1 (or the dominant category based on frequency if the post contained more than one image). We investigated how the transmitted information themes and emotions correlated with (1) whether crisis images were used or not, and (2) which types of crisis images were used. Particularly, we compared different image types on the information or emotion categories of posts to investigate their differences in informational and emotional goals. We also performed chi-square tests~\cite{mchugh2013chi} to validate the statistical significance of such differences, i.e., whether different image types had significantly different information and emotion distributions.

\subsection{RQ3: Understanding Strategies of Crisis Images with Inductive Coding}\label{method: RQ3}

RQ1 and RQ2 leveraged computational approaches across visual and linguistic modalities to comprehensively unpack the rich \textit{themes} and the nuanced informational and emotional \textit{goals} of crisis imagery. The comparison of crisis communication goals among visual themes helped to statistically depict the \textit{inter-modality correlation} between visual and text. Moving a further step, a more in-depth investigation of such inter-modality correlation, especially the strategic use of visuals to enhance crisis communication as a whole, could deepen the understanding of crisis image use. Therefore, we applied inductive coding to qualitatively look into the \textit{visual strategies} in crisis communication, i.e., how crisis images \textit{facilitated} information sharing and sentiment expression in the specific context. 

Specifically, we sampled 1200 distinct images (200 images $\times$ 6 visual theme categories) identified in Section \ref{method: RQ1}, and situated them within the original posts. We conducted inductive thematic analysis~\cite{fereday2006demonstrating} and inductive emotion coding~\cite{perovich2022self} on the data. Two authors (1) inductively coded the information themes and emotions expressed in both the image samples and corresponding post text; (2) compared visual-based and language-based information and emotions; (3) figured out \textit{how they were related}; and finally (4) recorded how crisis visuals strategically \textit{facilitated} the information and emotion sharing. Two coders took an iterative process of coding, comparing, and discussing to resolve the difference and finalize the codebook. The coding reached saturation given the sample size, so we did not code more data. This step of qualitative analysis helped to unpack how visual narratives manifested unique values and complemented text in crisis communication. After qualitatively identifying specific visual strategies, we also presented relevant quantitative evidence to illustrate their prevalence and potential correlations with user engagement, affording a more thorough understanding of the result.

\section{FINDINGS}

This work provides a comprehensive description of the themes, goals, and strategies of crisis imagery during a COVID-19 local outbreak, promoting the understanding of \textit{visual crisis communication} in general social media crisis images. In Section \ref{findings: RQ1}, we describe the taxonomy of crisis visual themes obtained from image clustering, showing distinct features of different visuals in the event. We further demonstrate how these crisis image themes were substantially correlated with different informational and emotional objectives in Section \ref{findings: RQ2}, unearthing the specialized use of crisis images in crisis communication. In Section \ref{findings: RQ3}, we depict several unique strategies of visuals that contributed to effective crisis communication. Overall, these findings uncover not only the rich features of social media crisis imagery, but also its special values in promoting situational awareness and establishing emotional connections.

\subsection{RQ1: Visual Themes}\label{findings: RQ1}

\begin{figure}
	\centering
		\includegraphics[scale=.4]{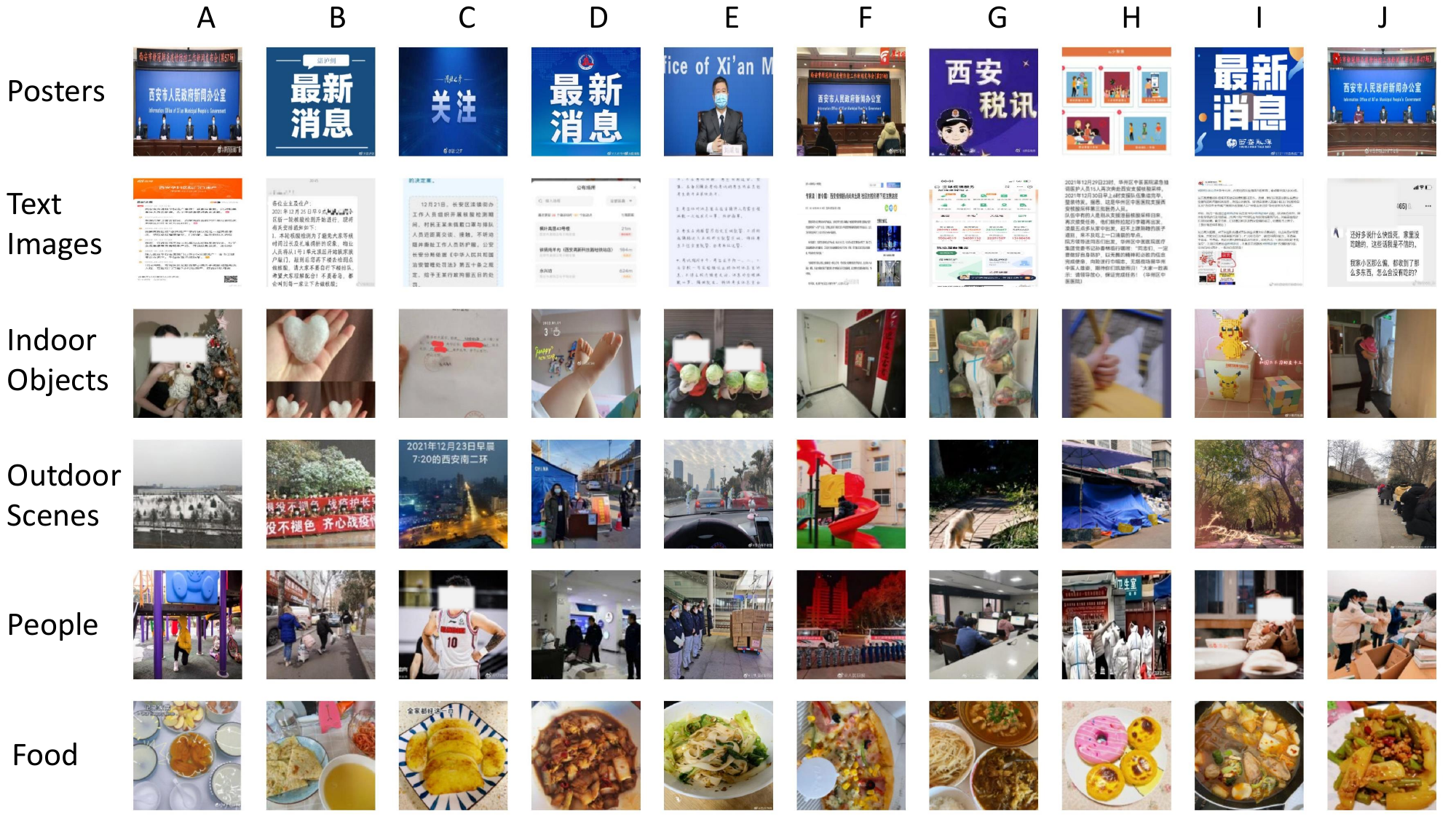}
	\caption{Image content themes during the Xi'an COVID-19 local outbreak based on image clustering (each row represents one image cluster). The image examples of each cluster (N=10) were picked from random samples (N=20), instead of crisis images nearest to cluster centers, to demonstrate the in-cluster diversity.}
	\label{FIG: imageExample}
\end{figure}

The image clustering yielded six representative visual themes as shown in Figure \ref{FIG: imageExample}, including two text-embedded image categories~\cite{manikonda2017modeling}: \textit{posters} and \textit{text images}; and four ``visual diary'' image categories that recorded life during the outbreak: \textit{indoor objects}, \textit{outdoor scenes}, \textit{people}, and \textit{food}. These crisis image themes communicated distinct visual elements in the local outbreak to enhance situational awareness and establish emotional connections. We detailed descriptive analysis of different visual themes, and started with two text-embedded image categories:

\begin{itemize}
  \item \textbf{Posters} (22.7\%, N=14,836): Images embedding some big-character text, especially with a solid-color background. Different from images with rich text about crisis information (text images), posters generally only had meta information (e.g., ``latest news'' in Figure \ref{FIG: imageExample} \textit{Posters, B}) of the post. Some poster-style photos, especially photos of COVID-19-related press conferences with the agency name embedded in a solid-color background (e.g., Figure \ref{FIG: imageExample} \textit{Posters, A and E}), also characterized this category. The highlighted text information of posters helped to convey the core message and catch users' eyes at a glance. In a similar visual style, text-embedded visualizations and memes (e.g., Figure \ref{FIG: imageExample} \textit{Posters, H}) were also categorized into this visual theme though not in a great amount.
  
  \item \textbf{Text images} (16.0\%, N=10,434): Images with plentiful embedded text. We noticed that plenty of authoritative regulations (e.g., the lockdown measures of the city) and community guidelines (e.g., communities' COVID-19 test notice) were shared by local residents in the form of images such as screenshots of government-posted articles, potentially due to the convenience of forwarding images in and outside the Weibo platform. Screenshots of chat messages (e.g., Figure \ref{FIG: imageExample} \textit{Text Images, J}) were another type of text image that provided situational information with original sources, such as the latest suggestions posted by community officers in a community group chat. Meanwhile, text images were also observed to convey some sensitive messages, such as conflicts with supply providers and rumors faked as chat logs with ``political insiders'', to circumvent moderation.
\end{itemize}

Below are the four ``visual diary'' image categories:

\begin{itemize}
  \item \textbf{Indoor objects} (17.6\%, N=11,480): Images capturing indoor objects that recorded everyday life during the lockdown. This category reflected people's resilience during this isolated lifestyle, such as making handicrafts (Figure \ref{FIG: imageExample} \textit{Indoor Objects, I}) and decorations (Figure \ref{FIG: imageExample} \textit{Indoor Objects, A}). Self-disclosure of mental status in quarantine such as anxiety or hopefulness was typical along with this type of visual diary, which is further demonstrated in Section \ref{findings: RQ2}. According to the model pretrained on specific indoor scenes~\cite{ViT-indoor}, the top 5 most frequent indoor items include (1)\textit{buffet} (9.4\%, N=1082): Food supply and cooking\footnote{We note the contextual interpretations of each category after observing a series of examples with the given label.}; (2) \textit{artstudio} (9.3\%, N=1072): Handicrafts, paintings, calligraphy, and homework; (3) \textit{inside\_subway} (7.8\%, N=894): Actually as close-range photos of indoor family activities (e.g., playing with pets); (4) \textit{airport\_inside} (4.8\%, N=553): Actually as spacious lobbies (sometimes with people queuing for the nucleic acid test); (5) \textit{greenhouse} (4.8\%, N=550): Potted plants, vegetables, fruits.
  
  \item \textbf{Outdoor scenes} (17.2\%, N=11,234): Outdoor photos such as streets, parks and attractions. Many images in this category demonstrated the emptiness of the city during the COVID-19 lockdown (e.g., Figure \ref{FIG: imageExample} \textit{Outdoor Scenes, C}), expressing the wish for a normal life. Also, some outdoor photos served as real-time proof of situations in the local community, such as the location of checkpoints of movement pass (Figure \ref{FIG: imageExample} \textit{Outdoor Scenes, D}) and the crowdedness of a specific COVID-19 test point that reminded community members to come later (Figure \textit{Outdoor Scenes, J}). According to the scene-based pre-trained model~\cite{gkallia2017keras_places365}, the top 5 most frequent outdoor items are (1) \textit{staircase} (16.0\%, N=1795): Actually as Chinese buildings especially ancient towers with storeyed shapes, such as Figure \ref{FIG: theme4_empathy} B; (2) \textit{artstudio} (14.0\%, N=1574): Supply distribution points; (3) \textit{ice\_cream\_parlor} (7.7\%, N=868): Actually as the lockdown checkpoints or COVID-19 test points; (4) \textit{campus} (5.1\%, N=576): (empty) campus grounds and buildings; (5) \textit{hospital} (4.0\%, N=447): Outdoor scenes with doctors or workers in biohazard suites.
  
  \item \textbf{People} (16.9\%, N=11,040): Images of humans in the outbreak. Two important sub-themes of this category were (1) selfies and photos of families (e.g., Figure \ref{FIG: imageExample} \textit{People, A and I}), notable types of social media images to show their lives to others and connect with audiences~\cite{manikonda2017modeling,hu2014we, marwick2015instafame}, which might be valuable for social connections during lockdown; and (2) photos of medical staff and community workers, some recording and appreciating their contributions and sacrifices during the crisis (e.g., Figure \ref{FIG: imageExample} \textit{People, F}), and some showing conflicts between community workers and residents and criticizing specific workers on their abuse of power during the lockdown (e.g., Figure \ref{FIG: imageExample} \textit{People, H}). Note that the theme \textit{People} did not necessarily contradict with \textit{indoor objects} or \textit{outdoor themes}; both the two categories also contained a noteworthy proportion of photos with humans. However, we noticed that compared to \textit{people}, human images categorized into \textit{outdoor scenes} typically suggested significant environmental knowledge (e.g., surroundings of the lockdown checkpoint and COVID-19 test queue in Figure \ref{FIG: imageExample} \textit{Outdoor Scenes, D \& J}), while those classified as \textit{indoor objects} often ``spotlighted'' items of quarantine life (e.g., Christmas tree in Figure \ref{FIG: imageExample} \textit{Indoor Objects, A} and supply in Figure \ref{FIG: imageExample} \textit{Indoor Objects, G}). Such subtle differences captured by the model were meaningful in distinguishing different crisis-related messages in the event.

  \item \textbf{Food} (9.7\%, N=6,532): plated food and drinks. In addition to recording everyday dishes, these food images also communicated information about the food supply of the local community during the lockdown to other affected citizens in the city.

\end{itemize}

\begin{figure}
	\centering
		\includegraphics[scale=.4]{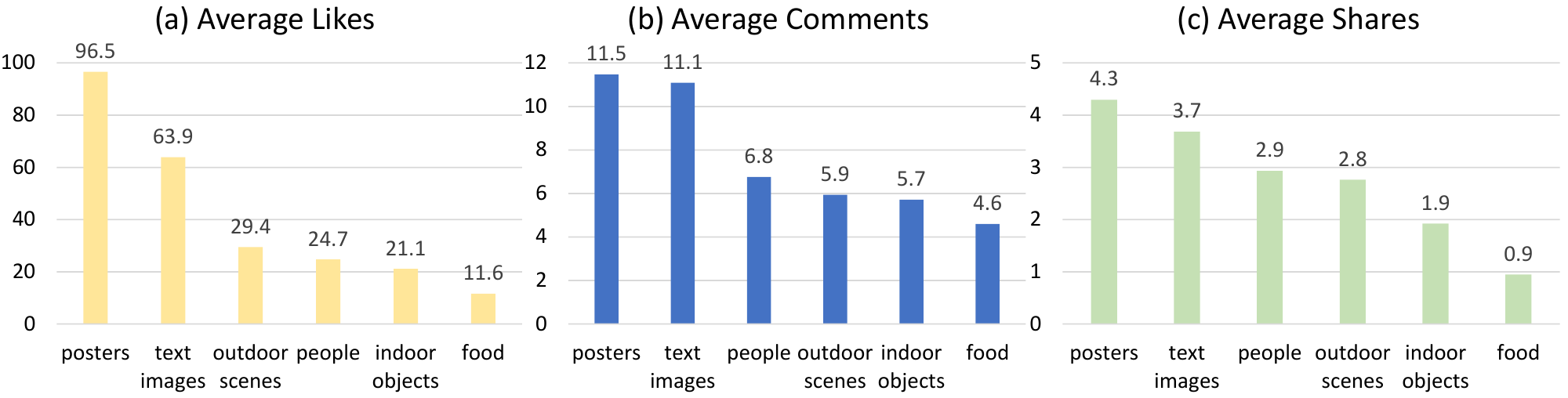}
	\caption{Comparison of different visual themes in public engagement indexes: (a) likes, (b) comments, and (c) shares.}
	\label{FIG: Comparison}
\end{figure}

Different visual themes exhibited substantial differences in public engagement indexes (i.e., likes, comments, and shares) as shown in Figure \ref{FIG: Comparison}. The differences in average comments and shares between visual themes were statistically significant under the one-way ANOVA test ($p<0.05$). Generally, users were most engaged in posts with \textit{posters} images which received the most likes, comments, and shares, followed by \textit{text images}. In comparison, users were less engaged in the other four ``visual diary'' categories. One potential reason might be that visual messages passed through posters and text images (e.g., screenshots of government-posted articles) were more likely to be generalized to different settings and useful to other citizens, while photos recording daily life might be more personal and related to specific scenarios.

\textbf{RQ1 Summary:} Through image clustering, we captured a comprehensive taxonomy of crisis visual themes during a COVID-19 local outbreak, including \textit{posters} and \textit{text images} as two text-embedded categories, and four ``visual diary'' types presenting life in lockdown. These visual themes attracted different levels of user engagement (e.g., \textit{posters} and \textit{text images} received more likes, comments, and shares). It demonstrated \textbf{the diversity of visual communication} during a local outbreak in Chinese social media, and implicitly reflected some specialized use of crisis images, such as the prevalence of \textit{posters} relating to the latest and authoritative information. In the next section, we explicitly described how these diverse visual themes were adopted for different informational and emotional goals, providing a quantitative view of visual-text correlation.

\subsection{RQ2: Visual Goals}\label{findings: RQ2}

Through text analysis of the posts that contained crisis images, we unveiled how different themes of crisis images were adopted for different objectives in information sharing and emotional expression during the COVID-19 outbreak. These findings provided a comprehensive picture of the crisis communication \textit{goals} of crisis images.

\subsubsection{Descriptive Statistics}

Using the information theme classifier described in Section \ref{OPAnalysis}, we captured the proportion of four information themes in crisis-related posts during the local outbreak. Specifically, \textit{situational information} (N=26,963, 37.6\%) and \textit{attitude disclosure} (N=25,291, 35.2\%) were the two most popular information categories. The prevalence of these two categories echoes prior works on crisis response~\cite{he2022help, qu2011microblogging, qu2009online}. \textit{Life recording under lockdown} accounted for 20.0\% crisis-related posts (N=14,389), and information about the \textit{latest policies and measures} had the smallest proportion (N=7.2\%, N=5,136).

The text emotion analysis revealed that in addition to the \textit{neutral} category, two negative emotions (i.e., \textit{annoyed} and \textit{anxious}) and two positive emotions (i.e., \textit{hopeful} and \textit{appreciative}) characterized emotions that users disclosed during the local outbreak. In the whole dataset, the total proportion of negative emotions (22.5\%, with $P_{anxious} = 12.4\%$ and $P_{annoyed} = 10.1\%$) was close to the total proportion of positive emotions (23.4\%, $P_{hopeful} = 20.0\%$ and $P_{appreciative} = 3.4\%$). Therefore, there was an upsurge in negative emotions in the aftermath of the pandemic compared to the initial COVID-19 outbreak in Wuhan in Chinese social media~\cite{wang2020covid}. It highlighted the prevalence of anxiety and annoyance, serving as a warning of unhealthy mental well-being that might be due to COVID-19 fatigue~\cite{williams2021variant,deng2023understanding} under repetitive and uncertain outbreaks.

\begin{figure}
	\centering
		\includegraphics[scale=.4]{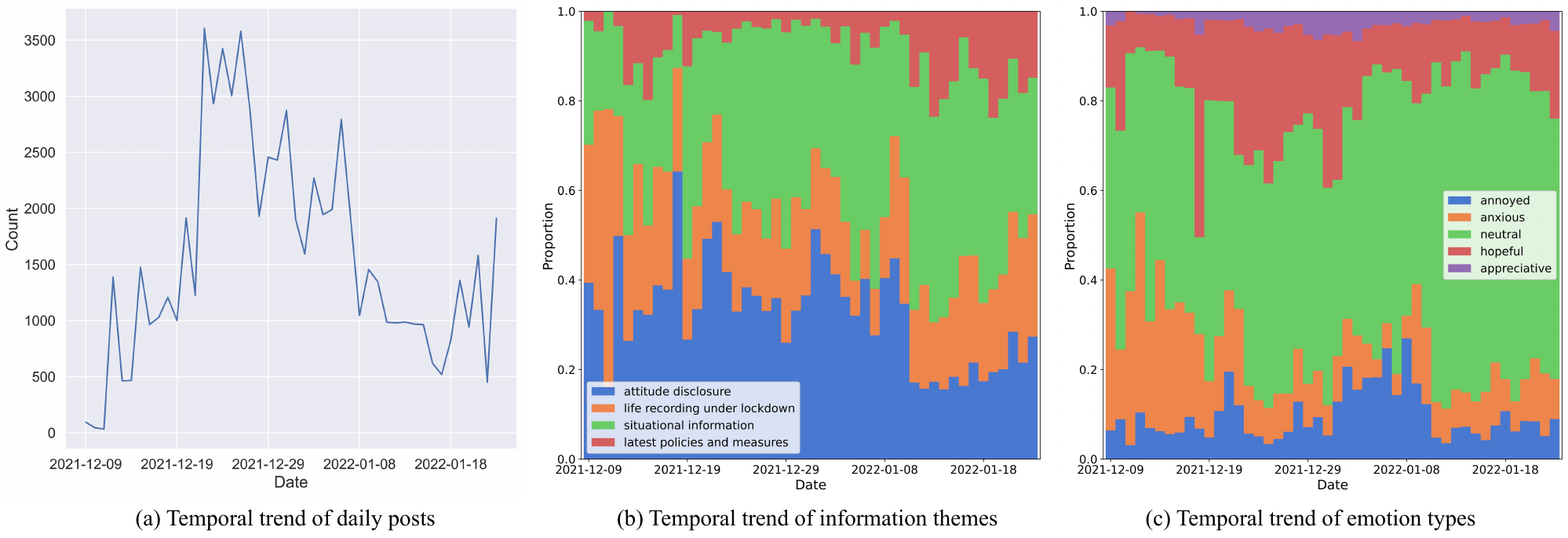}
	\caption{Temporal trend of the volume, information themes, and emotion types of crisis-related posts during Xi'an COVID-19 local outbreak.}
	\label{FIG: text_temporal}
\end{figure}

Figure \ref{FIG: text_temporal} described the temporal trends of post volume, information themes and emotion types during the event. The number of posts gradually increased at the early stage of the outbreak, and reached the first peak on Dec. 22, 2021 with the announcement of lockdown~\cite{XianlocalOutbreakWiki}. The discourse gradually tapered off around Jan. 10, 2022 when the social transmission was controlled with no new cases~\cite{XianlocalOutbreakWiki}. The information themes and emotion types exhibited notable temporal patterns along with the development of the local outbreak. Specifically, \textit{latest policies and measures} accounted for a significant proportion during the initial and final phases of the event. \textit{Situational information} gradually grew at the onset of the outbreak and became stable since the lockdown. Besides, \textit{attitude disclosure} was prevalent until the outbreak was brought under basic control. For emotion types, \textit{anxious} prevailed at the outset of the outbreak, and \textit{annoyed} manifested two peaks, one during the initial lockdown period and another after the lockdown had been in effect for some time. Meanwhile, \textit{hopeful} posts remained in a great amount during the lockdown period with the decrease of cases.

These descriptive statistics helped to contextualize the crisis imagery regarding information sharing and sentiment expression in the corpus. Next, we focused on how different types of crisis images were adopted for different informational and emotional goals.

\subsubsection{Information Sharing}\label{informationFinding}

\begin{figure}
	\centering
		\includegraphics[scale=.35]{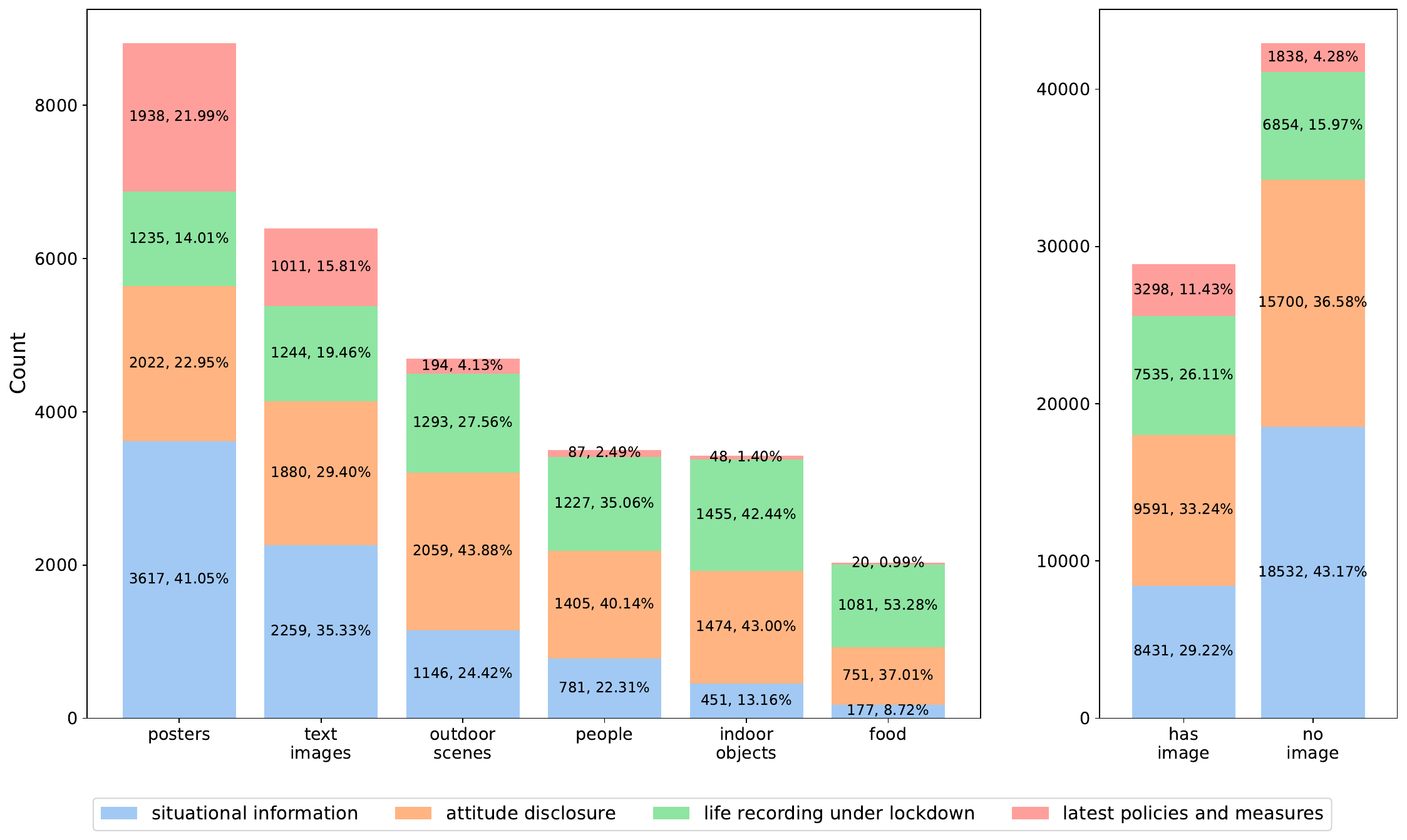}
	\caption{Distribution of crisis-related information themes in different types of crisis images. The chi-square test indicates a statistically significant difference in shared information among different crisis image types ($p < 0.001$).}
	\label{FIG: VisualGoals}
\end{figure}

Figure \ref{FIG: VisualGoals} demonstrates the proportions of crisis information themes in different types of crisis images. Through a chi-square test, we revealed that there was \textbf{a statistically significant difference in shared information among different crisis image types} ($p < 0.001$). It suggests that the types of images used in a post largely correlated with the messages the post aimed to target. We underscore the following critical findings: (1) Compared to crisis-related posts without images, image-attached posts were far more widely adopted to share the \textit{latest policies and measures}. It indicates that images were widely leveraged to signify or demonstrate formal and official information, which will be further illustrated in Section \ref{findings: RQ3}; (2)\textit{posters} contained the highest proportion of \textit{situational information} (41.05\%, N=3617) and the \textit{latest policies and measures} (21.99\%, N=1938), followed by \textit{text images}. In comparison, all remaining categories had less than 25\% \textit{situational information} and less than 5\% \textit{latest policies and measures}. This finding indicates that \textit{posters} and \textit{text images} had been broadly used to transmit policies and other situational information that were crucial in crisis response; (3) \textit{People}, \textit{outdoor scenes}, \textit{indoor objects} and \textit{food} all had a high proportion (greater than 70\%) of posts for \textit{life recording under lockdown} and \textit{attitude disclosure}. Therefore, these four crisis image categories largely captured and embodied the public livelihood and mentality, which is crucial when strict lockdown measures are enforced~\cite{sameer2020assessment,adams2020impact}.

\subsubsection{Sentiment Expression}\label{sentimentResult}

\begin{figure}
	\centering
		\includegraphics[scale=.35]{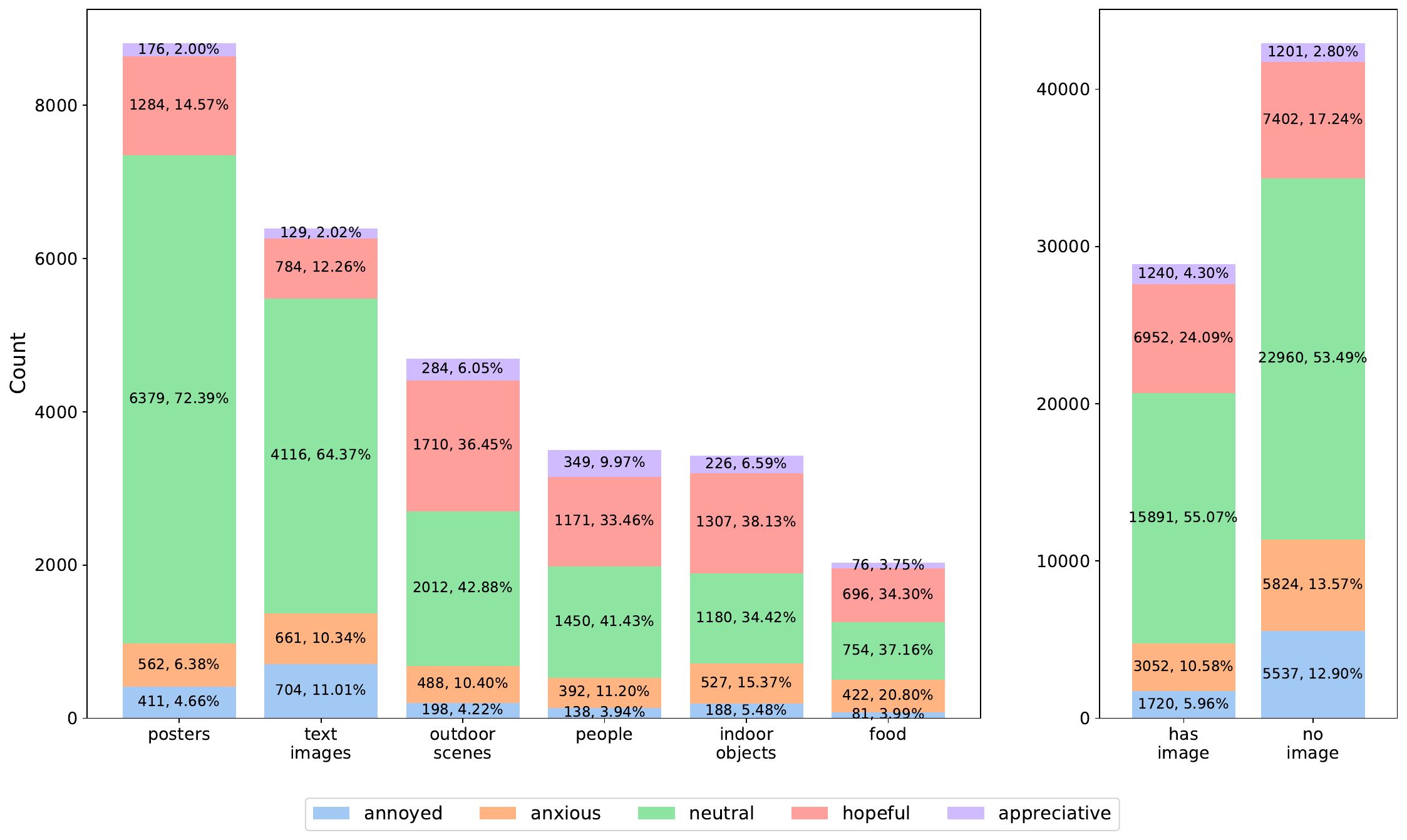}
	\caption{Distribution of crisis-related emotion categories in different types of crisis images. The chi-square test indicates a statistically significant difference in vented emotions among different crisis image types ($p < 0.001$).}
	\label{FIG: VisualGoalEmotion}
\end{figure}

Figure \ref{FIG: VisualGoalEmotion} demonstrates the proportions of crisis-related emotions in different types of crisis images. The chi-square test denoted that there was \textbf{a statistically significant difference in expressed emotions among different crisis image types} ($p < 0.001$). We note the following important findings: (1) crisis-related posts containing images were generally more positive than posts without images; (2) \textit{posters} was the most emotionally-neutral category containing 72.4\% neutral posts (N=6379), followed by \textit{text images}, which might be related to their nature for illustrating or signifying crisis-related information. Nevertheless, \textit{text images} contained a notable proportion of negative emotions. After manually reviewing a set of samples, we realized that users frequently posted screenshots (as \textit{text images}) to disclose their anxiety when the COVID-19 situation got worse (e.g., screenshots of an authoritative article on newly confirmed cases), or annoyance at the negligence of some staff (e.g., screenshots of chat revealing some staff's abuse of power); (3) \textit{outdoor scenes} and \textit{people} were strongly correlated with positive emotions, e.g., using outdoor scenery to express the wish for a normal life. In particular, the \textit{people} category had the highest proportion of \textit{appreciative} emotion, when some users posted photos of medical staff to express their gratitude; (4) \textit{food} was surprisingly the most negative category, especially with the highest \textit{anxiety} percentage. We observed that some users posted \textit{food} images to express their worry about food shortages or dissatisfaction with the food supply under lockdown; (5) \textit{indoor objects}, recording the isolated quarantine life, was the most emotional category with high volumes of both positive and negative emotions.

\textbf{RQ2 Summary:} Through text classification, we unveiled four dominant information types (i.e., \textit{Situational Information, Attitude Disclosure, Life Recording under Lockdown,} and \textit{Latest Policies and Measures}) and five emotion categories (\textit{Hopeful, Appreciative, Neutral, Annoyed}, and \textit{Anxious}) in the corpus, and quantified their correlations with crisis visual themes. Results indicated statistically significant differences in shared information and expressed emotions among different crisis image types, e.g., the wide use of \textit{posters} in disseminating the \textit{latest policies and measures}, and the prevalence of \textit{anxious} emotions regarding \textit{food} images. These findings demonstrate \textbf{inter-modality correlations} in crisis communication. In the next section, we provide more nuances on such inter-modality correlation, revealing how the strategic use of visuals complemented text narratives and contributed to effective crisis communication.

\subsection{RQ3: Visual Strategies}\label{findings: RQ3}

In this section, we demonstrate representative strategies of social media images to facilitate crisis communication, including \textit{images as signs of authority}, \textit{images as visual-based information enhancement}, \textit{images as evidence to improve credibility}, and \textit{images as triggers for empathy}.

\subsubsection{Images as Signs of Authority}\label{authority}

\begin{figure}
	\centering
		\includegraphics[scale=.4]{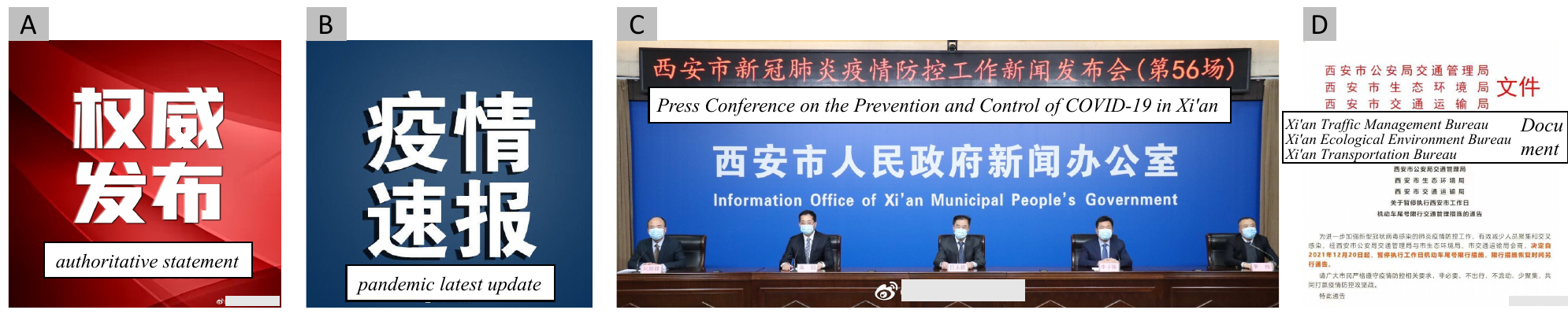}
	\caption{Crisis visual strategy 1: images signifying authority. Images with this strategy can be (1) big-character text embedded in a solid-color background, e.g., \textit{``authoritative statement''} in A and \textit{``pandemic latest updates''} in B; (2) photos of officials such as C, or (3) government documents such as D. They might not contain rich authoritative information within visual representations, but attract attention to and signify the authority of the post.}
	\label{FIG: theme1_authority}
\end{figure}

Adopting crisis images to signify authority surprisingly characterized visual crisis communication during the local outbreak in China as shown in Figure \ref{FIG: theme1_authority}. These signposting images, typically as \textit{posters} or \textit{text images}, might not contain rich authoritative information in themselves, yet helped to capture public attention, and attracted users to read the post text with authoritative situation updates or guidelines. Typical examples of this strategy are \textit{posters} with some big-character text (e.g., \textit{``authoritative statement''} in Figure \ref{FIG: theme1_authority}a, and \textit{``pandemic latest updates''} in Figure \ref{FIG: theme1_authority}b) embedded in a solid-color background, a clear indicator of official information in Chinese social media. \textit{Poster}-style photos of officials in the epidemic prevention press conference (Figure \ref{FIG: theme1_authority}c) were another kind of sign that implied an authoritative message in the text. Some users also forwarded \textit{text images} of government documents in COVID-19 management (Figure \ref{FIG: theme1_authority}d) to signify the authority of the transmitted text. Original posts along with this crisis image type mostly communicated the \textit{latest policies and measures}. As the quantitative evidence, about 90\% image-attached posts of \textit{latest policies and measures} had dominant visual types as \textit{posters} (58.8\%) and \textit{text images} (30.7\%), accounting for also about 60\% of all \textit{latest policies and measures} posts. Besides, \textit{latest policies and measures} posts using \textit{posters} or \textit{text images}, typically as authority signs, also attracted more likes ($\bar{X}_{likes}$=45.2), comments ($\bar{X}_{comments}$=6.8) and shares ($\bar{X}_{shares}$=3.0) on average compared to those with other visuals ($\bar{X}_{likes}$=14.2, $\bar{X}_{comments}$=4.4, $\bar{X}_{shares}$=2.5) or without images ($\bar{X}_{likes}$=29.6, $\bar{X}_{comments}$=4.6, $\bar{X}_{shares}$=1.2). These images shed light on how official, organizational, and individual users leveraged visuals to denote and strengthen the authority of posts, and thus promote the transmission and acceptance of the communicated crisis information during the local outbreak.

\subsubsection{Images as Visual-based Information Enhancement}\label{enhancement}

\begin{figure}
	\centering
		\includegraphics[scale=.4]{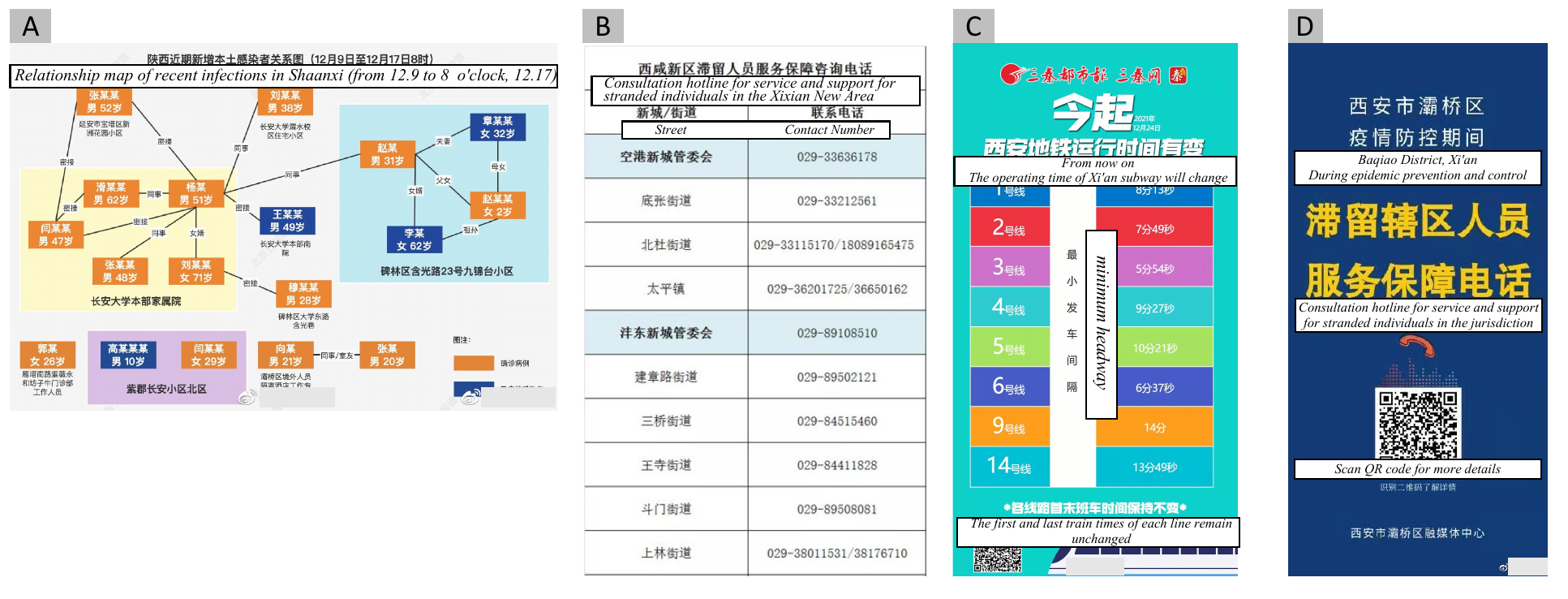}
	\caption{Crisis visual strategy 2: images as visual-based information enhancement. Images with this strategy can be (a) a diagram demonstrating epidemiological investigation, (b) a table summarizing helplines, (c) a colored table updating public transport information with a QR code, or (d) a QR code linking to helplines. These images, as ``visualization'' in a broad sense, efficiently convey situational information extending beyond textual narratives.}
	\label{FIG: theme2_enhancement}
\end{figure}

How experts and citizens use data visualization language (e.g., chart-based temporal visualization and map-based geospatial visualization) to better present crisis information has been a focus in crisis informatics literature~\cite{perovich2022self,zhang2021mapping,bica2019communicating}. This work revealed that users leveraged diverse visual representations, as ``visualizations'' in a broad sense, to convey what was difficult or obscure to express through plain language. Such crisis images well supplemented and enhanced situational crisis information in the post text. For example, Figure \ref{FIG: theme2_enhancement}a applied a diagram to elucidate the relationship and location of the infected population through epidemiological investigation, and Figure \ref{FIG: theme2_enhancement}b used a table to clearly summarize helplines in different districts during the local outbreak. QR codes, a special pattern to condense and route useful crisis information, were also frequently observed to achieve information augmentation, such as in Figure \ref{FIG: theme2_enhancement}c (public transport update during lockdown) and Figure \ref{FIG: theme2_enhancement}d (helplines during lockdown). These visualizations were mostly classified into \textit{posters} with a text-embedded visual style, which comprised 42.9\% visual use in \textit{situational information}. Indeed, \textit{posters} were not only more likely to be adopted in \textit{situational information} posts  (compared to the 30.5\% \textit{posters} use among all posts), but also achieved higher user engagement in \textit{situational information} ($\bar{X}_{likes}$=189.9, $\bar{X}_{comments}$=18.7, $\bar{X}_{shares}$=7.0) than other visuals ($\bar{X}_{likes}$=79.7, $\bar{X}_{comments}$=11.9, $\bar{X}_{shares}$=6.7). These examples point to ways in which multimodal communication with crisis images could enrich the presented situational knowledge in local outbreaks, going beyond what plain language could capture.

\subsubsection{Images as Evidence to Improve Credibility}\label{credibility}

\begin{figure}
	\centering
		\includegraphics[scale=.4]{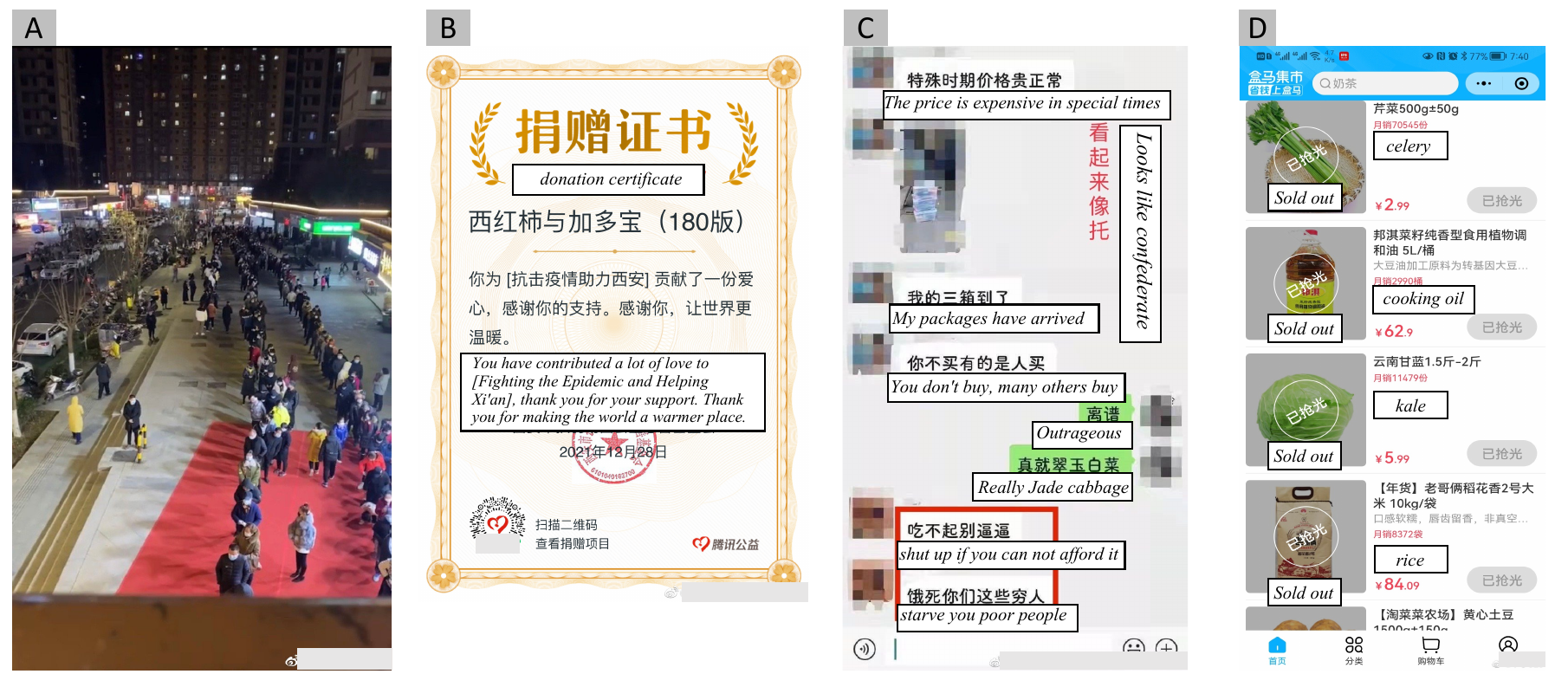}
	\caption{Crisis visual strategy 3: images as evidence to improve credibility. Images with this strategy can be (a) a physical photo proving the crowdedness of the COVID-19 test venues, (b) a digital picture proving donation, (c) a chat screenshot proving price gouging (\textit{``high price is normal during hard times''}), and (d) a screenshot of an online ordering platform (every item ``\textit{sold out}'') proving supply challenges. With visuals as evidence, these images manage to improve credibility and convince others when communicating situational information.}
	\label{FIG: theme3_evidence}
\end{figure}

Information credibility on social media has been a concern during health crises when rumors, conspiracy, and misinformation abound~\cite{zarocostas2020fight,sommariva2018spreading,kou2017conspiracy}. We found that social media users widely took images as evidence to improve the credibility of the shared information during the local break. Such visual evidence can either be physical photos (e.g., Figure \ref{FIG: theme3_evidence}a proving the crowdedness of COVID-19 test in a local community), digital pictures (e.g., Figure \ref{FIG: theme3_evidence}b proving poster's donation), or screenshots of mobile applications (e.g., Figure \ref{FIG: theme3_evidence}c and Figure \ref{FIG: theme3_evidence}d). Such visual evidence was a powerful way to demonstrate one's difficulties and challenges, especially when social media users in non-affected areas had no concrete idea of the local outbreak situation. For example, Figure \ref{FIG: theme3_evidence}c used a chat screenshot to reveal the price gouging during the local outbreak, and Figure \ref{FIG: theme3_evidence}d used a screenshot of an online shopping and ordering platform to give evidence of the supply shortage, both of which were persuasive ways to deliver particular challenges in the crisis. These images provide empirical evidence of how users took advantage of crisis images to convince others on social media. Nevertheless, we warn that visual-enhanced persuasion might potentially be more detrimental when containing unverified information, which is detailed in Section \ref{challenges}.

\subsubsection{Images as Triggers for Empathy}\label{empathy}

\begin{figure}
	\centering
		\includegraphics[scale=.4]{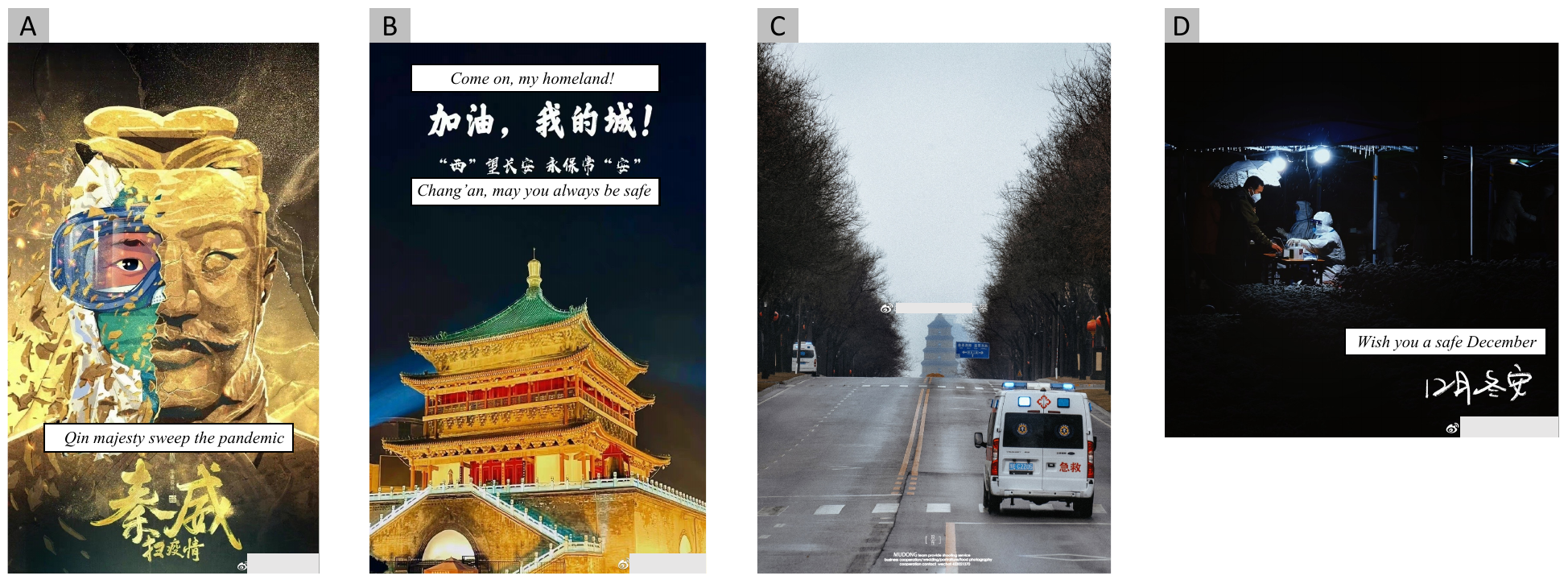}
	\caption{Crisis visual strategy 4: images as triggers for empathy. Images with this strategy can be local cultural symbols (Terracotta Warriors in A and ancient buildings in B) and local field photos (an empty street in C and a COVID-19 test at night in D). These visual elements could raise empathy among affected citizens with a common ground and transmit specific emotions such as encouragement.}
	\label{FIG: theme4_empathy}
\end{figure}

Through inductive coding, we identified that images were widely used as triggers to gain empathy and build emotional connections during the outbreak. Such images typically contained cultural, societal, or situational visual constructs that captured the sharing experience of the affected local population. For instance, Figure \ref{FIG: theme4_empathy}a and Figure \ref{FIG: theme4_empathy}b represent Terracotta Warriors and ancient buildings respectively, which were cultural symbols of the Xi'an city. The cultural symbols were integrated into the local challenge of the COVID-19 crisis, serving a role to call for unity and courage to overcome the difficulty. Figure \ref{FIG: theme4_empathy}c and Figure \ref{FIG: theme4_empathy}d used field photos of an empty street and a COVID-19 test at night to raise empathy among local residents under quarantine, expressing the wish for normal life and admiration of medical staff. In particular, we also found that visual use was more prevalent among \textit{positive (hopeful, appreciative)} posts (48.8\%) rather than \textit{neutral} (40.9\%) or \textit{negative (annoyed, anxious)} ones (29.6\%), potentially indicating the wider adoption of affective visuals in triggering and transmitting \textit{positive energy}~\cite{lu2021positive}. Besides, when both \textit{neutral} or \textit{negative} posts had \textit{text images} and \textit{posters} as the two most popular visual themes, \textit{hopeful} posts were mostly characterized by \textit{outdoor scenes} (24.6\%), potentially due to their use as symbols building emotional proximity; similarly, \textit{appreciative} posts had \textit{people} as the most dominant visual theme (28.1\%), when photos of medical staffs were given emotional values in collectively expressing gratitude. These examples shed light on how users exploited the power of the emotional contagion of visual symbols for collective emoting during a crisis.

\textbf{RQ3 Summary:} In this section, we described rich and nuanced strategies of crisis image use in contributing to effective crisis communication, including \textit{images as signs of authority}, \textit{images as visual-based information enhancement}, \textit{images as evidence to improve credibility}, and \textit{images as triggers for empathy}. These findings not only revealed the unique values of visual narratives in engaging, persuasion, and emoting, but also reflected how the strategic use of crisis images supplemented and enhanced text narratives to fulfill informational and emotional needs.

\section{DISCUSSION}

By investigating themes, goals, and strategies of crisis images in a COVID-19 local outbreak, this work enriches the understanding of the unique patterns and values of social media crisis imagery. In this section, we discuss the opportunities and challenges along with the proliferation of crisis imagery, and rethink the complementary roles of crisis images and language in crisis communication. We finally provide design implications for image-based searching and image-combined moderation to facilitate accurate and effective visual crisis communication.

\subsection{Unpacking Visual Crisis Communication on Social Media: Opportunities and Challenges}\label{challenges}

Existing work largely focuses on textual crisis communication to comprehend how people collaboratively cultivate situational awareness, develop risk perceptions, and provide informational and emotional support on social media under health crises (e.g., ~\cite{kou2017conspiracy, gui2018multidimensional, tsai2021help}). This work contributes to the literature on crisis communication by comprehensively unpacking the visual use on social media during a COVID-19 local outbreak. We reveal how crisis imagery could be rich and nuanced in themes, goals, and strategies, and play a significant role in fulfilling diverse communication needs in crises. In this section, we situate our findings within the existing literature, uncovering opportunities and challenges of visual communication.

Users developed diverse crisis visual use to adapt to specific needs during local outbreaks. For example, \textit{text images} embedded rich situational information in a single image, providing a convenient way for cross-platform sharing; in comparison, \textit{posters} strengthened the timeliness and credibility with highlighted meta information, managing to capture the audience's attention to the post text. Even different visual symbols of ``visual diary'' such as selfies and food, common in general social media for self-disclosure and social connection~\cite{manikonda2017modeling, hu2014we}, could convey crisis-specific information that visually enhances mutual understanding, cultivates situational awareness, and facilitates decision making (e.g., photos of long queues waiting for COVID-19 tests implying the crowdedness, and food images indicating community supply). These findings reflect the \textit{diversity and malleability of crisis imagery in conveying situational information with public wisdom}. It provides new evidence of self-organized civic participation and collective intelligence for crisis response~\cite{palen2010vision,dailey2015s} through social media images, and adds nuances to the understanding of strategic communication in collectively cultivating risk perceptions and assessments~\cite{demuth2018sometimes,gui2017managing}. In addition, these findings demonstrate that images could be a rich resource for public health researchers to \textit{uncover public understanding and attitudes towards health crises and crisis-related policies}, which resonates with prior work~\cite{he2021people,garimella2016social}. When images carry significant crisis communication roles, excluding image-based information could potentially lead to biases in understanding public opinions.

This work also revealed simple yet powerful emotional narratives~\cite{wang2019emotional,perovich2022self} in crisis images that could raise empathy among those sharing similar experiences and identities as described in Section \ref{empathy}. A photo of an empty fridge might express more anxiety than a large paragraph describing the food shortage, and a motivational poster integrating local cultural symbols might evoke more unity and courage than one thousand words. To this end, visual cues would not only serve as \textit{a vehicle for emotional disclosure}~\cite{manikonda2017modeling}, but also \textit{establish emotional connections and stimulate solidarity} among the affected populations in the crisis. It provides a new perspective on how visual narratives manage to build collective memory~\cite{porter2020visual} through cultural, societal, and situational constructs to transmit specific emotions during crises. We also discovered the strong correlation between visual cues and the expressed emotions given the specific context in Section \ref{findings: RQ2} (e.g., the prevalence of anxiety in \textit{food} images). These findings provided empirical evidence on how different visual symbols might be given different emotional values in the crisis setting, shedding light on the potential of \textit{affective visualization for crisis response}~\cite{perovich2022self}. 

On the other hand, given that crisis response and politics are tightly intertwined~\cite{jasanoff2021comparative}, we warn that such emotion-embedded crisis images might be exploited to \textit{weaponize emotion for political purposes}~\cite{boler2021affective, seo2014visual} \textit{or conspiracy theories}~\cite{kou2017conspiracy}. When particular emotions like fear or anger may influence people's crisis assessment and response~\cite{demuth2018sometimes,lindell2016immediate}, weaponized emotion may lead to biased perceptions and even exacerbate rumors and conflicts during crises. Such a hidden danger was particularly notable when prioritizing information valence over veracity characterizes crisis communication in Chinese social media~\cite{lu2021positive}. As such, it is significant for future work to situate crisis images in the sociopolitical context and examine the dynamics of affective visual narratives surrounding political objectives, persuasive communication, and information credibility.

This work also enriches the understanding of how \textit{visual-based misinformation}~\cite{brennen2021beyond} may go viral during local outbreaks. The information gap between local and non-local populations (or even different local communities during lockdown) naturally characterizes local outbreaks~\cite{tsai2021help}, in which crisis images shared by local stakeholders can be particularly persuasive. Section \ref{credibility} indicates that images, whether field photos or digital screenshots, were frequently used as evidence to improve credibility when ``pictures don't lie''. It vastly increases the misinformation risk when pictures do lie - for example, rumors expressed in faked screenshots of chat history~\cite{fakeScreenshot}, or appropriated photos from other scenarios~\cite{bica2017visual}. The emotions embedded in visual narratives may further promote the spread of visual-based misinformation. Therefore, we suggest future researchers look into the dynamics of how image-based misinformation affects local/non-local user perceptions and online/offline crisis response~\cite{brennen2021beyond,dewan2017towards,tomonto2019calamitous}, pay attention to the debunking effort of individuals and communities, and develop AI-supported or crowdsourcing-based countermeasures to cope.

\subsection{Multimodal Crisis Communication: Rethinking the Complementary Roles of Crisis Images and Language}

Images and language, two common modalities in social media, have been considered as two supplementary vehicles for communication and jointly provide affecting and persuasive narratives~\cite{manikonda2017modeling}. Nonetheless, how visual and linguistic elements complement each other to fulfill informational and emotional needs in crisis communication has been less investigated. This work reveals the nuances of the complementary roles of the two modalities in crisis communication.

Through qualitative analysis of crisis images within the context of a post, we revealed the distinct strategies of images to spotlight, enhance, and enrich situational information in the text. Images could either serve as simple yet powerful signs of authority (Section \ref{authority}), catching public attention on the text-based authoritative guidelines; work as an auxiliary component to support and verify the linguistic argument (Section \ref{credibility}), visually improving the information credibility; or embody rich situational information within images themselves through various visualization approaches (Section \ref{enhancement}), thus breaking the limit that text can express. These findings enrich the understanding of multimodal information sharing~\cite{osatuyi2013information} by \textbf{unpacking dynamics of visual-language correlations in crises}. Crisis images are more than a standalone component with visualized information, but connect with text and enhance the entire information through attention attraction, credibility indication~\cite{osatuyi2013information} and information embodiment~\cite{perovich2022self}. These visual-language correlations exhibit the potential to correspondingly alleviate the critical challenges in \textit{heterogeneity}, \textit{credibility}, and \textit{quality} of crisis information on social media~\cite{palen2010vision}. Moreover, the inter-modality correlations also afford user empowerment with evolving user-developed strategies for effective crisis communication, which resonates with prior work~\cite{pulos2020covid,he2021beyond}. To this end, we call for more fine-grained investigations of public strategies in establishing visual-language correlations to cope with specific challenges in crisis communication (e.g., attracting attention to critical posts among heterogeneous crisis information~\cite{he2022help}), and how they influence risk perceptions and emergency responses.

Social media images also play an irreplaceable role that complements language in communicating emotions during crises. On the one hand, visual elements afford particular values in emotion expression through emotion embodiment~\cite{perovich2022self} and emotion contagion through constructing collective memory~\cite{bica2017visual,porter2020visual}, which extend beyond linguistic narratives. On the other hand, when image-based emotion venting might suffer from ambiguity and subjectivity in interpretation~\cite{bica2017visual}, language is critical for contextualizing the visual constructs and facilitating emotion comprehension, achieving complementarity in emotional communication. On this note, it is warranted for future work to examine the strategic use and misuse of inter-modality emotional connections in crisis settings. For example, it is significant in understanding how conspiracy theorists may build malicious inter-modality connections~\cite{micallef2022cross}, e.g., situating unrelated but emotional crisis images within misinformation, to promote the spread of crisis-related rumors.

Generally, this work enriches the understanding of inter-modality correlation that empowers users for strategic crisis communication. Unpacking such inter-modality correlation would be as important as understanding information in each modality. Therefore, we suggest \textit{a holistic view} that takes the text and visual as an organic whole for further researchers to get a comprehensive understanding of multimodal crisis communication~\cite{unal2022visual}, focusing on not only the uniqueness of one modality, but also the correlation and complementarity between modalities.

\subsection{Design Implications}

\subsubsection{Implications for Crisis Information Seeking}

This work unpacks the significant role of crisis imagery, from authoritative guidelines through text images to situation descriptions through on-site photos, in communicating situational crisis information. Such public-generated visual knowledge, as supplements to textual information, is meaningful for both local victims who may be subject to movement restrictions and non-local people who may lack a comprehensive understanding of crisis~\cite{bica2019communicating}. That warrants an overarching design goal in \textit{supporting multimodal crisis information seeking}.

First, this work indicated the feasibility of automatic algorithms in classifying diverse crisis visuals, which were correlated with different informational and emotional goals. It sheds light on the potential of \textit{AI-supported visual categorization} to enhance crisis visual seeking. For example, the affordance of filtering based on AI-generated crisis visual categories could empower users to navigate all visuals categorized as ``text images'' to quickly find authoritative documents in crisis management.

Besides, we suggest designers broadly \textit{leverage the natural connection of multimodal data} to help users sift through crisis information. For instance, during crises with movement restrictions (e.g., lockdowns or natural disasters), the navigation of crisis visuals could largely benefit from the incorporation of geolocation information. This work revealed users' wide adoption of in-situ photos, such as outdoor scenes of COVID-19 test queues or supply locations, in spreading situational information. Nonetheless, other users may not easily capture such context-specific visual information under the overwhelming amount of crisis-related data~\cite{imran2020using}. In this scenario, allowing users to view the ``outdoor'' visual category and sort based on location proximity could empower them to easily acquire situations in the local community.

Finally, we suggest designers actively \textit{situate crisis information systems within the sociocultural and sociopolitical context} to propose specific interfaces for crisis communication. For example, when it is a norm to deliver authoritative guidelines along with visual authority signs in Chinese social media (as shown in Section \ref{authority}), automatically identifying authority signs and affording the interface to browse relevant posts could support the seeking of authoritative crisis information.

\subsubsection{Implications for Crisis Information Moderation}

The findings on crisis visual strategies in Section \ref{findings: RQ3} indicated that crisis images were often endowed with additional roles for crisis communication such as signifying authority, enhancing information, or triggering public emotions. Such visual use was typically correlated with higher public engagement indexes. These visual strategies could enhance crisis communication when properly delivered, but might also contribute to the spread of misinformation if the veracity was unchecked. Therefore, we suggest design interventions particularly pay attention to such critical visual use in crisis communication. For example, an AI-supported tool that \textit{prioritizes images with critical roles (e.g., images signifying authority) in fact-checking} might help capture high-risk misinformation before its wide dissemination. Also, such \textit{visual-role reminders} could be embedded into before-posting nudges~\cite{jahanbakhsh2021exploring} or during-exposure warnings~\cite{jia2022understanding} (e.g., remind sharers by ``This image indicates the authority of the post text. Please double-check the content'').

\subsection{Technical Implications}

This study also revealed the limitations of current methodologies that would be valuable to note for further multimodal analysis of crisis communication. First, this work reflected how a lack of \textit{cross-cultural and in-domain datasets} may bring unexpected algorithmic misinterpretations for multimodal analysis in the crisis setting. For example, deep learning models pretrained on scene-specific datasets wrongly categorized COVID-19 lockdown checkpoints in China as ice cream parlors. Such constraints also naturally limited the generalization of more fine-grained crisis-related types through image clustering. Therefore, it warrants the development of large-scale multimodal datasets for specific crisis settings, which echoed prior work~\cite{imran2020using,alam2018crisismmd}. A more high-level challenge was that mainstream object-centered analytic tools (e.g., image classification and object detection) may fail to unpack the nuanced roles of multimodal crisis communication. For example, recognizing an image as a ``chat screenshot'' might be less meaningful compared to revealing its actual goal as ``evidencing challenges mentioned in the post''. Therefore, content analysis might be only the starting point in understanding multimodal crisis communication rather than the ultimate goal.

On the other hand, the analytical approach employed in this study has highlighted potential avenues for conducting multimodal analyses of crisis-related data. We provide empirical evidence on the feasibility of transfer learning in understanding multimodal crisis communication, which could largely reduce the burden of manually curating labels. Also, this work integrated CV-based visual analysis and NLP-based text analysis to unpack the visual-language correlations for crisis communication. It enlightens a promising direction to adopt multimodal deep learning models~\cite{xu2023multimodal} to process multimodal crisis data as an organic whole.

\subsection{Limitations}

This work investigated the themes, goals, and strategies of social media crisis images during a COVID-19 local outbreak, deepening the understanding of visual crisis communication in HCI and CSCW. Nonetheless, this work has the following limitations: (1) this work focuses on a COVID-19 local outbreak in China in the aftermath of the pandemic. Therefore, some findings may not be generalized to other health crisis settings; (2) the image clustering method managed to capture visual elements such as people and objects in crisis images, but was still limited in discerning detailed visual themes, such as topics of text image; (3) as a secondary analysis, we did not investigate how the affected population perceived crisis images, and understanding users' perceptions on crisis images might provide more in-depth findings of the crisis visual strategies. When multimodal crisis communication has been an increasingly prevalent practice, we call for broad investigations into different crisis communication settings to reveal the full potential of crisis imagery. As a further step, it is also of great significance for future work to propose and evaluate proof-of-concept interfaces to facilitate and enhance visual crisis communication as well as maximally reduce visual-based misinformation.

\section{CONCLUSION}

This work makes the first investigation into how crisis communication is characterized and facilitated by social media crisis images during a COVID-19 local outbreak. Focusing on the Xi'an local outbreak in China, we collected 345,423 crisis-related posts and 65,376 original images on a popular Chinese social media platform Weibo, and conducted a mixed-methods study to understand visual themes, goals, and strategies in crisis communication. Through an image clustering approach, we unpacked the diversity of crisis imagery with two text-embedded visual categories (i.e., posters and text images) and four ``visual diary'' types that recorded life during the event (e.g., outdoor scenes during quarantine). We demonstrated how different visual types were leveraged to fulfill various informational and emotional communication goals, e.g., using text-highlighted posters to signify the latest policies and embedding anxiety into visual diaries of quarantine life. Users developed strategic use of crisis images as signs of authority, visual-based information enhancement, evidence to improve credibility, and triggers for empathy. We discuss opportunities and challenges of visual crisis communication, reflect on the inter-modality correlation and complementarity, and propose design implications to facilitate effective and accurate visual crisis communication.

\bibliographystyle{ACM-Reference-Format}
\bibliography{main}


\begin{thebibliography}{143}


\ifx \showCODEN    \undefined \def \showCODEN     #1{\unskip}     \fi
\ifx \showDOI      \undefined \def \showDOI       #1{#1}\fi
\ifx \showISBNx    \undefined \def \showISBNx     #1{\unskip}     \fi
\ifx \showISBNxiii \undefined \def \showISBNxiii  #1{\unskip}     \fi
\ifx \showISSN     \undefined \def \showISSN      #1{\unskip}     \fi
\ifx \showLCCN     \undefined \def \showLCCN      #1{\unskip}     \fi
\ifx \shownote     \undefined \def \shownote      #1{#1}          \fi
\ifx \showarticletitle \undefined \def \showarticletitle #1{#1}   \fi
\ifx \showURL      \undefined \def \showURL       {\relax}        \fi
\providecommand\bibfield[2]{#2}
\providecommand\bibinfo[2]{#2}
\providecommand\natexlab[1]{#1}
\providecommand\showeprint[2][]{arXiv:#2}

\bibitem[Abdullah et~al\mbox{.}(2015)]%
        {abdullah2015collective}
\bibfield{author}{\bibinfo{person}{Saeed Abdullah}, \bibinfo{person}{Elizabeth~L Murnane}, \bibinfo{person}{Jean~MR Costa}, {and} \bibinfo{person}{Tanzeem Choudhury}.} \bibinfo{year}{2015}\natexlab{}.
\newblock \showarticletitle{Collective smile: Measuring societal happiness from geolocated images}. In \bibinfo{booktitle}{\emph{Proceedings of the 18th ACM Conference on Computer Supported Cooperative Work \& Social Computing}}. \bibinfo{pages}{361--374}.
\newblock


\bibitem[Abouk and Heydari(2021)]%
        {abouk2021immediate}
\bibfield{author}{\bibinfo{person}{Rahi Abouk} {and} \bibinfo{person}{Babak Heydari}.} \bibinfo{year}{2021}\natexlab{}.
\newblock \showarticletitle{The immediate effect of COVID-19 policies on social-distancing behavior in the United States}.
\newblock \bibinfo{journal}{\emph{Public health reports}} \bibinfo{volume}{136}, \bibinfo{number}{2} (\bibinfo{year}{2021}), \bibinfo{pages}{245--252}.
\newblock


\bibitem[Adams-Prassl et~al\mbox{.}(2020)]%
        {adams2020impact}
\bibfield{author}{\bibinfo{person}{Abi Adams-Prassl}, \bibinfo{person}{Teodora Boneva}, \bibinfo{person}{Marta Golin}, {and} \bibinfo{person}{Christopher Rauh}.} \bibinfo{year}{2020}\natexlab{}.
\newblock \showarticletitle{The impact of the coronavirus lockdown on mental health: evidence from the US}.
\newblock  (\bibinfo{year}{2020}).
\newblock


\bibitem[Alam et~al\mbox{.}(2018)]%
        {alam2018crisismmd}
\bibfield{author}{\bibinfo{person}{Firoj Alam}, \bibinfo{person}{Ferda Ofli}, {and} \bibinfo{person}{Muhammad Imran}.} \bibinfo{year}{2018}\natexlab{}.
\newblock \showarticletitle{Crisismmd: Multimodal twitter datasets from natural disasters}. In \bibinfo{booktitle}{\emph{Twelfth international AAAI conference on web and social media}}.
\newblock


\bibitem[Andalibi et~al\mbox{.}(2017)]%
        {andalibi2017sensitive}
\bibfield{author}{\bibinfo{person}{Nazanin Andalibi}, \bibinfo{person}{Pinar Ozturk}, {and} \bibinfo{person}{Andrea Forte}.} \bibinfo{year}{2017}\natexlab{}.
\newblock \showarticletitle{Sensitive Self-disclosures, Responses, and Social Support on Instagram: the case of\# depression}. In \bibinfo{booktitle}{\emph{Proceedings of the 2017 ACM conference on computer supported cooperative work and social computing}}. \bibinfo{pages}{1485--1500}.
\newblock


\bibitem[Asif et~al\mbox{.}(2021)]%
        {asif2021automatic}
\bibfield{author}{\bibinfo{person}{Amna Asif}, \bibinfo{person}{Shaheen Khatoon}, \bibinfo{person}{Md~Maruf Hasan}, \bibinfo{person}{Majed~A Alshamari}, \bibinfo{person}{Sherif Abdou}, \bibinfo{person}{Khaled~Mostafa Elsayed}, {and} \bibinfo{person}{Mohsen Rashwan}.} \bibinfo{year}{2021}\natexlab{}.
\newblock \showarticletitle{Automatic analysis of social media images to identify disaster type and infer appropriate emergency response}.
\newblock \bibinfo{journal}{\emph{Journal of Big Data}} \bibinfo{volume}{8}, \bibinfo{number}{1} (\bibinfo{year}{2021}), \bibinfo{pages}{1--28}.
\newblock


\bibitem[Bakhshi et~al\mbox{.}(2015)]%
        {bakhshi2015we}
\bibfield{author}{\bibinfo{person}{Saeideh Bakhshi}, \bibinfo{person}{David Shamma}, \bibinfo{person}{Lyndon Kennedy}, {and} \bibinfo{person}{Eric Gilbert}.} \bibinfo{year}{2015}\natexlab{}.
\newblock \showarticletitle{Why we filter our photos and how it impacts engagement}. In \bibinfo{booktitle}{\emph{Proceedings of the International AAAI Conference on Web and Social Media}}, Vol.~\bibinfo{volume}{9}. \bibinfo{pages}{12--21}.
\newblock


\bibitem[Bica et~al\mbox{.}(2019)]%
        {bica2019communicating}
\bibfield{author}{\bibinfo{person}{Melissa Bica}, \bibinfo{person}{Julie~L Demuth}, \bibinfo{person}{James~E Dykes}, {and} \bibinfo{person}{Leysia Palen}.} \bibinfo{year}{2019}\natexlab{}.
\newblock \showarticletitle{Communicating hurricane risks: Multi-method examination of risk imagery diffusion}. In \bibinfo{booktitle}{\emph{Proceedings of the 2019 CHI Conference on Human Factors in Computing Systems}}. \bibinfo{pages}{1--13}.
\newblock


\bibitem[Bica et~al\mbox{.}(2017)]%
        {bica2017visual}
\bibfield{author}{\bibinfo{person}{Melissa Bica}, \bibinfo{person}{Leysia Palen}, {and} \bibinfo{person}{Chris Bopp}.} \bibinfo{year}{2017}\natexlab{}.
\newblock \showarticletitle{Visual representations of disaster}. In \bibinfo{booktitle}{\emph{Proceedings of the 2017 ACM conference on computer supported cooperative work and social computing}}. \bibinfo{pages}{1262--1276}.
\newblock


\bibitem[Boler and Davis(2021)]%
        {boler2021affective}
\bibfield{author}{\bibinfo{person}{Megan Boler} {and} \bibinfo{person}{E Davis}.} \bibinfo{year}{2021}\natexlab{}.
\newblock \bibinfo{booktitle}{\emph{Affective politics of digital media}}.
\newblock \bibinfo{publisher}{New York: Routledge}.
\newblock


\bibitem[Bowe et~al\mbox{.}(2020)]%
        {bowe2020learning}
\bibfield{author}{\bibinfo{person}{Emily Bowe}, \bibinfo{person}{Erin Simmons}, {and} \bibinfo{person}{Shannon Mattern}.} \bibinfo{year}{2020}\natexlab{}.
\newblock \showarticletitle{Learning from lines: Critical COVID data visualizations and the quarantine quotidian}.
\newblock \bibinfo{journal}{\emph{Big data \& society}} \bibinfo{volume}{7}, \bibinfo{number}{2} (\bibinfo{year}{2020}), \bibinfo{pages}{2053951720939236}.
\newblock


\bibitem[Brantner et~al\mbox{.}(2020)]%
        {brantner2020memes}
\bibfield{author}{\bibinfo{person}{Cornelia Brantner}, \bibinfo{person}{Katharina Lobinger}, {and} \bibinfo{person}{Miriam Stehling}.} \bibinfo{year}{2020}\natexlab{}.
\newblock \showarticletitle{Memes against sexism? A multi-method analysis of the feminist protest hashtag\# distractinglysexy and its resonance in the mainstream news media}.
\newblock \bibinfo{journal}{\emph{Convergence}} \bibinfo{volume}{26}, \bibinfo{number}{3} (\bibinfo{year}{2020}), \bibinfo{pages}{674--696}.
\newblock


\bibitem[Brennen et~al\mbox{.}(2021)]%
        {brennen2021beyond}
\bibfield{author}{\bibinfo{person}{J~Scott Brennen}, \bibinfo{person}{Felix~M Simon}, {and} \bibinfo{person}{Rasmus~Kleis Nielsen}.} \bibinfo{year}{2021}\natexlab{}.
\newblock \showarticletitle{Beyond (mis) representation: Visuals in COVID-19 misinformation}.
\newblock \bibinfo{journal}{\emph{The International Journal of Press/Politics}} \bibinfo{volume}{26}, \bibinfo{number}{1} (\bibinfo{year}{2021}), \bibinfo{pages}{277--299}.
\newblock


\bibitem[Brownson et~al\mbox{.}(2020)]%
        {brownson2020reimagining}
\bibfield{author}{\bibinfo{person}{Ross~C Brownson}, \bibinfo{person}{Thomas~A Burke}, \bibinfo{person}{Graham~A Colditz}, {and} \bibinfo{person}{Jonathan~M Samet}.} \bibinfo{year}{2020}\natexlab{}.
\newblock \showarticletitle{Reimagining public health in the aftermath of a pandemic}.
\newblock \bibinfo{journal}{\emph{American journal of public health}} \bibinfo{volume}{110}, \bibinfo{number}{11} (\bibinfo{year}{2020}), \bibinfo{pages}{1605--1610}.
\newblock


\bibitem[Chen et~al\mbox{.}(2021)]%
        {chen2021exploring}
\bibfield{author}{\bibinfo{person}{Yixin Chen}, \bibinfo{person}{Ke-Rou Wang}, \bibinfo{person}{Weikai Xu}, {and} \bibinfo{person}{Yun Huang}.} \bibinfo{year}{2021}\natexlab{}.
\newblock \showarticletitle{Exploring Commenting Behavior in the COVID-19 Super-Topic on Weibo}. In \bibinfo{booktitle}{\emph{Extended Abstracts of the 2021 CHI Conference on Human Factors in Computing Systems}}. \bibinfo{pages}{1--7}.
\newblock


\bibitem[Cho et~al\mbox{.}(2023)]%
        {cho2023bright}
\bibfield{author}{\bibinfo{person}{Hichang Cho}, \bibinfo{person}{Pengxiang Li}, \bibinfo{person}{Annabel Ngien}, \bibinfo{person}{Marion~Grace Tan}, \bibinfo{person}{Anfan Chen}, {and} \bibinfo{person}{Elmie Nekmat}.} \bibinfo{year}{2023}\natexlab{}.
\newblock \showarticletitle{The bright and dark sides of social media use during COVID-19 lockdown: Contrasting social media effects through social liability vs. social support}.
\newblock \bibinfo{journal}{\emph{Computers in Human Behavior}}  \bibinfo{volume}{146} (\bibinfo{year}{2023}), \bibinfo{pages}{107795}.
\newblock


\bibitem[Cooke(2005)]%
        {cooke2005visual}
\bibfield{author}{\bibinfo{person}{Lynne Cooke}.} \bibinfo{year}{2005}\natexlab{}.
\newblock \showarticletitle{A visual convergence of print, television, and the internet: charting 40 years of design change in news presentation}.
\newblock \bibinfo{journal}{\emph{New Media \& Society}} \bibinfo{volume}{7}, \bibinfo{number}{1} (\bibinfo{year}{2005}), \bibinfo{pages}{22--46}.
\newblock


\bibitem[Coombs(2020)]%
        {coombs2020conceptualizing}
\bibfield{author}{\bibinfo{person}{W~Timothy Coombs}.} \bibinfo{year}{2020}\natexlab{}.
\newblock \showarticletitle{Conceptualizing crisis communication}.
\newblock In \bibinfo{booktitle}{\emph{Handbook of risk and crisis communication}}. \bibinfo{publisher}{Routledge}, \bibinfo{pages}{99--118}.
\newblock


\bibitem[Corbin and Strauss(2014)]%
        {corbin2014basics}
\bibfield{author}{\bibinfo{person}{Juliet Corbin} {and} \bibinfo{person}{Anselm Strauss}.} \bibinfo{year}{2014}\natexlab{}.
\newblock \bibinfo{booktitle}{\emph{Basics of qualitative research: Techniques and procedures for developing grounded theory}}.
\newblock \bibinfo{publisher}{Sage publications}.
\newblock


\bibitem[Cui et~al\mbox{.}(2021)]%
        {cui2021pre}
\bibfield{author}{\bibinfo{person}{Yiming Cui}, \bibinfo{person}{Wanxiang Che}, \bibinfo{person}{Ting Liu}, \bibinfo{person}{Bing Qin}, {and} \bibinfo{person}{Ziqing Yang}.} \bibinfo{year}{2021}\natexlab{}.
\newblock \showarticletitle{Pre-training with whole word masking for chinese bert}.
\newblock \bibinfo{journal}{\emph{IEEE/ACM Transactions on Audio, Speech, and Language Processing}}  \bibinfo{volume}{29} (\bibinfo{year}{2021}), \bibinfo{pages}{3504--3514}.
\newblock


\bibitem[Dailey and Starbird(2015)]%
        {dailey2015s}
\bibfield{author}{\bibinfo{person}{Dharma Dailey} {and} \bibinfo{person}{Kate Starbird}.} \bibinfo{year}{2015}\natexlab{}.
\newblock \showarticletitle{" It's Raining Dispersants" Collective Sensemaking of Complex Information in Crisis Contexts}. In \bibinfo{booktitle}{\emph{Proceedings of the 18th ACM Conference Companion on Computer Supported Cooperative Work \& Social Computing}}. \bibinfo{pages}{155--158}.
\newblock


\bibitem[Daly and Thom(2016)]%
        {daly2016mining}
\bibfield{author}{\bibinfo{person}{Shannon Daly} {and} \bibinfo{person}{James~A Thom}.} \bibinfo{year}{2016}\natexlab{}.
\newblock \showarticletitle{Mining and Classifying Image Posts on Social Media to Analyse Fires.}. In \bibinfo{booktitle}{\emph{ISCRAM}}. Citeseer, \bibinfo{pages}{1--14}.
\newblock


\bibitem[Demuth et~al\mbox{.}(2018)]%
        {demuth2018sometimes}
\bibfield{author}{\bibinfo{person}{Julie~L Demuth}, \bibinfo{person}{Rebecca~E Morss}, \bibinfo{person}{Leysia Palen}, \bibinfo{person}{Kenneth~M Anderson}, \bibinfo{person}{Jennings Anderson}, \bibinfo{person}{Marina Kogan}, \bibinfo{person}{Kevin Stowe}, \bibinfo{person}{Melissa Bica}, \bibinfo{person}{Heather Lazrus}, \bibinfo{person}{Olga Wilhelmi}, {et~al\mbox{.}}} \bibinfo{year}{2018}\natexlab{}.
\newblock \showarticletitle{“Sometimes da\# beachlife ain't always da wave”: Understanding People’s Evolving Hurricane Risk Communication, Risk Assessments, and Responses Using Twitter Narratives}.
\newblock \bibinfo{journal}{\emph{Weather, climate, and society}} \bibinfo{volume}{10}, \bibinfo{number}{3} (\bibinfo{year}{2018}), \bibinfo{pages}{537--560}.
\newblock


\bibitem[Deng et~al\mbox{.}(2009)]%
        {deng2009imagenet}
\bibfield{author}{\bibinfo{person}{Jia Deng}, \bibinfo{person}{Wei Dong}, \bibinfo{person}{Richard Socher}, \bibinfo{person}{Li-Jia Li}, \bibinfo{person}{Kai Li}, {and} \bibinfo{person}{Li Fei-Fei}.} \bibinfo{year}{2009}\natexlab{}.
\newblock \showarticletitle{Imagenet: A large-scale hierarchical image database}. In \bibinfo{booktitle}{\emph{2009 IEEE conference on computer vision and pattern recognition}}. Ieee, \bibinfo{pages}{248--255}.
\newblock


\bibitem[Deng et~al\mbox{.}(2023)]%
        {deng2023understanding}
\bibfield{author}{\bibinfo{person}{Yue Deng}, \bibinfo{person}{Changyang He}, {and} \bibinfo{person}{Bo Li}.} \bibinfo{year}{2023}\natexlab{}.
\newblock \showarticletitle{Understanding Emotional Disclosure via Diary-keeping in Quarantine on Social Media}. In \bibinfo{booktitle}{\emph{The Eleventh International Symposium of Chinese CHI}}.
\newblock


\bibitem[Devlin et~al\mbox{.}(2018)]%
        {devlin2018bert}
\bibfield{author}{\bibinfo{person}{Jacob Devlin}, \bibinfo{person}{Ming-Wei Chang}, \bibinfo{person}{Kenton Lee}, {and} \bibinfo{person}{Kristina Toutanova}.} \bibinfo{year}{2018}\natexlab{}.
\newblock \showarticletitle{Bert: Pre-training of deep bidirectional transformers for language understanding}.
\newblock \bibinfo{journal}{\emph{arXiv preprint arXiv:1810.04805}} (\bibinfo{year}{2018}).
\newblock


\bibitem[Dewan et~al\mbox{.}(2017)]%
        {dewan2017towards}
\bibfield{author}{\bibinfo{person}{Prateek Dewan}, \bibinfo{person}{Anshuman Suri}, \bibinfo{person}{Varun Bharadhwaj}, \bibinfo{person}{Aditi Mithal}, {and} \bibinfo{person}{Ponnurangam Kumaraguru}.} \bibinfo{year}{2017}\natexlab{}.
\newblock \showarticletitle{Towards understanding crisis events on online social networks through pictures}. In \bibinfo{booktitle}{\emph{Proceedings of the 2017 IEEE/ACM International Conference on Advances in Social Networks Analysis and Mining 2017}}. \bibinfo{pages}{439--446}.
\newblock


\bibitem[Dosovitskiy et~al\mbox{.}(2020)]%
        {dosovitskiy2020image}
\bibfield{author}{\bibinfo{person}{Alexey Dosovitskiy}, \bibinfo{person}{Lucas Beyer}, \bibinfo{person}{Alexander Kolesnikov}, \bibinfo{person}{Dirk Weissenborn}, \bibinfo{person}{Xiaohua Zhai}, \bibinfo{person}{Thomas Unterthiner}, \bibinfo{person}{Mostafa Dehghani}, \bibinfo{person}{Matthias Minderer}, \bibinfo{person}{Georg Heigold}, \bibinfo{person}{Sylvain Gelly}, {et~al\mbox{.}}} \bibinfo{year}{2020}\natexlab{}.
\newblock \showarticletitle{An image is worth 16x16 words: Transformers for image recognition at scale}.
\newblock \bibinfo{journal}{\emph{arXiv preprint arXiv:2010.11929}} (\bibinfo{year}{2020}).
\newblock


\bibitem[Eriksson(2018)]%
        {eriksson2018lessons}
\bibfield{author}{\bibinfo{person}{Mats Eriksson}.} \bibinfo{year}{2018}\natexlab{}.
\newblock \showarticletitle{Lessons for crisis communication on social media: A systematic review of what research tells the practice}.
\newblock \bibinfo{journal}{\emph{International Journal of Strategic Communication}} \bibinfo{volume}{12}, \bibinfo{number}{5} (\bibinfo{year}{2018}), \bibinfo{pages}{526--551}.
\newblock


\bibitem[Fereday and Muir-Cochrane(2006)]%
        {fereday2006demonstrating}
\bibfield{author}{\bibinfo{person}{Jennifer Fereday} {and} \bibinfo{person}{Eimear Muir-Cochrane}.} \bibinfo{year}{2006}\natexlab{}.
\newblock \showarticletitle{Demonstrating rigor using thematic analysis: A hybrid approach of inductive and deductive coding and theme development}.
\newblock \bibinfo{journal}{\emph{International journal of qualitative methods}} \bibinfo{volume}{5}, \bibinfo{number}{1} (\bibinfo{year}{2006}), \bibinfo{pages}{80--92}.
\newblock


\bibitem[Ferrara et~al\mbox{.}(2020)]%
        {ferrara2020misinformation}
\bibfield{author}{\bibinfo{person}{Emilio Ferrara}, \bibinfo{person}{Stefano Cresci}, {and} \bibinfo{person}{Luca Luceri}.} \bibinfo{year}{2020}\natexlab{}.
\newblock \showarticletitle{Misinformation, manipulation, and abuse on social media in the era of COVID-19}.
\newblock \bibinfo{journal}{\emph{Journal of Computational Social Science}} \bibinfo{volume}{3}, \bibinfo{number}{2} (\bibinfo{year}{2020}), \bibinfo{pages}{271--277}.
\newblock


\bibitem[Garimella et~al\mbox{.}(2016)]%
        {garimella2016social}
\bibfield{author}{\bibinfo{person}{Venkata Rama~Kiran Garimella}, \bibinfo{person}{Abdulrahman Alfayad}, {and} \bibinfo{person}{Ingmar Weber}.} \bibinfo{year}{2016}\natexlab{}.
\newblock \showarticletitle{Social media image analysis for public health}. In \bibinfo{booktitle}{\emph{Proceedings of the 2016 CHI Conference on Human Factors in Computing Systems}}. \bibinfo{pages}{5543--5547}.
\newblock


\bibitem[Gillies et~al\mbox{.}(2005)]%
        {gillies2005painting}
\bibfield{author}{\bibinfo{person}{Val Gillies}, \bibinfo{person}{Angela Harden}, \bibinfo{person}{Katherine Johnson}, \bibinfo{person}{Paula Reavey}, \bibinfo{person}{Vicky Strange}, {and} \bibinfo{person}{Carla Willig}.} \bibinfo{year}{2005}\natexlab{}.
\newblock \showarticletitle{Painting pictures of embodied experience: The use of nonverbal data production for the study of embodiment}.
\newblock \bibinfo{journal}{\emph{Qualitative research in psychology}} \bibinfo{volume}{2}, \bibinfo{number}{3} (\bibinfo{year}{2005}), \bibinfo{pages}{199--212}.
\newblock


\bibitem[Goldberger et~al\mbox{.}(2002)]%
        {goldberger2002unsupervised}
\bibfield{author}{\bibinfo{person}{Jacob Goldberger}, \bibinfo{person}{Hayit Greenspan}, {and} \bibinfo{person}{Shiri Gordon}.} \bibinfo{year}{2002}\natexlab{}.
\newblock \showarticletitle{Unsupervised image clustering using the information bottleneck method}. In \bibinfo{booktitle}{\emph{Joint Pattern Recognition Symposium}}. Springer, \bibinfo{pages}{158--165}.
\newblock


\bibitem[Grimmer and Stewart(2013)]%
        {grimmer2013text}
\bibfield{author}{\bibinfo{person}{Justin Grimmer} {and} \bibinfo{person}{Brandon~M Stewart}.} \bibinfo{year}{2013}\natexlab{}.
\newblock \showarticletitle{Text as data: The promise and pitfalls of automatic content analysis methods for political texts}.
\newblock \bibinfo{journal}{\emph{Political analysis}} \bibinfo{volume}{21}, \bibinfo{number}{3} (\bibinfo{year}{2013}), \bibinfo{pages}{267--297}.
\newblock


\bibitem[Gu{\'e}rin et~al\mbox{.}(2017)]%
        {guerin2017cnn}
\bibfield{author}{\bibinfo{person}{Joris Gu{\'e}rin}, \bibinfo{person}{Olivier Gibaru}, \bibinfo{person}{St{\'e}phane Thiery}, {and} \bibinfo{person}{Eric Nyiri}.} \bibinfo{year}{2017}\natexlab{}.
\newblock \showarticletitle{CNN features are also great at unsupervised classification}.
\newblock \bibinfo{journal}{\emph{arXiv preprint arXiv:1707.01700}} (\bibinfo{year}{2017}).
\newblock


\bibitem[Gui et~al\mbox{.}(2018)]%
        {gui2018multidimensional}
\bibfield{author}{\bibinfo{person}{Xinning Gui}, \bibinfo{person}{Yubo Kou}, \bibinfo{person}{Kathleen Pine}, \bibinfo{person}{Elisa Ladaw}, \bibinfo{person}{Harold Kim}, \bibinfo{person}{Eli Suzuki-Gill}, {and} \bibinfo{person}{Yunan Chen}.} \bibinfo{year}{2018}\natexlab{}.
\newblock \showarticletitle{Multidimensional risk communication: public discourse on risks during an emerging epidemic}. In \bibinfo{booktitle}{\emph{Proceedings of the 2018 CHI Conference on Human Factors in Computing Systems}}. \bibinfo{pages}{1--14}.
\newblock


\bibitem[Gui et~al\mbox{.}(2017)]%
        {gui2017managing}
\bibfield{author}{\bibinfo{person}{Xinning Gui}, \bibinfo{person}{Yubo Kou}, \bibinfo{person}{Kathleen~H Pine}, {and} \bibinfo{person}{Yunan Chen}.} \bibinfo{year}{2017}\natexlab{}.
\newblock \showarticletitle{Managing uncertainty: using social media for risk assessment during a public health crisis}. In \bibinfo{booktitle}{\emph{Proceedings of the 2017 CHI conference on human factors in computing systems}}. \bibinfo{pages}{4520--4533}.
\newblock


\bibitem[Guttman(2017)]%
        {guttman2017ethical}
\bibfield{author}{\bibinfo{person}{Nurit Guttman}.} \bibinfo{year}{2017}\natexlab{}.
\newblock \showarticletitle{Ethical issues in health promotion and communication interventions}.
\newblock In \bibinfo{booktitle}{\emph{Oxford research encyclopedia of communication}}.
\newblock


\bibitem[He et~al\mbox{.}(2022)]%
        {he2022help}
\bibfield{author}{\bibinfo{person}{Changyang He}, \bibinfo{person}{Yue Deng}, \bibinfo{person}{Wenjie Yang}, {and} \bibinfo{person}{Bo Li}.} \bibinfo{year}{2022}\natexlab{}.
\newblock \showarticletitle{" Help! Can You Hear Me?": Understanding How Help-Seeking Posts are Overwhelmed on Social Media during a Natural Disaster}.
\newblock \bibinfo{journal}{\emph{Proceedings of the ACM on Human-Computer Interaction}} \bibinfo{volume}{6}, \bibinfo{number}{CSCW2} (\bibinfo{year}{2022}), \bibinfo{pages}{1--25}.
\newblock


\bibitem[He et~al\mbox{.}(2021a)]%
        {he2021beyond}
\bibfield{author}{\bibinfo{person}{Changyang He}, \bibinfo{person}{Lu He}, \bibinfo{person}{Tun Lu}, {and} \bibinfo{person}{Bo Li}.} \bibinfo{year}{2021}\natexlab{a}.
\newblock \showarticletitle{Beyond Entertainment: Unpacking Danmaku and Comments' Role of Information Sharing and Sentiment Expression in Online Crisis Videos}.
\newblock \bibinfo{journal}{\emph{Proceedings of the ACM on Human-Computer Interaction}} \bibinfo{volume}{5}, \bibinfo{number}{CSCW2} (\bibinfo{year}{2021}), \bibinfo{pages}{1--27}.
\newblock


\bibitem[He et~al\mbox{.}(2016)]%
        {he2016deep}
\bibfield{author}{\bibinfo{person}{Kaiming He}, \bibinfo{person}{Xiangyu Zhang}, \bibinfo{person}{Shaoqing Ren}, {and} \bibinfo{person}{Jian Sun}.} \bibinfo{year}{2016}\natexlab{}.
\newblock \showarticletitle{Deep residual learning for image recognition}. In \bibinfo{booktitle}{\emph{Proceedings of the IEEE conference on computer vision and pattern recognition}}. \bibinfo{pages}{770--778}.
\newblock


\bibitem[He and He(2022)]%
        {he2022debunk}
\bibfield{author}{\bibinfo{person}{Lu He} {and} \bibinfo{person}{Changyang He}.} \bibinfo{year}{2022}\natexlab{}.
\newblock \showarticletitle{Help Me DebunkThis: Unpacking Individual and Community’s Collaborative Work in Information Credibility Assessment}.
\newblock \bibinfo{journal}{\emph{Proceedings of the ACM on Human-Computer Interaction}} \bibinfo{volume}{6}, \bibinfo{number}{CSCW2} (\bibinfo{year}{2022}).
\newblock


\bibitem[He et~al\mbox{.}(2021b)]%
        {he2021people}
\bibfield{author}{\bibinfo{person}{Lu He}, \bibinfo{person}{Changyang He}, \bibinfo{person}{Tera~L Reynolds}, \bibinfo{person}{Qiushi Bai}, \bibinfo{person}{Yicong Huang}, \bibinfo{person}{Chen Li}, \bibinfo{person}{Kai Zheng}, {and} \bibinfo{person}{Yunan Chen}.} \bibinfo{year}{2021}\natexlab{b}.
\newblock \showarticletitle{Why do people oppose mask wearing? A comprehensive analysis of US tweets during the COVID-19 pandemic}.
\newblock \bibinfo{journal}{\emph{Journal of the American Medical Informatics Association}} \bibinfo{volume}{28}, \bibinfo{number}{7} (\bibinfo{year}{2021}), \bibinfo{pages}{1564--1573}.
\newblock


\bibitem[Howard et~al\mbox{.}(2017)]%
        {howard2017mobilenets}
\bibfield{author}{\bibinfo{person}{Andrew~G Howard}, \bibinfo{person}{Menglong Zhu}, \bibinfo{person}{Bo Chen}, \bibinfo{person}{Dmitry Kalenichenko}, \bibinfo{person}{Weijun Wang}, \bibinfo{person}{Tobias Weyand}, \bibinfo{person}{Marco Andreetto}, {and} \bibinfo{person}{Hartwig Adam}.} \bibinfo{year}{2017}\natexlab{}.
\newblock \showarticletitle{Mobilenets: Efficient convolutional neural networks for mobile vision applications}.
\newblock \bibinfo{journal}{\emph{arXiv preprint arXiv:1704.04861}} (\bibinfo{year}{2017}).
\newblock


\bibitem[Hu et~al\mbox{.}(2014)]%
        {hu2014we}
\bibfield{author}{\bibinfo{person}{Yuheng Hu}, \bibinfo{person}{Lydia Manikonda}, {and} \bibinfo{person}{Subbarao Kambhampati}.} \bibinfo{year}{2014}\natexlab{}.
\newblock \showarticletitle{What we instagram: A first analysis of instagram photo content and user types}. In \bibinfo{booktitle}{\emph{Eighth International AAAI conference on weblogs and social media}}.
\newblock


\bibitem[Huang et~al\mbox{.}(2015)]%
        {huang2015connected}
\bibfield{author}{\bibinfo{person}{Y~Linlin Huang}, \bibinfo{person}{Kate Starbird}, \bibinfo{person}{Mania Orand}, \bibinfo{person}{Stephanie~A Stanek}, {and} \bibinfo{person}{Heather~T Pedersen}.} \bibinfo{year}{2015}\natexlab{}.
\newblock \showarticletitle{Connected through crisis: Emotional proximity and the spread of misinformation online}. In \bibinfo{booktitle}{\emph{Proceedings of the 18th ACM conference on computer supported cooperative work \& social computing}}. \bibinfo{pages}{969--980}.
\newblock


\bibitem[Imran et~al\mbox{.}(2020)]%
        {imran2020using}
\bibfield{author}{\bibinfo{person}{Muhammad Imran}, \bibinfo{person}{Ferda Ofli}, \bibinfo{person}{Doina Caragea}, {and} \bibinfo{person}{Antonio Torralba}.} \bibinfo{year}{2020}\natexlab{}.
\newblock \bibinfo{title}{Using AI and social media multimodal content for disaster response and management: Opportunities, challenges, and future directions}.
\newblock , \bibinfo{numpages}{102261}~pages.
\newblock


\bibitem[inspurer(2021)]%
        {SuperTopicCrawling}
\bibfield{author}{\bibinfo{person}{inspurer}.} \bibinfo{year}{2021}\natexlab{}.
\newblock \bibinfo{title}{WeiboSuperSpider}.
\newblock \bibinfo{howpublished}{\url{https://github.com/Python3Spiders/WeiboSuperSpider}}.
\newblock
\newblock
\shownote{Accessed: 2021-11-08}.


\bibitem[Jahanbakhsh et~al\mbox{.}(2021)]%
        {jahanbakhsh2021exploring}
\bibfield{author}{\bibinfo{person}{Farnaz Jahanbakhsh}, \bibinfo{person}{Amy~X Zhang}, \bibinfo{person}{Adam~J Berinsky}, \bibinfo{person}{Gordon Pennycook}, \bibinfo{person}{David~G Rand}, {and} \bibinfo{person}{David~R Karger}.} \bibinfo{year}{2021}\natexlab{}.
\newblock \showarticletitle{Exploring lightweight interventions at posting time to reduce the sharing of misinformation on social media}.
\newblock \bibinfo{journal}{\emph{Proceedings of the ACM on Human-Computer Interaction}} \bibinfo{volume}{5}, \bibinfo{number}{CSCW1} (\bibinfo{year}{2021}), \bibinfo{pages}{1--42}.
\newblock


\bibitem[Jasanoff et~al\mbox{.}(2021)]%
        {jasanoff2021comparative}
\bibfield{author}{\bibinfo{person}{Sheila Jasanoff}, \bibinfo{person}{Stephen Hilgartner}, \bibinfo{person}{J~Benjamin Hurlbut}, \bibinfo{person}{Onur Ozgode}, {and} \bibinfo{person}{Margarita Rayzberg}.} \bibinfo{year}{2021}\natexlab{}.
\newblock \showarticletitle{Comparative Covid response: crisis, knowledge, politics}.
\newblock \bibinfo{journal}{\emph{Ithaca: CompCoRe Network, Cornell University}} (\bibinfo{year}{2021}).
\newblock


\bibitem[Jia et~al\mbox{.}(2022)]%
        {jia2022understanding}
\bibfield{author}{\bibinfo{person}{Chenyan Jia}, \bibinfo{person}{Alexander Boltz}, \bibinfo{person}{Angie Zhang}, \bibinfo{person}{Anqing Chen}, {and} \bibinfo{person}{Min~Kyung Lee}.} \bibinfo{year}{2022}\natexlab{}.
\newblock \showarticletitle{Understanding Effects of Algorithmic vs. Community Label on Perceived Accuracy of Hyper-partisan Misinformation}.
\newblock \bibinfo{journal}{\emph{Proceedings of the ACM on Human-Computer Interaction}} \bibinfo{volume}{6}, \bibinfo{number}{CSCW2} (\bibinfo{year}{2022}), \bibinfo{pages}{1--27}.
\newblock


\bibitem[Joffe(2008)]%
        {joffe2008power}
\bibfield{author}{\bibinfo{person}{H{\'e}l{\`e}ne Joffe}.} \bibinfo{year}{2008}\natexlab{}.
\newblock \showarticletitle{The power of visual material: Persuasion, emotion and identification}.
\newblock \bibinfo{journal}{\emph{Diogenes}} \bibinfo{volume}{55}, \bibinfo{number}{1} (\bibinfo{year}{2008}), \bibinfo{pages}{84--93}.
\newblock


\bibitem[Kalliatakis(2017)]%
        {gkallia2017keras_places365}
\bibfield{author}{\bibinfo{person}{Grigorios Kalliatakis}.} \bibinfo{year}{2017}\natexlab{}.
\newblock \bibinfo{title}{Keras-VGG16-Places365}.
\newblock \bibinfo{howpublished}{\url{https://github.com/GKalliatakis/Keras-VGG16-places365}}.
\newblock


\bibitem[Khan et~al\mbox{.}(2014)]%
        {khan2014dbscan}
\bibfield{author}{\bibinfo{person}{Kamran Khan}, \bibinfo{person}{Saif~Ur Rehman}, \bibinfo{person}{Kamran Aziz}, \bibinfo{person}{Simon Fong}, {and} \bibinfo{person}{Sababady Sarasvady}.} \bibinfo{year}{2014}\natexlab{}.
\newblock \showarticletitle{DBSCAN: Past, present and future}. In \bibinfo{booktitle}{\emph{The fifth international conference on the applications of digital information and web technologies (ICADIWT 2014)}}. IEEE, \bibinfo{pages}{232--238}.
\newblock


\bibitem[Kou et~al\mbox{.}(2017)]%
        {kou2017conspiracy}
\bibfield{author}{\bibinfo{person}{Yubo Kou}, \bibinfo{person}{Xinning Gui}, \bibinfo{person}{Yunan Chen}, {and} \bibinfo{person}{Kathleen Pine}.} \bibinfo{year}{2017}\natexlab{}.
\newblock \showarticletitle{Conspiracy talk on social media: collective sensemaking during a public health crisis}.
\newblock \bibinfo{journal}{\emph{Proceedings of the ACM on Human-Computer Interaction}} \bibinfo{volume}{1}, \bibinfo{number}{CSCW} (\bibinfo{year}{2017}), \bibinfo{pages}{1--21}.
\newblock


\bibitem[Leavitt and Clark(2014)]%
        {leavitt2014upvoting}
\bibfield{author}{\bibinfo{person}{Alex Leavitt} {and} \bibinfo{person}{Joshua~A Clark}.} \bibinfo{year}{2014}\natexlab{}.
\newblock \showarticletitle{Upvoting hurricane Sandy: event-based news production processes on a social news site}. In \bibinfo{booktitle}{\emph{Proceedings of the SIGCHI conference on human factors in computing systems}}. \bibinfo{pages}{1495--1504}.
\newblock


\bibitem[Leavitt and Robinson(2017)]%
        {leavitt2017role}
\bibfield{author}{\bibinfo{person}{Alex Leavitt} {and} \bibinfo{person}{John~J Robinson}.} \bibinfo{year}{2017}\natexlab{}.
\newblock \showarticletitle{The role of information visibility in network gatekeeping: Information aggregation on Reddit during crisis events}. In \bibinfo{booktitle}{\emph{Proceedings of the 2017 ACM conference on computer supported cooperative work and social computing}}. \bibinfo{pages}{1246--1261}.
\newblock


\bibitem[Li et~al\mbox{.}(2019)]%
        {li2019using}
\bibfield{author}{\bibinfo{person}{Jing Li}, \bibinfo{person}{Keri~K Stephens}, \bibinfo{person}{Yaguang Zhu}, {and} \bibinfo{person}{Dhiraj Murthy}.} \bibinfo{year}{2019}\natexlab{}.
\newblock \showarticletitle{Using social media to call for help in Hurricane Harvey: Bonding emotion, culture, and community relationships}.
\newblock \bibinfo{journal}{\emph{International Journal of Disaster Risk Reduction}}  \bibinfo{volume}{38} (\bibinfo{year}{2019}), \bibinfo{pages}{101212}.
\newblock


\bibitem[Li et~al\mbox{.}(2020)]%
        {li2020data}
\bibfield{author}{\bibinfo{person}{Jiawei Li}, \bibinfo{person}{Qing Xu}, \bibinfo{person}{Raphael Cuomo}, \bibinfo{person}{Vidya Purushothaman}, \bibinfo{person}{Tim Mackey}, {et~al\mbox{.}}} \bibinfo{year}{2020}\natexlab{}.
\newblock \showarticletitle{Data mining and content analysis of the Chinese social media platform Weibo during the early COVID-19 outbreak: retrospective observational infoveillance study}.
\newblock \bibinfo{journal}{\emph{JMIR Public Health and Surveillance}} \bibinfo{volume}{6}, \bibinfo{number}{2} (\bibinfo{year}{2020}), \bibinfo{pages}{e18700}.
\newblock


\bibitem[Li et~al\mbox{.}(2021)]%
        {li2021hello}
\bibfield{author}{\bibinfo{person}{Xuyang Li}, \bibinfo{person}{Antara Bahursettiwar}, {and} \bibinfo{person}{Marina Kogan}.} \bibinfo{year}{2021}\natexlab{}.
\newblock \showarticletitle{Hello? Is There Anybody in There? Analysis of Factors Promoting Response From Authoritative Sources in Crisis}.
\newblock \bibinfo{journal}{\emph{Proceedings of the ACM on Human-Computer Interaction}} \bibinfo{volume}{5}, \bibinfo{number}{CSCW1} (\bibinfo{year}{2021}), \bibinfo{pages}{1--21}.
\newblock


\bibitem[Li et~al\mbox{.}(2018)]%
        {li2018localizing}
\bibfield{author}{\bibinfo{person}{Xukun Li}, \bibinfo{person}{Doina Caragea}, \bibinfo{person}{Huaiyu Zhang}, {and} \bibinfo{person}{Muhammad Imran}.} \bibinfo{year}{2018}\natexlab{}.
\newblock \showarticletitle{Localizing and quantifying damage in social media images}. In \bibinfo{booktitle}{\emph{2018 IEEE/ACM International Conference on Advances in Social Networks Analysis and Mining (ASONAM)}}. IEEE, \bibinfo{pages}{194--201}.
\newblock


\bibitem[Lindell et~al\mbox{.}(2016)]%
        {lindell2016immediate}
\bibfield{author}{\bibinfo{person}{Michael~K Lindell}, \bibinfo{person}{Carla~S Prater}, \bibinfo{person}{Hao~Che Wu}, \bibinfo{person}{Shih-Kai Huang}, \bibinfo{person}{David~M Johnston}, \bibinfo{person}{Julia~S Becker}, {and} \bibinfo{person}{Hideyuki Shiroshita}.} \bibinfo{year}{2016}\natexlab{}.
\newblock \showarticletitle{Immediate behavioural responses to earthquakes in Christchurch, New Zealand, and Hitachi, Japan}.
\newblock \bibinfo{journal}{\emph{Disasters}} \bibinfo{volume}{40}, \bibinfo{number}{1} (\bibinfo{year}{2016}), \bibinfo{pages}{85--111}.
\newblock


\bibitem[Lloyd(1982)]%
        {lloyd1982least}
\bibfield{author}{\bibinfo{person}{Stuart Lloyd}.} \bibinfo{year}{1982}\natexlab{}.
\newblock \showarticletitle{Least squares quantization in PCM}.
\newblock \bibinfo{journal}{\emph{IEEE transactions on information theory}} \bibinfo{volume}{28}, \bibinfo{number}{2} (\bibinfo{year}{1982}), \bibinfo{pages}{129--137}.
\newblock


\bibitem[Lowe(2004)]%
        {lowe2004distinctive}
\bibfield{author}{\bibinfo{person}{David~G Lowe}.} \bibinfo{year}{2004}\natexlab{}.
\newblock \showarticletitle{Distinctive image features from scale-invariant keypoints}.
\newblock \bibinfo{journal}{\emph{International journal of computer vision}} \bibinfo{volume}{60}, \bibinfo{number}{2} (\bibinfo{year}{2004}), \bibinfo{pages}{91--110}.
\newblock


\bibitem[Lu et~al\mbox{.}(2021)]%
        {lu2021positive}
\bibfield{author}{\bibinfo{person}{Zhicong Lu}, \bibinfo{person}{Yue Jiang}, \bibinfo{person}{Chenxinran Shen}, \bibinfo{person}{Margaret~C Jack}, \bibinfo{person}{Daniel Wigdor}, {and} \bibinfo{person}{Mor Naaman}.} \bibinfo{year}{2021}\natexlab{}.
\newblock \showarticletitle{" Positive Energy" Perceptions and Attitudes Towards COVID-19 Information on Social Media in China}.
\newblock \bibinfo{journal}{\emph{Proceedings of the ACM on human-computer interaction}} \bibinfo{volume}{5}, \bibinfo{number}{CSCW1} (\bibinfo{year}{2021}), \bibinfo{pages}{1--25}.
\newblock


\bibitem[Manikonda and De~Choudhury(2017)]%
        {manikonda2017modeling}
\bibfield{author}{\bibinfo{person}{Lydia Manikonda} {and} \bibinfo{person}{Munmun De~Choudhury}.} \bibinfo{year}{2017}\natexlab{}.
\newblock \showarticletitle{Modeling and understanding visual attributes of mental health disclosures in social media}. In \bibinfo{booktitle}{\emph{Proceedings of the 2017 CHI Conference on Human Factors in Computing Systems}}. \bibinfo{pages}{170--181}.
\newblock


\bibitem[Manikonda et~al\mbox{.}(2016)]%
        {manikonda2016tweeting}
\bibfield{author}{\bibinfo{person}{Lydia Manikonda}, \bibinfo{person}{Venkata~Vamsikrishna Meduri}, {and} \bibinfo{person}{Subbarao Kambhampati}.} \bibinfo{year}{2016}\natexlab{}.
\newblock \showarticletitle{Tweeting the mind and instagramming the heart: Exploring differentiated content sharing on social media}. In \bibinfo{booktitle}{\emph{Tenth international AAAI conference on web and social media}}.
\newblock


\bibitem[Marwick(2015)]%
        {marwick2015instafame}
\bibfield{author}{\bibinfo{person}{Alice~E Marwick}.} \bibinfo{year}{2015}\natexlab{}.
\newblock \showarticletitle{Instafame: Luxury selfies in the attention economy}.
\newblock \bibinfo{journal}{\emph{Public culture}} \bibinfo{volume}{27}, \bibinfo{number}{1 (75)} (\bibinfo{year}{2015}), \bibinfo{pages}{137--160}.
\newblock


\bibitem[Marzouki et~al\mbox{.}(2021)]%
        {marzouki2021understanding}
\bibfield{author}{\bibinfo{person}{Yousri Marzouki}, \bibinfo{person}{Fatimah~Salem Aldossari}, {and} \bibinfo{person}{Giuseppe~A Veltri}.} \bibinfo{year}{2021}\natexlab{}.
\newblock \showarticletitle{Understanding the buffering effect of social media use on anxiety during the COVID-19 pandemic lockdown}.
\newblock \bibinfo{journal}{\emph{Humanities and Social Sciences Communications}} \bibinfo{volume}{8}, \bibinfo{number}{1} (\bibinfo{year}{2021}).
\newblock


\bibitem[McHugh(2012)]%
        {mchugh2012interrater}
\bibfield{author}{\bibinfo{person}{Mary~L McHugh}.} \bibinfo{year}{2012}\natexlab{}.
\newblock \showarticletitle{Interrater reliability: the kappa statistic}.
\newblock \bibinfo{journal}{\emph{Biochemia medica}} \bibinfo{volume}{22}, \bibinfo{number}{3} (\bibinfo{year}{2012}), \bibinfo{pages}{276--282}.
\newblock


\bibitem[McHugh(2013)]%
        {mchugh2013chi}
\bibfield{author}{\bibinfo{person}{Mary~L McHugh}.} \bibinfo{year}{2013}\natexlab{}.
\newblock \showarticletitle{The chi-square test of independence}.
\newblock \bibinfo{journal}{\emph{Biochemia medica}} \bibinfo{volume}{23}, \bibinfo{number}{2} (\bibinfo{year}{2013}), \bibinfo{pages}{143--149}.
\newblock


\bibitem[Micallef et~al\mbox{.}(2022)]%
        {micallef2022cross}
\bibfield{author}{\bibinfo{person}{Nicholas Micallef}, \bibinfo{person}{Marcelo Sandoval-Casta{\~n}eda}, \bibinfo{person}{Adi Cohen}, \bibinfo{person}{Mustaque Ahamad}, \bibinfo{person}{Srijan Kumar}, {and} \bibinfo{person}{Nasir Memon}.} \bibinfo{year}{2022}\natexlab{}.
\newblock \showarticletitle{Cross-Platform Multimodal Misinformation: Taxonomy, Characteristics and Detection for Textual Posts and Videos}. In \bibinfo{booktitle}{\emph{Proceedings of the International AAAI Conference on Web and Social Media}}, Vol.~\bibinfo{volume}{16}. \bibinfo{pages}{651--662}.
\newblock


\bibitem[Mortensen et~al\mbox{.}(2017)]%
        {mortensen2017really}
\bibfield{author}{\bibinfo{person}{Tara~M Mortensen}, \bibinfo{person}{Kevin Hull}, {and} \bibinfo{person}{Kelli~S Boling}.} \bibinfo{year}{2017}\natexlab{}.
\newblock \showarticletitle{Really social disaster: an examination of photo sharing on twitter during the\# SCFlood}.
\newblock \bibinfo{journal}{\emph{Visual Communication Quarterly}} \bibinfo{volume}{24}, \bibinfo{number}{4} (\bibinfo{year}{2017}), \bibinfo{pages}{219--229}.
\newblock


\bibitem[News(2021a)]%
        {fakeScreenshot}
\bibfield{author}{\bibinfo{person}{China News}.} \bibinfo{year}{2021}\natexlab{a}.
\newblock \bibinfo{title}{Analysis of recent epidemic rumors: chat screenshots and short videos become the main source (in Chinese)}.
\newblock \bibinfo{howpublished}{\url{https://www.chinanews.com.cn/sh/2021/08-11/9541137.shtml}}.
\newblock
\newblock
\shownote{Accessed: 2022-09-09}.


\bibitem[News(2021b)]%
        {XianSource}
\bibfield{author}{\bibinfo{person}{China News}.} \bibinfo{year}{2021}\natexlab{b}.
\newblock \bibinfo{title}{The local epidemic in Xi'an, Shaanxi Province is caused by the delta variant, and there is hidden transmission (in Chinese)}.
\newblock \bibinfo{howpublished}{\url{https://www.chinanews.com.cn/sh/2021/12-22/9635743.shtml}}.
\newblock
\newblock
\shownote{Accessed: 2022-09-14}.


\bibitem[Nguyen et~al\mbox{.}(2017)]%
        {nguyen2017personalized}
\bibfield{author}{\bibinfo{person}{Hanh~TH Nguyen}, \bibinfo{person}{Martin Wistuba}, {and} \bibinfo{person}{Lars Schmidt-Thieme}.} \bibinfo{year}{2017}\natexlab{}.
\newblock \showarticletitle{Personalized tag recommendation for images using deep transfer learning}. In \bibinfo{booktitle}{\emph{Joint European Conference on Machine Learning and Knowledge Discovery in Databases}}. Springer, \bibinfo{pages}{705--720}.
\newblock


\bibitem[Nishikawa et~al\mbox{.}(2018)]%
        {nishikawa2018time}
\bibfield{author}{\bibinfo{person}{Shuji Nishikawa}, \bibinfo{person}{Nozomi Tanaka}, \bibinfo{person}{Keisuke Utsu}, {and} \bibinfo{person}{Osamu Uchida}.} \bibinfo{year}{2018}\natexlab{}.
\newblock \showarticletitle{Time trend analysis of “\# Rescue” tweets during and after the 2017 northern Kyushu heavy rain disaster}. In \bibinfo{booktitle}{\emph{2018 5th International Conference on Information and Communication Technologies for Disaster Management (ICT-DM)}}. IEEE, \bibinfo{pages}{1--4}.
\newblock


\bibitem[of~Public~Health(2022)]%
        {localOutbreak}
\bibfield{author}{\bibinfo{person}{California~Department of Public~Health}.} \bibinfo{year}{2022}\natexlab{}.
\newblock \bibinfo{title}{COVID-19 Outbreak Data}.
\newblock \bibinfo{howpublished}{\url{https://www.cdph.ca.gov/Programs/CID/DCDC/Pages/COVID-19/COVID-19-Outbreak-Data.aspx}}.
\newblock
\newblock
\shownote{Accessed: 2022-09-14}.


\bibitem[Olteanu et~al\mbox{.}(2015)]%
        {olteanu2015expect}
\bibfield{author}{\bibinfo{person}{Alexandra Olteanu}, \bibinfo{person}{Sarah Vieweg}, {and} \bibinfo{person}{Carlos Castillo}.} \bibinfo{year}{2015}\natexlab{}.
\newblock \showarticletitle{What to expect when the unexpected happens: Social media communications across crises}. In \bibinfo{booktitle}{\emph{Proceedings of the 18th ACM conference on computer supported cooperative work \& social computing}}. \bibinfo{pages}{994--1009}.
\newblock


\bibitem[Organization(2 07)]%
        {worldCases}
\bibfield{author}{\bibinfo{person}{World~Health Organization}.} \bibinfo{year}{2021-12-07}\natexlab{}.
\newblock \bibinfo{booktitle}{\emph{COVID-19 weekly epidemiological update, edition 69, 7 December 2021}}.
\newblock \bibinfo{type}{Technical documents}. \bibinfo{pages}{19 p.} pages.
\newblock


\bibitem[Osatuyi(2013)]%
        {osatuyi2013information}
\bibfield{author}{\bibinfo{person}{Babajide Osatuyi}.} \bibinfo{year}{2013}\natexlab{}.
\newblock \showarticletitle{Information sharing on social media sites}.
\newblock \bibinfo{journal}{\emph{Computers in Human Behavior}} \bibinfo{volume}{29}, \bibinfo{number}{6} (\bibinfo{year}{2013}), \bibinfo{pages}{2622--2631}.
\newblock


\bibitem[Palen et~al\mbox{.}(2010)]%
        {palen2010vision}
\bibfield{author}{\bibinfo{person}{Leysia Palen}, \bibinfo{person}{Kenneth~M Anderson}, \bibinfo{person}{Gloria Mark}, \bibinfo{person}{James Martin}, \bibinfo{person}{Douglas Sicker}, \bibinfo{person}{Martha Palmer}, {and} \bibinfo{person}{Dirk Grunwald}.} \bibinfo{year}{2010}\natexlab{}.
\newblock \showarticletitle{A vision for technology-mediated support for public participation \& assistance in mass emergencies \& disasters}.
\newblock \bibinfo{journal}{\emph{ACM-BCS Visions of Computer Science 2010}} (\bibinfo{year}{2010}), \bibinfo{pages}{1--12}.
\newblock


\bibitem[Pang et~al\mbox{.}(2015)]%
        {pang2015monitoring}
\bibfield{author}{\bibinfo{person}{Ran Pang}, \bibinfo{person}{Agustin Baretto}, \bibinfo{person}{Henry Kautz}, {and} \bibinfo{person}{Jiebo Luo}.} \bibinfo{year}{2015}\natexlab{}.
\newblock \showarticletitle{Monitoring adolescent alcohol use via multimodal analysis in social multimedia}. In \bibinfo{booktitle}{\emph{2015 IEEE International Conference on Big Data (Big Data)}}. IEEE, \bibinfo{pages}{1509--1518}.
\newblock


\bibitem[Peng(2021)]%
        {peng2021makes}
\bibfield{author}{\bibinfo{person}{Yilang Peng}.} \bibinfo{year}{2021}\natexlab{}.
\newblock \showarticletitle{What makes politicians’ Instagram posts popular? Analyzing social media strategies of candidates and office holders with computer vision}.
\newblock \bibinfo{journal}{\emph{The International Journal of Press/Politics}} \bibinfo{volume}{26}, \bibinfo{number}{1} (\bibinfo{year}{2021}), \bibinfo{pages}{143--166}.
\newblock


\bibitem[Perovich et~al\mbox{.}(2022)]%
        {perovich2022self}
\bibfield{author}{\bibinfo{person}{Laura~J Perovich}, \bibinfo{person}{Meryl Alper}, {and} \bibinfo{person}{Corey Cleveland}.} \bibinfo{year}{2022}\natexlab{}.
\newblock \showarticletitle{" Self-Quaranteens" Process COVID-19: Understanding Information Visualization Language in Memes}.
\newblock \bibinfo{journal}{\emph{Proceedings of the ACM on Human-Computer Interaction}} \bibinfo{volume}{6}, \bibinfo{number}{CSCW1} (\bibinfo{year}{2022}), \bibinfo{pages}{1--20}.
\newblock


\bibitem[Porter et~al\mbox{.}(2020)]%
        {porter2020visual}
\bibfield{author}{\bibinfo{person}{Emily Porter}, \bibinfo{person}{PM Krafft}, {and} \bibinfo{person}{Brian Keegan}.} \bibinfo{year}{2020}\natexlab{}.
\newblock \showarticletitle{Visual Narratives and Collective Memory across Peer-Produced Accounts of Contested Sociopolitical Events}.
\newblock \bibinfo{journal}{\emph{ACM Transactions on Social Computing}} \bibinfo{volume}{3}, \bibinfo{number}{1} (\bibinfo{year}{2020}), \bibinfo{pages}{1--20}.
\newblock


\bibitem[Powell et~al\mbox{.}(2015)]%
        {powell2015clearer}
\bibfield{author}{\bibinfo{person}{Thomas~E Powell}, \bibinfo{person}{Hajo~G Boomgaarden}, \bibinfo{person}{Knut De~Swert}, {and} \bibinfo{person}{Claes~H de Vreese}.} \bibinfo{year}{2015}\natexlab{}.
\newblock \showarticletitle{A clearer picture: The contribution of visuals and text to framing effects}.
\newblock \bibinfo{journal}{\emph{Journal of communication}} \bibinfo{volume}{65}, \bibinfo{number}{6} (\bibinfo{year}{2015}), \bibinfo{pages}{997--1017}.
\newblock


\bibitem[Preim and Lawonn(2020)]%
        {preim2020survey}
\bibfield{author}{\bibinfo{person}{Bernhard Preim} {and} \bibinfo{person}{Kai Lawonn}.} \bibinfo{year}{2020}\natexlab{}.
\newblock \showarticletitle{A survey of visual analytics for public health}. In \bibinfo{booktitle}{\emph{Computer Graphics Forum}}, Vol.~\bibinfo{volume}{39}. Wiley Online Library, \bibinfo{pages}{543--580}.
\newblock


\bibitem[Pulos(2020)]%
        {pulos2020covid}
\bibfield{author}{\bibinfo{person}{Rick Pulos}.} \bibinfo{year}{2020}\natexlab{}.
\newblock \showarticletitle{COVID-19 crisis memes, rhetorical arena theory and multimodality}.
\newblock \bibinfo{journal}{\emph{Journal of Science Communication}} \bibinfo{volume}{19}, \bibinfo{number}{7} (\bibinfo{year}{2020}), \bibinfo{pages}{A01}.
\newblock


\bibitem[Qu et~al\mbox{.}(2011)]%
        {qu2011microblogging}
\bibfield{author}{\bibinfo{person}{Yan Qu}, \bibinfo{person}{Chen Huang}, \bibinfo{person}{Pengyi Zhang}, {and} \bibinfo{person}{Jun Zhang}.} \bibinfo{year}{2011}\natexlab{}.
\newblock \showarticletitle{Microblogging after a major disaster in China: a case study of the 2010 Yushu earthquake}. In \bibinfo{booktitle}{\emph{Proceedings of the ACM 2011 conference on Computer supported cooperative work}}. \bibinfo{pages}{25--34}.
\newblock


\bibitem[Qu et~al\mbox{.}(2009)]%
        {qu2009online}
\bibfield{author}{\bibinfo{person}{Yan Qu}, \bibinfo{person}{Philip~Fei Wu}, {and} \bibinfo{person}{Xiaoqing Wang}.} \bibinfo{year}{2009}\natexlab{}.
\newblock \showarticletitle{Online community response to major disaster: A study of Tianya forum in the 2008 Sichuan earthquake}. In \bibinfo{booktitle}{\emph{2009 42nd Hawaii International Conference on System Sciences}}. IEEE, \bibinfo{pages}{1--11}.
\newblock


\bibitem[Quattoni and Torralba(2009)]%
        {quattoni2009recognizing}
\bibfield{author}{\bibinfo{person}{Ariadna Quattoni} {and} \bibinfo{person}{Antonio Torralba}.} \bibinfo{year}{2009}\natexlab{}.
\newblock \showarticletitle{Recognizing indoor scenes}. In \bibinfo{booktitle}{\emph{2009 IEEE conference on computer vision and pattern recognition}}. IEEE, \bibinfo{pages}{413--420}.
\newblock


\bibitem[Reynolds(2009)]%
        {reynolds2009gaussian}
\bibfield{author}{\bibinfo{person}{Douglas~A Reynolds}.} \bibinfo{year}{2009}\natexlab{}.
\newblock \showarticletitle{Gaussian mixture models.}
\newblock \bibinfo{journal}{\emph{Encyclopedia of biometrics}} \bibinfo{volume}{741}, \bibinfo{number}{659-663} (\bibinfo{year}{2009}).
\newblock


\bibitem[Ross et~al\mbox{.}(2021)]%
        {ross2021household}
\bibfield{author}{\bibinfo{person}{Stuart Ross}, \bibinfo{person}{George Breckenridge}, \bibinfo{person}{Mengdie Zhuang}, {and} \bibinfo{person}{Ed Manley}.} \bibinfo{year}{2021}\natexlab{}.
\newblock \showarticletitle{Household visitation during the COVID-19 pandemic}.
\newblock \bibinfo{journal}{\emph{Scientific reports}} \bibinfo{volume}{11}, \bibinfo{number}{1} (\bibinfo{year}{2021}), \bibinfo{pages}{1--11}.
\newblock


\bibitem[Rousseeuw(1987)]%
        {rousseeuw1987silhouettes}
\bibfield{author}{\bibinfo{person}{Peter~J Rousseeuw}.} \bibinfo{year}{1987}\natexlab{}.
\newblock \showarticletitle{Silhouettes: a graphical aid to the interpretation and validation of cluster analysis}.
\newblock \bibinfo{journal}{\emph{Journal of computational and applied mathematics}}  \bibinfo{volume}{20} (\bibinfo{year}{1987}), \bibinfo{pages}{53--65}.
\newblock


\bibitem[Salda{\~n}a(2021)]%
        {saldana2021coding}
\bibfield{author}{\bibinfo{person}{Johnny Salda{\~n}a}.} \bibinfo{year}{2021}\natexlab{}.
\newblock \showarticletitle{The coding manual for qualitative researchers}.
\newblock \bibinfo{journal}{\emph{The coding manual for qualitative researchers}} (\bibinfo{year}{2021}), \bibinfo{pages}{1--440}.
\newblock


\bibitem[Sameer et~al\mbox{.}(2020)]%
        {sameer2020assessment}
\bibfield{author}{\bibinfo{person}{AS Sameer}, \bibinfo{person}{MA Khan}, \bibinfo{person}{S Nissar}, {and} \bibinfo{person}{MZ Banday}.} \bibinfo{year}{2020}\natexlab{}.
\newblock \showarticletitle{Assessment of mental health and various coping strategies among general population living under imposed COVID-lockdown across world: a cross-sectional study}.
\newblock \bibinfo{journal}{\emph{Ethics, Medicine and Public Health}}  \bibinfo{volume}{15} (\bibinfo{year}{2020}), \bibinfo{pages}{100571}.
\newblock


\bibitem[Sandler et~al\mbox{.}(2018)]%
        {sandler2018mobilenetv2}
\bibfield{author}{\bibinfo{person}{Mark Sandler}, \bibinfo{person}{Andrew Howard}, \bibinfo{person}{Menglong Zhu}, \bibinfo{person}{Andrey Zhmoginov}, {and} \bibinfo{person}{Liang-Chieh Chen}.} \bibinfo{year}{2018}\natexlab{}.
\newblock \showarticletitle{Mobilenetv2: Inverted residuals and linear bottlenecks}. In \bibinfo{booktitle}{\emph{Proceedings of the IEEE conference on computer vision and pattern recognition}}. \bibinfo{pages}{4510--4520}.
\newblock


\bibitem[Sawyer and Chen(2012)]%
        {sawyer2012impact}
\bibfield{author}{\bibinfo{person}{Rebecca Sawyer} {and} \bibinfo{person}{Guo-Ming Chen}.} \bibinfo{year}{2012}\natexlab{}.
\newblock \showarticletitle{The impact of social media on intercultural adaptation}.
\newblock  (\bibinfo{year}{2012}).
\newblock


\bibitem[Schill(2012)]%
        {schill2012visual}
\bibfield{author}{\bibinfo{person}{Dan Schill}.} \bibinfo{year}{2012}\natexlab{}.
\newblock \showarticletitle{The visual image and the political image: A review of visual communication research in the field of political communication}.
\newblock \bibinfo{journal}{\emph{Review of communication}} \bibinfo{volume}{12}, \bibinfo{number}{2} (\bibinfo{year}{2012}), \bibinfo{pages}{118--142}.
\newblock


\bibitem[Seo(2014)]%
        {seo2014visual}
\bibfield{author}{\bibinfo{person}{Hyunjin Seo}.} \bibinfo{year}{2014}\natexlab{}.
\newblock \showarticletitle{Visual propaganda in the age of social media: An empirical analysis of Twitter images during the 2012 Israeli--Hamas conflict}.
\newblock \bibinfo{journal}{\emph{Visual Communication Quarterly}} \bibinfo{volume}{21}, \bibinfo{number}{3} (\bibinfo{year}{2014}), \bibinfo{pages}{150--161}.
\newblock


\bibitem[Simonyan and Zisserman(2014)]%
        {simonyan2014very}
\bibfield{author}{\bibinfo{person}{Karen Simonyan} {and} \bibinfo{person}{Andrew Zisserman}.} \bibinfo{year}{2014}\natexlab{}.
\newblock \showarticletitle{Very deep convolutional networks for large-scale image recognition}.
\newblock \bibinfo{journal}{\emph{arXiv preprint arXiv:1409.1556}} (\bibinfo{year}{2014}).
\newblock


\bibitem[Sivic and Zisserman(2003)]%
        {sivic2003video}
\bibfield{author}{\bibinfo{person}{Josef Sivic} {and} \bibinfo{person}{Andrew Zisserman}.} \bibinfo{year}{2003}\natexlab{}.
\newblock \showarticletitle{Video Google: A text retrieval approach to object matching in videos}. In \bibinfo{booktitle}{\emph{Computer Vision, IEEE International Conference on}}, Vol.~\bibinfo{volume}{3}. IEEE Computer Society, \bibinfo{pages}{1470--1470}.
\newblock


\bibitem[Sleigh et~al\mbox{.}(2021)]%
        {sleigh2021qualitative}
\bibfield{author}{\bibinfo{person}{Joanna Sleigh}, \bibinfo{person}{Julia Amann}, \bibinfo{person}{Manuel Schneider}, {and} \bibinfo{person}{Effy Vayena}.} \bibinfo{year}{2021}\natexlab{}.
\newblock \showarticletitle{Qualitative analysis of visual risk communication on twitter during the Covid-19 pandemic}.
\newblock \bibinfo{journal}{\emph{BMC public health}} \bibinfo{volume}{21}, \bibinfo{number}{1} (\bibinfo{year}{2021}), \bibinfo{pages}{1--12}.
\newblock


\bibitem[Sommariva et~al\mbox{.}(2018)]%
        {sommariva2018spreading}
\bibfield{author}{\bibinfo{person}{Silvia Sommariva}, \bibinfo{person}{Cheryl Vamos}, \bibinfo{person}{Alexios Mantzarlis}, \bibinfo{person}{Lillie Uy{\^e}n-Loan {\DJ}{\`a}o}, {and} \bibinfo{person}{Dinorah Martinez~Tyson}.} \bibinfo{year}{2018}\natexlab{}.
\newblock \showarticletitle{Spreading the (fake) news: exploring health messages on social media and the implications for health professionals using a case study}.
\newblock \bibinfo{journal}{\emph{American journal of health education}} \bibinfo{volume}{49}, \bibinfo{number}{4} (\bibinfo{year}{2018}), \bibinfo{pages}{246--255}.
\newblock


\bibitem[Sosea et~al\mbox{.}(2021)]%
        {sosea2021using}
\bibfield{author}{\bibinfo{person}{Tiberiu Sosea}, \bibinfo{person}{Iustin Sirbu}, \bibinfo{person}{Cornelia Caragea}, \bibinfo{person}{Doina Caragea}, {and} \bibinfo{person}{Traian Rebedea}.} \bibinfo{year}{2021}\natexlab{}.
\newblock \showarticletitle{Using the Image-Text Relationship to Improve Multimodal Disaster Tweet Classification}. In \bibinfo{booktitle}{\emph{The 18th International Conference on Information Systems for Crisis Response and Management (ISCRAM 2021)}}.
\newblock


\bibitem[Starbird(2013)]%
        {starbird2013delivering}
\bibfield{author}{\bibinfo{person}{Kate Starbird}.} \bibinfo{year}{2013}\natexlab{}.
\newblock \showarticletitle{Delivering patients to sacr{\'e} coeur: collective intelligence in digital volunteer communities}. In \bibinfo{booktitle}{\emph{Proceedings of the SIGCHI Conference on Human Factors in Computing Systems}}. \bibinfo{pages}{801--810}.
\newblock


\bibitem[Starbird et~al\mbox{.}(2014)]%
        {starbird2014rumors}
\bibfield{author}{\bibinfo{person}{Kate Starbird}, \bibinfo{person}{Jim Maddock}, \bibinfo{person}{Mania Orand}, \bibinfo{person}{Peg Achterman}, {and} \bibinfo{person}{Robert~M Mason}.} \bibinfo{year}{2014}\natexlab{}.
\newblock \showarticletitle{Rumors, false flags, and digital vigilantes: Misinformation on twitter after the 2013 boston marathon bombing}.
\newblock \bibinfo{journal}{\emph{IConference 2014 proceedings}} (\bibinfo{year}{2014}).
\newblock


\bibitem[Starbird and Palen(2011)]%
        {starbird2011voluntweeters}
\bibfield{author}{\bibinfo{person}{Kate Starbird} {and} \bibinfo{person}{Leysia Palen}.} \bibinfo{year}{2011}\natexlab{}.
\newblock \showarticletitle{" Voluntweeters" self-organizing by digital volunteers in times of crisis}. In \bibinfo{booktitle}{\emph{Proceedings of the SIGCHI conference on human factors in computing systems}}. \bibinfo{pages}{1071--1080}.
\newblock


\bibitem[Starbird et~al\mbox{.}(2012)]%
        {starbird2012promoting}
\bibfield{author}{\bibinfo{person}{Kate Starbird}, \bibinfo{person}{Leysia Palen}, \bibinfo{person}{Sophia~B Liu}, \bibinfo{person}{Sarah Vieweg}, \bibinfo{person}{Amanda Hughes}, \bibinfo{person}{Aaron Schram}, \bibinfo{person}{Kenneth~Mark Anderson}, \bibinfo{person}{Mossaab Bagdouri}, \bibinfo{person}{Joanne White}, \bibinfo{person}{Casey McTaggart}, {et~al\mbox{.}}} \bibinfo{year}{2012}\natexlab{}.
\newblock \showarticletitle{Promoting structured data in citizen communications during disaster response: an account of strategies for diffusion of the'Tweak the Tweet'syntax}.
\newblock In \bibinfo{booktitle}{\emph{Crisis Information Management}}. \bibinfo{publisher}{Elsevier}, \bibinfo{pages}{43--63}.
\newblock


\bibitem[Stefanone et~al\mbox{.}(2015)]%
        {stefanone2015image}
\bibfield{author}{\bibinfo{person}{Michael~A Stefanone}, \bibinfo{person}{Gregory~D Saxton}, \bibinfo{person}{Michael~J Egnoto}, \bibinfo{person}{Wayne Wei}, {and} \bibinfo{person}{Yun Fu}.} \bibinfo{year}{2015}\natexlab{}.
\newblock \showarticletitle{Image attributes and diffusion via Twitter: The case of\# guncontrol}. In \bibinfo{booktitle}{\emph{2015 48th Hawaii International Conference on System Sciences}}. IEEE, \bibinfo{pages}{1788--1797}.
\newblock


\bibitem[Tomonto(2019)]%
        {tomonto2019calamitous}
\bibfield{author}{\bibinfo{person}{Matthew Tomonto}.} \bibinfo{year}{2019}\natexlab{}.
\newblock \emph{\bibinfo{title}{A Calamitous Imagination: Disaster Images, Fake News, and Challenges to Journalistic Objectivity}}.
\newblock \bibinfo{thesistype}{Ph.\,D. Dissertation}. \bibinfo{school}{MA Thesis, Aristotle University of Thessaloniki. http://ikee. lib. auth. gr~…}.
\newblock


\bibitem[Torrey and Shavlik(2010)]%
        {torrey2010transfer}
\bibfield{author}{\bibinfo{person}{Lisa Torrey} {and} \bibinfo{person}{Jude Shavlik}.} \bibinfo{year}{2010}\natexlab{}.
\newblock \showarticletitle{Transfer learning}.
\newblock In \bibinfo{booktitle}{\emph{Handbook of research on machine learning applications and trends: algorithms, methods, and techniques}}. \bibinfo{publisher}{IGI global}, \bibinfo{pages}{242--264}.
\newblock


\bibitem[Tsai et~al\mbox{.}(2021)]%
        {tsai2021help}
\bibfield{author}{\bibinfo{person}{Chun-Hua Tsai}, \bibinfo{person}{Xinning Gui}, \bibinfo{person}{Yubo Kou}, {and} \bibinfo{person}{John~M Carroll}.} \bibinfo{year}{2021}\natexlab{}.
\newblock \showarticletitle{With Help from Afar: Cross-Local Communication in an Online COVID-19 Pandemic Community}.
\newblock \bibinfo{journal}{\emph{Proceedings of the ACM on Human-Computer Interaction}} \bibinfo{volume}{5}, \bibinfo{number}{CSCW2} (\bibinfo{year}{2021}), \bibinfo{pages}{1--24}.
\newblock


\bibitem[Unal et~al\mbox{.}(2022)]%
        {unal2022visual}
\bibfield{author}{\bibinfo{person}{Mesut~Erhan Unal}, \bibinfo{person}{Adriana Kovashka}, \bibinfo{person}{Wen-Ting Chung}, {and} \bibinfo{person}{Yu-Ru Lin}.} \bibinfo{year}{2022}\natexlab{}.
\newblock \showarticletitle{Visual persuasion in covid-19 social media content: A multi-modal characterization}. In \bibinfo{booktitle}{\emph{Companion Proceedings of the Web Conference 2022}}. \bibinfo{pages}{694--704}.
\newblock


\bibitem[Veil et~al\mbox{.}(2011)]%
        {veil2011work}
\bibfield{author}{\bibinfo{person}{Shari~R Veil}, \bibinfo{person}{Tara Buehner}, {and} \bibinfo{person}{Michael~J Palenchar}.} \bibinfo{year}{2011}\natexlab{}.
\newblock \showarticletitle{A work-in-process literature review: Incorporating social media in risk and crisis communication}.
\newblock \bibinfo{journal}{\emph{Journal of contingencies and crisis management}} \bibinfo{volume}{19}, \bibinfo{number}{2} (\bibinfo{year}{2011}), \bibinfo{pages}{110--122}.
\newblock


\bibitem[Vieweg et~al\mbox{.}(2010)]%
        {vieweg2010microblogging}
\bibfield{author}{\bibinfo{person}{Sarah Vieweg}, \bibinfo{person}{Amanda~L Hughes}, \bibinfo{person}{Kate Starbird}, {and} \bibinfo{person}{Leysia Palen}.} \bibinfo{year}{2010}\natexlab{}.
\newblock \showarticletitle{Microblogging during two natural hazards events: what twitter may contribute to situational awareness}. In \bibinfo{booktitle}{\emph{Proceedings of the SIGCHI conference on human factors in computing systems}}. \bibinfo{pages}{1079--1088}.
\newblock


\bibitem[vincentclaes(2022)]%
        {ViT-indoor}
\bibfield{author}{\bibinfo{person}{vincentclaes}.} \bibinfo{year}{2022}\natexlab{}.
\newblock \bibinfo{title}{mit-indoor-scenes (ViT)}.
\newblock \bibinfo{howpublished}{\url{https://huggingface.co/vincentclaes/mit-indoor-scenes}}.
\newblock
\newblock
\shownote{Accessed: 2023-07-01}.


\bibitem[Wang et~al\mbox{.}(2020b)]%
        {wang2020concerns}
\bibfield{author}{\bibinfo{person}{Junze Wang}, \bibinfo{person}{Ying Zhou}, \bibinfo{person}{Wei Zhang}, \bibinfo{person}{Richard Evans}, {and} \bibinfo{person}{Chengyan Zhu}.} \bibinfo{year}{2020}\natexlab{b}.
\newblock \showarticletitle{Concerns expressed by Chinese social media users during the COVID-19 pandemic: content analysis of Sina Weibo microblogging data}.
\newblock \bibinfo{journal}{\emph{Journal of medical Internet research}} \bibinfo{volume}{22}, \bibinfo{number}{11} (\bibinfo{year}{2020}), \bibinfo{pages}{e22152}.
\newblock


\bibitem[Wang et~al\mbox{.}(2020a)]%
        {wang2020covid}
\bibfield{author}{\bibinfo{person}{Tianyi Wang}, \bibinfo{person}{Ke Lu}, \bibinfo{person}{Kam~Pui Chow}, {and} \bibinfo{person}{Qing Zhu}.} \bibinfo{year}{2020}\natexlab{a}.
\newblock \showarticletitle{COVID-19 sensing: negative sentiment analysis on social media in China via BERT model}.
\newblock \bibinfo{journal}{\emph{Ieee Access}}  \bibinfo{volume}{8} (\bibinfo{year}{2020}), \bibinfo{pages}{138162--138169}.
\newblock


\bibitem[Wang et~al\mbox{.}(2019)]%
        {wang2019emotional}
\bibfield{author}{\bibinfo{person}{Yun Wang}, \bibinfo{person}{Adrien Segal}, \bibinfo{person}{Roberta Klatzky}, \bibinfo{person}{Daniel~F Keefe}, \bibinfo{person}{Petra Isenberg}, \bibinfo{person}{J{\"o}rn Hurtienne}, \bibinfo{person}{Eva Hornecker}, \bibinfo{person}{Tim Dwyer}, {and} \bibinfo{person}{Stephen Barrass}.} \bibinfo{year}{2019}\natexlab{}.
\newblock \showarticletitle{An emotional response to the value of visualization}.
\newblock \bibinfo{journal}{\emph{IEEE computer graphics and applications}} \bibinfo{volume}{39}, \bibinfo{number}{5} (\bibinfo{year}{2019}), \bibinfo{pages}{8--17}.
\newblock


\bibitem[Website(2021)]%
        {localOutbreakChina}
\bibfield{author}{\bibinfo{person}{Chinese~Government Website}.} \bibinfo{year}{2021}\natexlab{}.
\newblock \bibinfo{title}{The COVID-19 pandemic is in a state of local outbreaks and sporadic distribution in many places (in Chinese)}.
\newblock \bibinfo{howpublished}{\url{http://www.gov.cn/xinwen/2021-01/13/content_5579636.htm}}.
\newblock
\newblock
\shownote{Accessed: 2022-09-14}.


\bibitem[Weibo(2020)]%
        {WeiboVisibility}
\bibfield{author}{\bibinfo{person}{Weibo}.} \bibinfo{year}{2020}\natexlab{}.
\newblock \bibinfo{title}{Adjustment of Weibo Visibility Range Setting (in Chinese)}.
\newblock \bibinfo{howpublished}{\url{https://weibo.com/ttarticle/p/show?id=2309404485675293474819}}.
\newblock
\newblock
\shownote{Accessed: 2023-06-20}.


\bibitem[Welhausen(2015)]%
        {welhausen2015visualizing}
\bibfield{author}{\bibinfo{person}{Candice~A Welhausen}.} \bibinfo{year}{2015}\natexlab{}.
\newblock \showarticletitle{Visualizing a non-pandemic: Considerations for communicating public health risks in intercultural contexts}.
\newblock \bibinfo{journal}{\emph{Technical Communication}} \bibinfo{volume}{62}, \bibinfo{number}{4} (\bibinfo{year}{2015}), \bibinfo{pages}{244--257}.
\newblock


\bibitem[Wikipedia(2022)]%
        {XianlocalOutbreakWiki}
\bibfield{author}{\bibinfo{person}{Wikipedia}.} \bibinfo{year}{2022}\natexlab{}.
\newblock \bibinfo{title}{2021-2022 Xi'an COVID-19 outbreak (in Chinese)}.
\newblock \bibinfo{howpublished}{\url{https://zh.wikipedia.org/zh-cn/2021\%EF\%BC\%8D2022\%E5\%B9\%B4\%E8\%A5\%BF\%E5\%AE\%89\%E5\%B8\%822019\%E5\%86\%A0\%E7\%8A\%B6\%E7\%97\%85\%E6\%AF\%92\%E7\%97\%85\%E8\%81\%9A\%E9\%9B\%86\%E6\%80\%A7\%E7\%96\%AB\%E6\%83\%85}}.
\newblock
\newblock
\shownote{Accessed: 2022-09-14}.


\bibitem[Williams et~al\mbox{.}(2020)]%
        {williams2020images}
\bibfield{author}{\bibinfo{person}{Nora~Webb Williams}, \bibinfo{person}{Andreu Casas}, {and} \bibinfo{person}{John~D Wilkerson}.} \bibinfo{year}{2020}\natexlab{}.
\newblock \bibinfo{booktitle}{\emph{Images as data for social science research: An introduction to convolutional neural nets for image classification}}.
\newblock \bibinfo{publisher}{Cambridge University Press}.
\newblock


\bibitem[WIlliams and Dienes(2021)]%
        {williams2021variant}
\bibfield{author}{\bibinfo{person}{Simon~Nicholas WIlliams} {and} \bibinfo{person}{Kimberly Dienes}.} \bibinfo{year}{2021}\natexlab{}.
\newblock \showarticletitle{‘Variant fatigue’? Public attitudes to COVID-19 18 months into the pandemic: A qualitative study}.
\newblock  (\bibinfo{year}{2021}).
\newblock


\bibitem[Wu et~al\mbox{.}(2021)]%
        {wu2021characterizing}
\bibfield{author}{\bibinfo{person}{Jiang Wu}, \bibinfo{person}{Kaili Wang}, \bibinfo{person}{Chaocheng He}, \bibinfo{person}{Xiao Huang}, {and} \bibinfo{person}{Ke Dong}.} \bibinfo{year}{2021}\natexlab{}.
\newblock \showarticletitle{Characterizing the patterns of China's policies against COVID-19: A bibliometric study}.
\newblock \bibinfo{journal}{\emph{Information Processing \& Management}} \bibinfo{volume}{58}, \bibinfo{number}{4} (\bibinfo{year}{2021}), \bibinfo{pages}{102562}.
\newblock


\bibitem[Xu et~al\mbox{.}(2023)]%
        {xu2023multimodal}
\bibfield{author}{\bibinfo{person}{Peng Xu}, \bibinfo{person}{Xiatian Zhu}, {and} \bibinfo{person}{David~A Clifton}.} \bibinfo{year}{2023}\natexlab{}.
\newblock \showarticletitle{Multimodal learning with transformers: A survey}.
\newblock \bibinfo{journal}{\emph{IEEE Transactions on Pattern Analysis and Machine Intelligence}} (\bibinfo{year}{2023}).
\newblock


\bibitem[Xu et~al\mbox{.}(2020)]%
        {xu2020characterizing}
\bibfield{author}{\bibinfo{person}{Qing Xu}, \bibinfo{person}{Ziyi Shen}, \bibinfo{person}{Neal Shah}, \bibinfo{person}{Raphael Cuomo}, \bibinfo{person}{Mingxiang Cai}, \bibinfo{person}{Matthew Brown}, \bibinfo{person}{Jiawei Li}, \bibinfo{person}{Tim Mackey}, {et~al\mbox{.}}} \bibinfo{year}{2020}\natexlab{}.
\newblock \showarticletitle{Characterizing Weibo social media posts from Wuhan, China during the early stages of the COVID-19 pandemic: qualitative content analysis}.
\newblock \bibinfo{journal}{\emph{JMIR public health and surveillance}} \bibinfo{volume}{6}, \bibinfo{number}{4} (\bibinfo{year}{2020}), \bibinfo{pages}{e24125}.
\newblock


\bibitem[Yang et~al\mbox{.}(2021)]%
        {yang2021know}
\bibfield{author}{\bibinfo{person}{Wenjie Yang}, \bibinfo{person}{Sitong Wang}, \bibinfo{person}{Zhenhui Peng}, \bibinfo{person}{Chuhan Shi}, \bibinfo{person}{Xiaojuan Ma}, {and} \bibinfo{person}{Diyi Yang}.} \bibinfo{year}{2021}\natexlab{}.
\newblock \showarticletitle{Know it to Defeat it: Exploring Health Rumor Characteristics and Debunking Efforts on Chinese Social Media during COVID-19 Crisis}.
\newblock \bibinfo{journal}{\emph{arXiv preprint arXiv:2109.12372}} (\bibinfo{year}{2021}).
\newblock


\bibitem[Yang et~al\mbox{.}(2022)]%
        {yang2022save}
\bibfield{author}{\bibinfo{person}{Wenjie Yang}, \bibinfo{person}{Zhiyang Wu}, \bibinfo{person}{Nga~Yiu Mok}, {and} \bibinfo{person}{Xiaojuan Ma}.} \bibinfo{year}{2022}\natexlab{}.
\newblock \showarticletitle{How to Save Lives with Microblogs? Lessons From the Usage of Weibo for Requests for Medical Assistance During COVID-19}. In \bibinfo{booktitle}{\emph{CHI Conference on Human Factors in Computing Systems}}. \bibinfo{pages}{1--18}.
\newblock


\bibitem[Yang et~al\mbox{.}(2017)]%
        {yang2017harvey}
\bibfield{author}{\bibinfo{person}{Zhou Yang}, \bibinfo{person}{Long~Hoang Nguyen}, \bibinfo{person}{Joshua Stuve}, \bibinfo{person}{Guofeng Cao}, {and} \bibinfo{person}{Fang Jin}.} \bibinfo{year}{2017}\natexlab{}.
\newblock \showarticletitle{Harvey flooding rescue in social media}. In \bibinfo{booktitle}{\emph{2017 IEEE International Conference on Big Data (Big Data)}}. IEEE, \bibinfo{pages}{2177--2185}.
\newblock


\bibitem[Yi et~al\mbox{.}(2022)]%
        {yi2022depicting}
\bibfield{author}{\bibinfo{person}{Jingjing Yi}, \bibinfo{person}{Jiayu Gina~Qu}, {and} \bibinfo{person}{Wanjiang~Jacob Zhang}.} \bibinfo{year}{2022}\natexlab{}.
\newblock \showarticletitle{Depicting the Emotion Flow: Super-Spreaders of Emotional Messages on Weibo During the COVID-19 Pandemic}.
\newblock \bibinfo{journal}{\emph{Social Media+ Society}} \bibinfo{volume}{8}, \bibinfo{number}{1} (\bibinfo{year}{2022}), \bibinfo{pages}{20563051221084950}.
\newblock


\bibitem[Zade et~al\mbox{.}(2018)]%
        {zade2018situational}
\bibfield{author}{\bibinfo{person}{Himanshu Zade}, \bibinfo{person}{Kushal Shah}, \bibinfo{person}{Vaibhavi Rangarajan}, \bibinfo{person}{Priyanka Kshirsagar}, \bibinfo{person}{Muhammad Imran}, {and} \bibinfo{person}{Kate Starbird}.} \bibinfo{year}{2018}\natexlab{}.
\newblock \showarticletitle{From situational awareness to actionability: Towards improving the utility of social media data for crisis response}.
\newblock \bibinfo{journal}{\emph{Proceedings of the ACM on human-computer interaction}} \bibinfo{volume}{2}, \bibinfo{number}{CSCW} (\bibinfo{year}{2018}), \bibinfo{pages}{1--18}.
\newblock


\bibitem[Zarocostas(2020)]%
        {zarocostas2020fight}
\bibfield{author}{\bibinfo{person}{John Zarocostas}.} \bibinfo{year}{2020}\natexlab{}.
\newblock \showarticletitle{How to fight an infodemic}.
\newblock \bibinfo{journal}{\emph{The lancet}} \bibinfo{volume}{395}, \bibinfo{number}{10225} (\bibinfo{year}{2020}), \bibinfo{pages}{676}.
\newblock


\bibitem[Zerman(1995)]%
        {zerman1995crisis}
\bibfield{author}{\bibinfo{person}{David Zerman}.} \bibinfo{year}{1995}\natexlab{}.
\newblock \showarticletitle{Crisis communication: Managing the mass media}.
\newblock \bibinfo{journal}{\emph{Information Management \& Computer Security}} (\bibinfo{year}{1995}).
\newblock


\bibitem[Zhang and Peng(2021)]%
        {zhang2021image}
\bibfield{author}{\bibinfo{person}{Han Zhang} {and} \bibinfo{person}{Yilang Peng}.} \bibinfo{year}{2021}\natexlab{}.
\newblock \showarticletitle{Image clustering: An unsupervised approach to categorize visual data in social science research}.
\newblock \bibinfo{journal}{\emph{Sociological Methods \& Research}} (\bibinfo{year}{2021}), \bibinfo{pages}{00491241221082603}.
\newblock


\bibitem[Zhang and Rui(2021)]%
        {XianOutbreakSCMP}
\bibfield{author}{\bibinfo{person}{Phoebe Zhang} {and} \bibinfo{person}{Guo Rui}.} \bibinfo{year}{2021}\natexlab{}.
\newblock \bibinfo{title}{Chinese Terracotta Warriors city Xian in lockdown as Covid-19 outbreak grows}.
\newblock \bibinfo{howpublished}{\url{https://www.scmp.com/news/china/science/article/3160816/chinese-terracotta-warriors-city-xian-lockdown-covid-19-outbreak}}.
\newblock
\newblock
\shownote{Accessed: 2022-11-28}.


\bibitem[Zhang et~al\mbox{.}(2021)]%
        {zhang2021mapping}
\bibfield{author}{\bibinfo{person}{Yixuan Zhang}, \bibinfo{person}{Yifan Sun}, \bibinfo{person}{Lace Padilla}, \bibinfo{person}{Sumit Barua}, \bibinfo{person}{Enrico Bertini}, {and} \bibinfo{person}{Andrea~G Parker}.} \bibinfo{year}{2021}\natexlab{}.
\newblock \showarticletitle{Mapping the landscape of covid-19 crisis visualizations}. In \bibinfo{booktitle}{\emph{Proceedings of the 2021 CHI Conference on Human Factors in Computing Systems}}. \bibinfo{pages}{1--23}.
\newblock


\bibitem[Zhao et~al\mbox{.}(2020)]%
        {zhao2020online}
\bibfield{author}{\bibinfo{person}{Xiaoman Zhao}, \bibinfo{person}{Ju Fan}, \bibinfo{person}{Iccha Basnyat}, \bibinfo{person}{Baijing Hu}, {et~al\mbox{.}}} \bibinfo{year}{2020}\natexlab{}.
\newblock \showarticletitle{Online health information seeking using “\# COVID-19 patient seeking help” on Weibo in Wuhan, China: descriptive study}.
\newblock \bibinfo{journal}{\emph{Journal of Medical Internet Research}} \bibinfo{volume}{22}, \bibinfo{number}{10} (\bibinfo{year}{2020}), \bibinfo{pages}{e22910}.
\newblock


\bibitem[Zhou et~al\mbox{.}(2017)]%
        {zhou2017places}
\bibfield{author}{\bibinfo{person}{Bolei Zhou}, \bibinfo{person}{Agata Lapedriza}, \bibinfo{person}{Aditya Khosla}, \bibinfo{person}{Aude Oliva}, {and} \bibinfo{person}{Antonio Torralba}.} \bibinfo{year}{2017}\natexlab{}.
\newblock \showarticletitle{Places: A 10 million image database for scene recognition}.
\newblock \bibinfo{journal}{\emph{IEEE transactions on pattern analysis and machine intelligence}} \bibinfo{volume}{40}, \bibinfo{number}{6} (\bibinfo{year}{2017}), \bibinfo{pages}{1452--1464}.
\newblock


\end{thebibliography}

\appendix

\section{Image Clustering Model Performance based on Within-cluster Consistency}\label{ModelPerformance}

\begin{table}[h]
\centering
\caption{Image Clustering Model Performance based on Within-cluster Consistency.}
\label{tab:model-performance}
\small
\begin{tabular}{p{2cm}|p{3.3cm}|p{3.3cm}|p{1.5cm}|p{2cm}}
\hline
\textbf{Experiment} & \textbf{Feature Extraction Model} & \textbf{Clustering Model} & \textbf{Splitting and Merging} & \textbf{Averaged Within-Cluster Consistency} \\ 

\hline
Determining Feature Extraction Model & bag-of-visual-words~\cite{sivic2003video} & K-Means~\cite{lloyd1982least} & No & 0.64 \\ 

\cline{2-5} &VGG16~\cite{simonyan2014very} & K-Means~\cite{lloyd1982least} & No & 0.71 \\
\cline{2-5} & VGG19~\cite{simonyan2014very} & K-Means~\cite{lloyd1982least} & No & 0.70 \\ 
\cline{2-5} & ResNet~\cite{he2016deep} & K-Means~\cite{lloyd1982least} & No & 0.72 \\ 
\cline{2-5} & MobileNet~\cite{howard2017mobilenets} & K-Means~\cite{lloyd1982least} & No & 0.67 \\ 
\cline{2-5} & MobileNetV2~\cite{sandler2018mobilenetv2} & K-Means~\cite{lloyd1982least} & No & 0.72 \\

 \hline
Determining Clustering Model & MobileNetV2~\cite{sandler2018mobilenetv2} & K-Means~\cite{lloyd1982least} & No & 0.72 \\ 
\cline{2-5} & MobileNetV2~\cite{sandler2018mobilenetv2} & DBSCAN~\cite{khan2014dbscan} & No & 0.65 \\ \cline{2-5} & MobileNetV2~\cite{sandler2018mobilenetv2} & Gaussian Mixture Model~\cite{reynolds2009gaussian} & No & 0.69 \\ \hline

Final Model & \textbf{MobileNetV2}~\cite{sandler2018mobilenetv2} & \textbf{K-Means}~\cite{lloyd1982least} & \textbf{Yes} & \textbf{0.80} \\ \cline{2-5} \hline
\end{tabular}
\end{table}

\end{document}